%% file: canonicalization.tex
\documentclass[12pt]{article}
\usepackage[T1]{fontenc}
\usepackage{textcomp}
\usepackage{algpseudocode}
\usepackage[ruled,section]{algorithm}
\usepackage{amssymb,amsmath}
\usepackage{datetime}
\usepackage{graphicx}
\usepackage[svgnames]{xcolor}
\usepackage{tikzexternal}
\usepackage[font=it, labelfont={bf, up}]{caption}
\usepackage[labelformat=simple]{subcaption}
\usepackage{enumitem}
\usepackage{booktabs}
\usepackage{bbm}
\usepackage[vcentermath]{youngtab}
\usepackage[only,llbracket,rrbracket]{stmaryrd}

\usepackage[toc]{appendix}

\usepackage{datetime}
\usepackage[
		colorlinks=false,
		linkcolor=darkblue,  
		urlcolor=blue,    
		filecolor=blue,     
		citecolor=red,
		linktocpage=true,
		pdfstartview=FitV,
		bookmarksopen=true    
	]{hyperref}

\usepackage{benmacros}

\numberwithin{equation}{section}

\newcommand{\access}[1]{\llbracket #1 \rrbracket}

\newcommand{\software}[1]{\textsf{\textit{\textmd{#1}}}}
\newcommand{\code}[1]{\texttt{\textup{\textmd{#1}}}}
\newcommand{\variable}[1]{\textsf{\textup{\textmd{#1}}}}
\newcommand{\function}[2]{\textsc{#1}\big(#2\big)}

\renewcommand{\textproc}[1]{\textsc{\textmd{#1}}}

\newcommand{\set}[1]{\{{#1}\}}
\newcommand{\seq}[1]{\langle {#1} \rangle}

\newcommand{\free}{\varnothing}
\newcommand{\none}{\varnothing}
\newcommand{\component}{\mathord{\sim}}
\newcommand{\sdummy}{\mathord{+}}
\newcommand{\adummy}{\mathord{-}}
\newcommand{\ldummy}{\mathord{\bot}}
\newcommand{\udummy}{\mathord{\top}}

\newenvironment{definition}{}{}

\DeclareCaptionType{definition}[Definition]

\newcommand{\Input}{\textbf}
\let\Input\undefined

\newcommand{\Output}{\textbf}
\let\Output\undefined

\newcommand{\ForEach}{\textbf}
\let\ForEach\undefined

\newcommand{\InlineComment}{\textbf}
\let\InlineComment\undefined

\algnewcommand\algorithmicinput{\textbf{Input:}}
\algnewcommand\Input{\item[\algorithmicinput]}

\algnewcommand\algorithmicoutput{\textbf{Output:}}
\algnewcommand\Output{\item[\algorithmicoutput]}

\algnewcommand\algorithmicforeach{\textbf{for each}}
\algdef{S}[FOR]{ForEach}[1]{\algorithmicforeach\ #1\ \algorithmicdo}

\makeatletter
\algnewcommand{\InlineComment}[1]{\Statex \hskip\ALG@thistlm #1}
\makeatother

\newcommand{\Skip}{\textbf{next}}
\let\Skip\undefined

\algnewcommand\algorithmicskip{\textbf{skip}}
\algnewcommand\Skip{\algorithmicskip{} }

\tikzsetexternalprefix{graphics/}
\begin{document}  
%%%%%%%%%%%%%%%%%%%%%%%%%%%%%%

% surely there is a title page style that does this better...

%\begin{titlepage}
 
\bigskip
\bigskip
\bigskip
\bigskip
\begin{center} 
{\Large \bf  Faster Tensor Canonicalization}

\medskip
\bigskip
\bigskip
{\bf Benjamin~E.~Niehoff} \\
\bigskip
Institute for Theoretical Physics, KU Leuven\\
Celestijnenlaan 200D, B-3001 Leuven, Belgium\\
\bigskip
\bigskip
{\rm ben.niehoff@kuleuven.be} \\

%\bigskip
%\textcolor{red}{\today}

\bigskip
\bigskip 

\end{center}

\begin{abstract}

\noindent The Butler-Portugal algorithm for obtaining the canonical form of a tensor expression with respect to slot symmetries and dummy-index renaming suffers, in certain cases with a high degree of symmetry, from $O(n!)$ explosion in both computation time and memory.  We present a modified algorithm which alleviates this problem in the most common cases---tensor expressions with subsets of indices which are totally symmetric or totally antisymmetric---in polynomial time.  We also present an implementation of the label-renaming mechanism which improves upon that of the original Butler-Portugal algorithm, thus providing a significant speed increase for the average case as well as the highly-symmetric special case.  The worst-case behavior remains $O(n!)$, although it occurs in more limited situations unlikely to appear in actual computations.  We comment on possible strategies to take if the nature of a computation should make these situations more likely.

\end{abstract}

%\end{titlepage}

%%%%%%%%%%%%%%%%%%%%%%%%%%%%%%%%%%%%%

\tableofcontents

%%%%%%%%%%%%%%
\section{Introduction}
%%%%%%%%%%%%%%

Computer algebra systems have become quite powerful at simplifying expressions with polynomials, integrals, derivatives, special functions, etc.  All of these fall under the category of what we will call \emph{scalar} algebra, meaning that they deal with objects that are functions from $\RR \to \RR$ or $\CC \to \CC$.  A field that is still quite in development is the extension of this computational power to \emph{tensor} algebra, by which we mean the manipulation of objects with ``indices'' such as vectors $v^i$, matrices $M_{ij}$, and higher-rank tensors $T^i{}_{jk\ell}$.  Such \emph{indexed objects} are a way to denote multidimensional arrays (and thus represent objects of \emph{multilinear algebra}).  The indices refer to the components of the object in a basis of some vector space like $\RR^n$; for example, one can write:
\begin{equation} \label{contraction demo}
\mathbf{v} = v^i \, \mathbf{e}_i \equiv v^1 \, \mathbf{e}_1 + v^2 \, \mathbf{e}_2 + \dotsc + v^n \, \mathbf{e}_n,
\end{equation}
where $\mathbf{v} \in \RR^n$ is a vector and $\set{\mathbf{e}_i}$ is a given basis.\footnotemark{}  Where multiple indices occur on a single tensor, these refer to the component along the tensor product of the corresponding basis elements, as in $\mathbf{T} = T^i{}_{jk\ell} \, \mathbf{e}_i \otimes \mathbf{e}^j \otimes \mathbf{e}^k \otimes \mathbf{e}^\ell$.  Alternatively, we can choose to think of indices \emph{abstractly}, referring merely to the \emph{structure} of a tensor and its contractions, rather than literally referring to the components of the underlying multidimensional array \cite{Penrose:1987uia}.

\footnotetext{\,``Upstairs'' (superscript) and ``downstairs'' (subscript) indices are to be distinguished, and when the same index is repeated both upstairs and downstairs, it is meant to be \emph{contracted} or summed over, as shown in \eqref{contraction demo}.}

In either case, whether indices are thought of as literal or abstract, the algebra of indexed objects is as follows:  We can think of each index as a \emph{slot} which has been filled with a \emph{label} such as $i,j,k,\ell$, etc.  The slots are, for our purposes, an ordered sequence of ``blanks'' in a tensor expression into which labels can go,
\begin{equation}
T^\circ{}_{\circ \circ \circ},
\end{equation}
whereas the labels $i,j,k,\ell$ are a means to ``name'' these slots, and either indicate that a pair of slots have been contracted (in the case that a label is repeated), or indicate that a slot is \emph{free}; that is, available to be contracted.\footnotemark{}  The positional order of the labels has meaning, and can be used to effectively indicate notions like the transpose of matrices:
\begin{equation}
(M^\top)_{ij} \equiv M_{ji}.
\end{equation}
The free labels must be ``balanced'' in any sum or equation in order for it to be a valid mathematical sentence.  For repeated labels which indicate a contraction, one may use any available symbol; one calls these ``dummy'' labels, as the particular symbol used is unimportant.  A final use of labels is one we will call \emph{component labels}, which refer to particular components of a tensor (in a particular concrete basis); for example,
\begin{equation}
T^1{}_{2 2 3}
\end{equation}
refers to the component of $\mathbf{T}$ along the tensor product basis element $\mathbf{e}_1 \otimes \mathbf{e}^2 \otimes \mathbf{e}^2 \otimes \mathbf{e}^3$.

\footnotetext{Thus in $T^k{}_{i k j}$, the first and third slots are contracted, while the second and fourth slots are free.  We will frequently refer to the slots by their names, saying that $i,j$ are free, while $k$ is contracted.  Note that a given label may only be repeated twice, once upstairs and once downstairs, to indicate a single contraction.}

The task of a \emph{tensor} computer algebra system---that is, to manipulate algebraic expressions with indexed objects---is complicated by two types of symmetries: \emph{re-labelling} symmetries, which we have already hinted at, and \emph{intrinsic} symmetries, which we will define in a moment.  Re-labelling symmetries arise when there are dummy labels present, as illustrated by
\begin{equation} \label{dummy demo}
T_{ab}{}^{ab} = T_{bc}{}^{bc}.
\end{equation}
Since the labels used to indicate each contraction are unimportant, the two sides of \eqref{dummy demo} are equal, even though they do not look the same.  In order to see that they are equal, the system must either \emph{canonicalize} the sequence of dummy labels used (thus putting each monomial into a standard form), or use some sort of \emph{isomorphism}-detection algorithm (in order to match one monomial to the other).

With only dummy-relabelling to contend with, this problem would be rather simple.  However, a tensor can also have \emph{intrinsic} symmetries, which are symmetries that involve permuting the tensor's \emph{slots}.  For example, a symmetric matrix $M_{ij} = M_{ji}$ is symmetric irrespective of the particular labels used; one can think of this symmetry as a property of $\mathbf{M}$ itself.  Intrinsic symmetries come in two distinct forms, which we will call \emph{mono-term} symmetries and \emph{multi-term} symmetries.  Some simple examples of mono-term symmetries are the slot-exchange symmetries of the Riemann curvature tensor,
\begin{equation}
R_{abcd} = - R_{bacd}, \qquad R_{abcd} = R_{cdab},
\end{equation}
whereas an example of a multi-term symmetry is the algebraic Bianchi identity:
\begin{equation} \label{bianchi}
R_{abcd} + R_{bcad} + R_{cabd} = 0.
\end{equation}

In the context of a calculation (that is, when labels have been placed in the slots), the intrinsic symmetries of a tensor monomial can interact with its re-labelling symmetries in a complicated way.  For example, a tensor $T_{abcd}$ which has \emph{no} intrinsic symmetries may acquire the \emph{appearance} of a slot symmetry when contracted,
\begin{equation}
T_{ab}{}^{ab} = T_{ba}{}^{ba},
\end{equation}
because we are free to exchange dummy labels.  Similarly, if a tensor $A_{ab} = -A_{ba}$ is intrinsically antisymmetric, then placing \emph{component} labels into its slots can have non-trivial results:
\begin{equation}
A_{11} = A_{22} = 0, \qquad A_{12} = - A_{21}.
\end{equation}
Here we can think of the first equalities as arising from the contradiction between the antisymmetry of $A$ and the symmetry of exchanging identical component labels.  The same sort of situation can happen with dummy labels, if say we take $M^{ab} A_{ab}$ where $M$ is symmetric and $A$ is antisymmetric.  In general, all of these complications may be mixed within a single expression: a tensor monomial may have intrinsic slot symmetries (of both single- and multi-term types), some free indices, some dummy indices, and some concrete component labels, all at the same time.

Thus any CAS with the power to simplify tensor expressions is faced with a challenge to efficiently manipulate index symmetries.  The greatest amount of progress has been made on mono-term symmetries, whereas multi-term symmetries are considerably more difficult to manage.  For mono-term symmetries, there are two main approaches:  index canonicalization and index isomorphism:

The \emph{index canonicalization} approach seeks to reduce each monomial in an expression to its canonical form, which is defined as the \emph{least} possible arrangement of its indices, modulo slot symmetries and label exchanges, where \emph{least} is defined by lexicographical order.  Simplification can then proceed as it does in a standard (i.e. non-tensorial) CAS, as all objects which are equal have been reduced to the same form.  While processing a given monomial, the same algorithm can easily detect if the monomial is itself equal to zero due to its own internal symmetries (say, by having a pair of symmetric slots which are contracted against a pair of antisymmetric ones); thus, the canonicalization approach can also pre-emptively simplify some parts of expressions.  The canonicalization strategy is well-suited for use as a plugin or package for a standard CAS such as \software{Maple} or \software{Mathematica}, since it reduces terms to a form which can then be manipulated by pre-existing routines.

By contrast, the \emph{index isomorphism} approach works by attempting to \emph{match} the various tensor monomials of an expression onto each other by constructing a bijection between their free indices (subject to symmetries).  With every successful bijection, the expression can be simplified by combining the corresponding terms.  This approach has two advantages:  First, the algorithm to make a single match could conceivably run faster than a canonicalization algorithm, since one has a definite goal in mind rather than needing to search the space of equivalent configurations for the ``least'' one.\footnote{At the time of writing, however, we do not know of any software which actually implements such an algorithm which is efficient in the presence of slot symmetries.}  Second, a matching algorithm can treat compound expressions such as $T^{ac}(S^b{}_{bcd} + V_{cd})$ as a single monomial $(\cdot)^a{}_d$, and thus simplify the expression tree without flattening it; whereas the canonical form of such an expression is not obvious, and may be ill-defined.\footnotemark{}  However, the disadvantages of the isomorphism approach may outweigh the advantages:  First, while a single matching test might run faster than canonicalization, one still has to make up to $O(n^2)$ matching tests, vs. $O(n)$ canonicalizations (where $n$ here is the number of terms in an expression).  Second, a matching algorithm alone has no good way of detecting whether a \emph{single} monomial vanishes due to its own symmetry, so in the end one has to run a separate test for that.  And finally, attempting to use the index isomorphism approach within an existing (scalar) CAS is cumbersome and requires circumventing what that system already knows about expression simplification; it is an approach better suited to standalone software.

\footnotetext{However, the notion of canonical-ness of compound expressions has not been explored in the literature, nor implemented in any software package to our knowledge, and it may well be that one can come up with a workable definition!}

There are many implementations of the index canonicalization strategy in currently-available CAS software.\footnotemark{}  Some of the more popular ones include the open-source \software{Canon} package \cite{2004CoPhC.157..173M} for \software{Maple}, which is also included in the \software{GRtensorII} \cite{GRtensorII} suite of packages; the open-source \software{xPerm} package \cite{MartinGarcia2008597} of the package suite \software{xAct} \cite{citeulike:13127953} for \software{Mathematica}; the standalone open-source tensor CAS \software{Cadabra} \cite{Peeters:2006kp} (which in fact uses the same code from \software{xPerm}); and the tensor subsystem of the standalone open-source Python CAS \software{SymPy} \cite{10.7717/peerj-cs.103}.  Since Version 9, \software{Mathematica} also includes a built-in implementation of tensor canonicalization, which is not open-source, but we can nevertheless draw conclusions about its algorithm from testing.

\footnotetext{There is currently only one software package which uses the index isomorphism strategy: the standalone tensor CAS \software{Redberry} \cite{Bolotin:2013qgr}.}

The most efficient algorithm in the literature for index canonicalization is the Butler-Portugal algorithm \cite{DBLP:Butler91,2002IJMPC..13..859M}, which is used in all of the above-mentioned software packages, and has been shown to have polynomial complexity (in the number of indices) in \emph{most} common cases.  However, in tensor calculus in higher dimensions (of particular interest to the high-energy theoretical physics community), there are certain types of simplification problems which thwart the Butler-Portugal algorithm and cause it to consume $O(n!)$ memory and take $O(n!)$ time.  These are problems where tensors with many indices and a high degree of symmetry are contracted with each other; thus the slots of these highly-symmetric tensors are filled with dummy indices and are subject to a high degree of re-labelling ambiguity.  The overlap of slot symmetries with re-labelling symmetries inhibits the algorithm's ability to deduce whether two arrangements of indices are actually different.

In this article we present a modification which improves upon the Butler-Portugal algorithm by automatically recognizing (in polynomial time) these worst-case situations, and adapting to avoid processing $O(n!)$ redundant steps.  Thus for tensor monomials with total symmetry or total antisymmetry over some subset(s) of slots, our improved algorithm can canonicalize them in polynomial time, even subject to index re-labelling symmetries.\footnotemark{}  In the course of implementing these modifications for these worst-case situations, we introduce a number of data structures which significantly increase the algorithm's average-case performance as well.

\footnotetext{We note that some attempt has been made at improving the canonicalization of (anti)symmetric tensors in \cite{sym:Backdahl}, although it is far from complete.}

Aside from (subsets of indices with) total symmetry or total antisymmetry, there are still situations that can lead to $O(n!)$ behavior (such as a symmetric group which acts on \emph{pairs} of indices rather than individuals), and there is no efficient means to identify all potential sources of combinatorial explosion.  Thus our algorithm still has worst-case time and space complexities of $O(n!)$, although the cases in which this behavior occurs are much less frequent in practical calculations.  However, in the event that one \emph{does} expect these situations to occur, we discuss some possible strategies based upon the concepts introduced in this work.

This paper is structured as follows: in Section~\ref{sec:symmetries} we review tensor symmetries and permutation groups, and precisely define the problem to be solved; in Section~\ref{sec:bp algorithm}, we review the Butler-Portugal algorithm; in Section~\ref{sec:improved} we present the improved algorithm; in Section~\ref{sec:testing} we give the results of performance testing on a few different types of tensor inputs; and in Section~\ref{sec:discussion} we discuss these results and potential further improvements.

%%%%%%%%%%%%%%
\section{Tensor symmetries and permutation groups}
\label{sec:symmetries}
%%%%%%%%%%%%%%

In this work we will focus on the mono-term symmetries of indexed objects and their relation to permutation groups.  A \emph{mono-term symmetry} of a tensor can be described by a \emph{signed permutation}\footnote{One could also consider promoting $\varepsilon$ to be a root of unity.} $(\varepsilon, \pi) \in \set{-1,1} \times S_n$, where $S_n$ is the symmetric group on the tensor's slots:
\begin{equation}
\label{monoterm-def}
T_{a_1 a_2 \ldots a_n} = \varepsilon \, T_{\pi(a_1 a_2 \ldots a_n)}.
\end{equation}
That is, the tensor returns to itself, up to a sign, after permuting its slots in some specific way.  This type of symmetry encompasses symmetric and antisymmetric matrices, and most of the examples discussed in the Introduction.\footnotemark{}

\footnotetext{By contrast, to describe a multi-term symmetry of a tensor (such as the one in the algebraic Bianchi identity \eqref{bianchi}) requires \emph{several} signed permutations $\set{(\varepsilon_1, \pi_1), (\varepsilon_2, \pi_2), \ldots, (\varepsilon_k, \pi_k)}$, and takes the form
	\begin{equation*}
	\label{multiterm-def}
	T_{a_1 a_2 \ldots a_n} - \varepsilon_1 \, T_{\pi_1(a_1 a_2 \ldots a_n)} - \varepsilon_2 \, T_{\pi_2(a_1 a_2 \ldots a_n)} - \ldots - \varepsilon_k \, T_{\pi_k(a_1 a_2 \ldots a_n)} = 0.
	\end{equation*}
	Written in this way, we see that \eqref{monoterm-def} arises as a special case.  Dealing with multi-term symmetries efficiently is an important open problem in the symbolic computation literature, which we will not address here.  One approach used in the \software{Cadabra} system \cite{Peeters:2006kp} is to enforce the multi-term symmetry by projection onto Young tableaux, but this has the tendency to introduce many extra terms into an expression.  Another approach is described in \cite{2017arXiv170108487L} and is based on graph isomorphism.}

%%%%%%%%%%%%%%
\subsection{Specification of the problem}
%%%%%%%%%%%%%%

For mono-term symmetries, one can address the canonicalization problem purely from the standpoint of permutation groups \cite{2002IJMPC..13..859M}, which are well-studied \cite{DBLP:Butler91,holt2005handbook,seress2003permutation}.  Thus using this well-honed tool, we will focus on the mono-term symmetries.

First let us formalize the definition of the problem.  It is useful first of all to think of a tensor monomial
\begin{equation}
\label{T-example}
T_{bdafce}
\end{equation}
as a \emph{map} from a space of slots into a space of labels, which therefore informs us which label should go into which slot.  We will refer to this map as a \emph{configuration} $g : \cS \to \cL$, where $\cS$ is the space of slots and $\cL$ is the space of labels.  The configuration corresponding to \eqref{T-example} can then be written as
\begin{equation}
\label{g-example}
g = \begin{pmatrix}
1 & 2 & 3 & 4 & 5 & 6 \\
b & d & a & f & c & e
\end{pmatrix}.
\end{equation}
We note that $g$ is essentially a permutation---if one instead replaces the labels $a$ through $f$ by numbers 1 to 6, then the notation \eqref{g-example} becomes the standard Cauchy two-line notation for a permutation map \cite{Cauchy1815}.  However, we will find it useful to maintain a distinction between the slot and label spaces.

If a tensor monomial has mono-term symmetries which permute the slots, we can think of these as acting on the first row of $g$.  Thus if we act first with a permutation (in cyclic notation) to exchange slots $(1,2)$, and then afterwards act with one to exchange $(2,3)$, we obtain
\begin{equation}
g' = \begin{pmatrix}
3 & 1 & 2 & 4 & 5 & 6 \\
b & d & a & f & c & e
\end{pmatrix} = 
\begin{pmatrix}
1 & 2 & 3 & 4 & 5 & 6 \\
d & a & b & f & c & e
\end{pmatrix} \quad \Rightarrow \quad T_{dabfce}.
\end{equation}
The slot symmetries of a tensor monomial form a group $S$ which acts on the slot space $\cS$.

If a tensor monomial has multiple repeated labels, such as when there are dummy pairs or concrete component labels, we may clarify what happens by treating the repeated labels as distinct.  We can then implement their ``sameness'' via symmetries which exchange the labels.  For example, we could have
\begin{equation}
T_{11ab}{}^{bc} = T_{1_1 1_2 a b_1}{}^{b_2 c},
\end{equation}
where we attach a subscript to some of the labels to distinguish identical ones\footnote{For dummy pairs we adopt the convention that the ``downstairs'' label comes first.  This is opposite the convention of \cite{2002IJMPC..13..859M,MartinGarcia2008597}.}.  Then the configuration could be given as
\begin{equation}
\label{label example}
\begin{pmatrix}
1 & 2 & 3 & 4 & 5 & 6 \\
1_1 & 1_2 & a & b_1 & b_2 & c
\end{pmatrix}.
\end{equation}
Due to the repetition of labels, even if $T$ has no slot symmetries, the following configurations are equivalent to \eqref{label example} (assuming the presence of a metric tensor):
\begin{equation}
\begin{pmatrix}
1 & 2 & 3 & 4 & 5 & 6 \\
1_2 & 1_1 & a & b_1 & b_2 & c
\end{pmatrix}, \qquad
\begin{pmatrix}
1 & 2 & 3 & 4 & 5 & 6 \\
1_1 & 1_2 & a & b_2 & b_1 & c
\end{pmatrix}, \qquad
\begin{pmatrix}
1 & 2 & 3 & 4 & 5 & 6 \\
1_2 & 1_1 & a & b_2 & b_1 & c
\end{pmatrix}.
\end{equation}
Thus we see that repeated labels of either type give us a symmetry group $L$ that acts on label space $\cL$.  These symmetries are not intrinsic to the tensor, but arise from the context; i.e., having placed labels in the slots of the tensor.

The mono-term symmetries of a tensor monomial then consist of the union of two distinct symmetry actions:  those symmetries, intrinsic to the definition of $T$, which act on the slots $\cS$; and those symmetries, arising from context, which act on the labels $\cL$.  Again treating the configuration $g$ as a permutation, the group of slot symmetries $S$ act from the right of $g$, and the group of label symmetries $L$ act from the left.  Thus the set of all equivalent configurations of the indices of a tensor monomial is the \emph{double coset} $L g S$, and the task of index canonicalization is precisely that of finding the \emph{canonical representative} of a double coset.

In all of this discussion thus far, we have ignored the fact that tensor symmetries are \emph{signed} permutations.  In fact, both groups $S$ and $L$ can have signed permutations; the ones in $L$ can come from ``antisymmetric metrics'', used in spinor formalisms in certain dimensions, which can introduce minus signs when raising/lowering contracted indices, e.g.
\begin{equation}
\psi^\alpha \chi_\alpha = - \psi_\alpha \chi^\alpha.
\end{equation}
This corresponds, in label space, to the signed permutation $-(\alpha_1, \alpha_2)$ that exchanges the two (distinguished) $\alpha$'s.  It is not difficult to expand our formalism to include signed permutations.  As was observed in \cite{MartinGarcia2008597}, the sign is just an element of a $\ZZ_2$ group, which can be represented as a permutation group.  One just needs to add an extra two columns at the end of $g$:
\begin{equation}
\label{g signed}
g = \begin{pmatrix}
1 & 2 & 3 & 4 & 5 & 6 & 7 & 8 \\
b & d & a & f & c & e & + & -
\end{pmatrix},
\end{equation}
and in label space, allocate two arbitrary symbols for tracking the order of the last two columns.  The action of the signed permutation $-(1,2)$ on \eqref{g signed} then gives
\begin{equation}
g' = \begin{pmatrix}
1 & 2 & 3 & 4 & 5 & 6 & 7 & 8 \\
d & b & a & f & c & e & - & +
\end{pmatrix},
\end{equation}
and thus, one can work out whether the sign of any expression has flipped by checking the last pair of columns.

We point out that $g$ has a fairly obvious computer implementation which makes clear our notions of left- and right-actions.  If we store the configuration of \eqref{g-example}, \eqref{g signed} in computer memory as the array\footnotemark{}
\begin{equation}
g = \seq{b, d, a, f, c, e, +, -},
\end{equation}
\footnotetext{Of course, to encode the symbols of the label alphabet, we should use numbers in the same range as the slot numbers, so this becomes $g = \seq{2,4,1,6,3,5,7,8}$.  So in the end, $g$ is truly just a permutation map.}
then the map $g$ can be implemented as array access $i \mapsto g \access{i}$, for $i \in \cS$.  One can easily extend this notion to the group multiplication of permutations, and thus for $\ell \in L$ and $s \in S$, one has schematically
\begin{equation}
g' = \ell \access{g \access{s}},
\end{equation}
which corresponds to the stated notions of left and right.\footnotemark{}  Finally, one of the main advantages of treating the configuration as a slots-to-labels map, rather than labels-to-slots, is that it becomes trivial to \emph{order} configurations lexicographically; for example one has that $\seq{b,a,f,c,e,d,+,-} \prec \seq{b,c,a,d,f,e,+,-}$, etc., for the lexicographical ordering $\prec$.  Our choice to append the sign columns to the end rather than the beginning is motived by this lexicographic ordering: if a set of configurations contains both $+g$ and $-g$, then these two configurations will be \emph{adjacent} when the set is sorted lexicographically.  Thus it becomes easy to detect when the only tensor consistent with a given set of index permutations is the zero tensor.

\footnotetext{Alternatively, if one wants to use cyclic notation for permutations, it is more convenient to work with the right-action $i \mapsto i^g$, in which case the notions of left and right are reversed: $g' = s \cdot g \cdot \ell$.}

%%%%%%%%%%%%%%
\section{The Butler-Portugal algorithm}
\label{sec:bp algorithm}
%%%%%%%%%%%%%%

In \cite{2002IJMPC..13..859M}, Manssur and Portugal adapt Butler's algorithm \cite{DBLP:Butler91} for finding the canonical representative of a double coset to the specific problem of simplifying tensor monomials with dummy indices.  Their main modification, which results in a highly-efficient algorithm for most tensor monomials in practice, is to take advantage of the simple and predictable nature of the label permutation group $L$ for dummy indices.  In \cite{MartinGarcia2008597}, Mart\'in-Garc\'ia extends the algorithm of \cite{2002IJMPC..13..859M} to include dummy indices drawn from multiple separate pools (in the case that a tensor has indices belonging to different vector bundles), as well as component labels.  The label permutation group $L$ retains a simple structure, which allows these modifications to run equally efficiently.

Like many algorithms in finite group theory, the Butler-Portugal algorithm makes use of the concepts of a \emph{base} and \emph{strong generating set} to deal with the permutation groups in an efficient way.  We refer the reader to Appendix~\ref{app:groups} for a brief presentation of these concepts.

%%%%%%%%%%%%%%
\subsubsection*{Notational conventions}
\label{sec:notation}
%%%%%%%%%%%%%%

Before delving into algorithms, we briefly mention some notational conventions.  We will use curly braces $\set{\ldots}$ to denote sets, in the strict sense of being unordered.  To denote arrays or lists, whose elements are ordered by their position, we use angle brackets $\seq{\ldots}$.  To denote array access, we use double square brackets; thus $g\access{i}$ is the $i$-th entry of $g$.  Double square brackets can also denote the action of a permutation, if their argument is itself a permutation map rather than a single number; thus $g\access{s}$ denotes the \emph{permutation} that results by acting with $g$ on $s$.  The convention for $\access{\cdot}$ is \emph{left} multiplication; whereas in cyclic notation, the equivalent action is \emph{right} multiplication.  Thus the array $g\access{s}$ is the same as the array corresponding to $s \cdot g$ in cyclic notation; that is the action on some element $i$ is ``first $s$, then $g$'':
\begin{equation}
g\access{s}\access{i} = g\access{s\access{i}} = i^{s \cdot g}.
\end{equation}

In the pseudocode of the algorithms we present, we will adopt a few conventions on typefaces.  Variables will be indicated by \variable{sans-serif} type, and arrays will be indicated by \variable{Sans-Serif-with-Caps}.  Function names will take \textproc{Small-Caps-Camel-Case}.

Finally, we will frequently refer to ``lexicographical ordering'', by which we mean an ordering $\prec$ in which all of the free indices come first, and then the component indices and dummy indices of various types.  For dummy indices, the ordering is defined such that they come in pairs, lower first and then upper.  Therefore the indices of the tensor $T^{ab}{}_{fgc}{}^c{}_d{}^d{}_e{}^e$ are in lexicographical order.

%%%%%%%%%%%%%%
\subsection{The label symmetry group}
%%%%%%%%%%%%%%

The observation of \cite{2002IJMPC..13..859M} which contributes most to the \emph{speed} of its algorithm is that the label symmetry group $L$ takes a simple form.  As an example, take the tensor monomial
\begin{equation}
T_a{}^a{}_b{}^b{}_c{}^c,
\end{equation}
whose indices come only in contracted pairs.  Acting on the label space $\cL = \set{a_1, a_2, b_1, b_2, c_1, c_2}$, we can immediately see that the label symmetry group is given by the signed permutations
\begin{equation}
\label{3 pairs}
L = \set{ +(1,2), \; +(3,4), \; +(5,6), \; +(1,3)(2,4), \; +(3,5)(4,6) },
\end{equation}
and moreover, this is a strong generating set with respect to the base $B = \seq{1,3,5}$.  More importantly, however, it is simple to \emph{re-order the base}:  If instead we want a strong generating set with respect to the base $B = \seq{3,1,5}$, then we need only \emph{conjugate} the set \eqref{3 pairs} by the signed permutation $+(1,3)(2,4)$, which obtains
\begin{equation}
L = \set{ +(3,4), \; +(1,2), \; +(5,6), \; +(1,3)(2,4), \; +(1,5)(2,6) }.
\end{equation}
Butler's algorithm \cite{DBLP:Butler91} for the canonical representative of a double coset relies on repeatedly changing the base order for one of the two subgroups.  The best known algorithms for an arbitrary base rearrangement on generic groups require $O(n^3)$ time \cite{brown1989new,Cooperman:1992:FCB:143242.143316}.  The conjugation operation, however, is much faster.  Portugal observed \cite{2002IJMPC..13..859M} that for tensor canonicalization, the fixed, simple structure of the label symmetry group always allows for a much-streamlined base-change-by-conjugation.

A similar situation occurs with component labels.  This time take the tensor monomial
\begin{equation}
T_{1111},
\end{equation}
with label space $\cL = \set{1_1, 1_2, 1_3, 1_4}$.  The label symmetry group is just the symmetric group on four elements, whose strong generating set we can easily write down:
\begin{equation}
L = \set{+(1,2), \; +(2,3), \; +(3,4)},
\end{equation}
with respect to the base $B = \seq{1,2,3,4}$.  Again, re-ordering the base can be done by conjugation, thus avoiding any more expensive group theory algorithms.

%%%%%%%%%%%%%%
\subsection{The slot symmetry group}
%%%%%%%%%%%%%%

The slot symmetry group $S$ is less well-behaved, because in principle it can be any permutation group.  However, Butler's algorithm to find the canonical double coset representative only requires changing the base order of \emph{one} of the subgroups.  Therefore we can leave the base of the slot symmetry group intact, and change only the base of the label symmetry group.

We will adopt a further convenience that the base of the slot symmetry group will be chosen to be a \emph{complete, position-ordered base}.  Thus if a tensor has 6 slots, the base chosen for its slot symmetry group will always be $B = \seq{1,2,3,4,5,6}$ (or, including the two columns for the sign, $B = \seq{1,2,3,4,5,6,7,8}$).  Typically this results in some redundancy; for example, if a slot symmetry group only moves points $\set{1,4,5}$, then there is no reason to include points $\set{2,3}$ in the base.  There is some small overhead cost in skipping past the redundant base entries; however, the tradeoff here is simplicity in describing the algorithm.  We will always canonicalize with respect to strict lexicographic order $\prec$, as it is more human-readable than alternatives.\footnotemark{}

\footnotetext{We note that this convention is a departure from Portugal's, which uses an ordering $\prec_B$ \emph{defined} by the base order.  So if the base is given as $B = \seq{1,4,5}$, one first extends this to a complete ordering by appending the remaining slot numbers, $B = \seq{1,4,5,2,3,6}$, and then $\prec_B$ is defined relative to this ordering (and hence $5 \prec_B 3$, for example).  The running time of an algorithm can be sensitive to the \emph{length} of the base, which in turn is influenced by the choice of base order, and in some cases it may be worth finding a more optimal choice (however, optimizing the base order is itself a costly operation!).  We prefer to keep the positional/lexicographic order for simplicity, although it is not hard to modify the algorithms here for base-ordering.}

%%%%%%%%%%%%%%
\subsection{The algorithm}
%%%%%%%%%%%%%%

%%%%%%%%%%%%%%%
% Butler-Portugal algorithm
%%%%%%%%%%%%%%%
\begin{algorithm}
	\caption{The Butler-Portugal algorithm for tensor canonicalization}
	\label{alg:butler-portugal}
	\vspace{0.5ex}
	\begin{algorithmic}
		\Input Initial configuration $g_\text{init}$; ~ slot symmetry group $S$; ~ label symmetry group $L$
		\Output Canonical configuration $g_\text{can}$ or zero if tensor vanishes
	\end{algorithmic}
	\vspace{0.5ex} \hrule \vspace{0.5ex}
	\begin{algorithmic}[1]
		\Procedure{Butler-Portugal}{$g_\text{init}, S, L$}
			\State $\variable{Configs} \gets \set{ g_\text{init} }$
			\label{compound 1}
			\For{$i \gets 1$ to $n$}
				\State $\Delta^S_i \gets$ \set{orbit of slot $i$ under $S^{(i)}$}
				\label{leastloop start}
				\State $\variable{global-least-label} \gets n$; ~ $\variable{Config-Slots-of-Least-Label} \gets \set{}$
				\label{compound 2}
				\ForEach{$g \in \variable{Configs}$}
					\label{configs loop start}
					\ForEach{$j \in \Delta^S_i$}
						\State $\Delta^{L}_{g\access{j}} \gets$ \set{orbit of label $g\access{j}$ under $L^{(i)}$}
						\label{label orbit}
						\State $\variable{least-label-in-orbit} \gets$ lexicographically-least label in $\Delta^{L}_{g\access{j}}$
						\label{least in orbit}
						\If{$\big(\variable{least-label-in-orbit} < \variable{global-least-label}\big)$}
							\State $\variable{global-least-label} \gets \variable{least-label-in-orbit}$
							\State $\variable{Config-Slots-of-Least-Label} \gets \set{}$
						\EndIf
						\If{$\big(\variable{least-label-in-orbit} = \variable{global-least-label}\big)$}
							\State $\variable{Config-Slots-of-Least-Label} \gets \function{Append}{\variable{Config-Slots-of-Least-Label}, (g,j)}$
						\EndIf
					\EndFor
				\EndFor
				\label{leastloop end}
				\State $L \gets \function{Re-order-Base}{L, i, \variable{global-least-label}}$
				\label{reorder L}
				\label{nextloop start}
				\State $\variable{Next-Configs} \gets \set{}$
				\ForEach{$(g, j) \in \variable{Config-Slots-of-Least-Label}$}
					\State $s \gets \function{Coset-Rep}{j,S^{(i)}}$
					\label{coset-rep s}
					\State $\ell \gets \function{Permutation-Inverse}{\function{Coset-Rep}{g\access{j}, L^{(i)}}}$
					\label{coset-rep l}
					\State $\variable{Next-Configs} \gets \function{Append}{\variable{Next-Configs}, \ell\access{g\access{s}}}$
					\label{configs append}
				\EndFor
				\label{nextloop end}
				\State $\variable{Configs} \gets \function{Remove-Duplicates}{\function{Sort}{\variable{Next-Configs}}}$
				\If{\big(\variable{Configs} contains equal configurations $+g$ and $-g$ of opposite sign\big)}
					\State \Return 0
				\EndIf
			\EndFor
			\State \Return $g_\text{can} = $ first element of \variable{Configs}
		\EndProcedure
	\end{algorithmic}
\end{algorithm}

We are now prepared to describe the Butler-Portugal algorithm for tensor canonicalization.  The basic strategy is straightforward:  Proceed slot by slot from 1 to $n$ and find, for each slot, the least (lexicographically) label which can be moved into that slot via a combination of slot and label symmetries.  Once a slot has been filled with the least available label, that slot is \emph{frozen} and we move on to the next slot.  Since the slots are consumed in position-order and the base for the slot symmetry group is \emph{also} in position-order, freezing a slot is the same as moving to the next $S^{(i)}$ in the stabilizer chain of $S$.  The labels, on the other hand, are consumed in an order which cannot be predicted before running the algorithm.  Thus we must re-order the base of the label symmetry group at every step.  At each step the least possible label is placed at the first available position, and thus at the end we are guaranteed to have the arrangement of labels, out of all elements in $LgS$, that is lexicographically first.

The main complication of the algorithm is that at each step, there may be \emph{multiple different ways} in which the least label can be moved into the current slot, thus bifurcating the search tree.  We are forced to track each possibility, because only at some future step will we be able to deduce which branch of the tree results in the lowest overall configuration of labels.  For example, suppose that some tensor has a cyclic symmetry,
\begin{equation}
\label{cyclic example}
T_{abc} = T_{bca} = T_{cab},
\end{equation}
and we wish to canonicalize an expression like
\begin{equation}
T_{cb}{}^c \, U^{ba}{}_a,
\end{equation}
where $U$ is some other tensor that has no symmetries.  We can find three different ways to put the label $a$ into the first slot; one for each of the cyclic configurations of \eqref{cyclic example}, while simultaneously exchanging the labels $(a,c)$ or $(a,b)$ as appropriate,
\begin{equation}
\label{cyclic example 2}
T_{ab}{}^a \, U^{bc}{}_c, \qquad T_a{}^c{}_c \, U^{ab}{}_b, \quad \text{and} \quad T_a{}^a{}_{b} \, U^{bc}{}_c,
\end{equation}
and in the last case we have also applied the metric tensor to swap ${}^a{}_a \mapsto {}_a{}^a$.  Since we must now freeze the first slot, it is clear that only the last configuration in \eqref{cyclic example 2} will lead to the least possible arrangement by lexicographic order; however, one cannot see which of these three possibilities should be kept until we examine the second slot.  It is not hard to imagine situations that frustrate the decision process for arbitrarily many steps before finally allowing us to resolve the correct choice, and thus we must potentially store many intermediate configurations as we traverse the search tree.  Thus in addition to iterating over the slots, we must also iterate over the set of intermediate configurations $\set{g_i}$.  We store the set of configurations at the current level of the search tree in a variable \variable{Configs}, and prepare the set of configurations at the next level of the search tree in \variable{Next-Configs}.

In Algorithm~\ref{alg:butler-portugal} we present a high-level view of the algorithm of \cite{2002IJMPC..13..859M}, glossing over some details of implementation.  The groups $S$ and $L$ are presumed to be stored via their bases and strong generating sets, and the base for $S$ is ordered by slot position.  The algorithm will internally modify $L$ by re-ordering its base in line \ref{reorder L} once for each slot, when a label is selected to be placed in that slot.  Each step of the main iteration over slots is split into two phases:  First, in lines \ref{leastloop start}--\ref{leastloop end}, we obtain the least available label that can be moved into the current slot under the action of the symmetry groups $S^{(i)}$ and $L^{(i)}$, which respectively stabilize the already-consumed slots and the already-consumed labels.  At the same time, we populate the set \variable{Config-Slots-of-Least-Label} with ordered pairs $(g, j)$ where $g \in \variable{Configs}$ is one of the nodes at this level of the search tree, and $j$ is a slot reachable from $i$ (via some $s \in S^{(i)}$) wherein it is possible to encounter the least-available label (via some $\ell \in L^{(i)}$).  In the second phase, from lines \ref{nextloop start}--\ref{nextloop end}, we iterate through each of the $(g, j)$ just collected and populate the set \variable{Next-Configs} with the next level of the search tree, given by $\ell\access{g\access{s}}$, where $s\access{i} = j$ and $\ell\access{g\access{j}} = \variable{global-least-label}$.

The function $\function{Re-order-Base}{L, i, \variable{global-least-label}}$ in line \ref{reorder L} re-orders the base of $L$ such that the $i$-th base point becomes \variable{global-least-label}, while keeping the first $i-1$ points unmodified.  The main insight of \cite{2002IJMPC..13..859M} is that \textproc{Re-order-Base} can be implemented for the label symmetry group via conjugation by the appropriate label swap.  The function $\function{Coset-Rep}{j,S^{(i)}}$ in line \ref{coset-rep s} returns a group element of $S^{(i)}$ which maps slot $i$ to slot $j$  (there may be several such group elements; it does not matter which is returned).  Similarly, the call to $\function{Coset-Rep}{g\access{j}, L^{(i)}}$ in line \ref{coset-rep l} returns a group element of $L^{(i)}$ which maps the $i$-th base point (which was earlier set to \variable{global-least-label}) onto the label $g\access{j}$.  The functions \textproc{Permutation-Inverse} and \textproc{Append} are straightforward.

We now comment briefly on the data structures needed to represent the various quantities in Algorithm~\ref{alg:butler-portugal}.  In order to implement \textproc{Coset-Rep} we presume that some sort of \emph{Schreier tree structure} (various examples are described in \cite{brown1989new,Cooperman:1992:FCB:143242.143316,JERRUM198660,Knuth1991}) has been constructed in the course of enumerating the orbits $\Delta^S_i$ and $\Delta^{L}_{g\access{j}}$; \textproc{Coset-Rep} then constructs a group element by tracing the Schreier tree from the specified leaf back to the root.  In fact, since the base of $S$ is never re-ordered, it is possible to compute $\Delta^S_i$ ahead-of-time, if computational time is at a higher premium than memory space.  There are a few different options to store all of the $\Delta^S_i$ together with their Schreier trees, such as a \emph{labelled branching} \cite{JERRUM198660} or a \emph{Schreier vector structure} \cite{Knuth1991}.\footnotemark{}  There are also other small optimizations, such as for the intermediate array \variable{Config-Slots-of-Least-Label} to store not the actual configuration elements $g$, but rather an index into the \variable{Configs} array, etc.

\footnotetext{In our particular implementation of the algorithm in Section~\ref{sec:full algorithm}, we have chosen a ``cube Schreier tree'' structure as detailed in \cite{Cooperman:1992:FCB:143242.143316}.  This structure requires more time to create than a standard Schreier tree, but the tree is guaranteed to be balanced and thus minimizes the time used by \textproc{Coset-Rep}.}

%%%%%%%%%%%%%%
\subsection{Worst-case complexity analysis}
\label{sec:bp complexity}
%%%%%%%%%%%%%%

In \cite{2002IJMPC..13..859M}, Portugal shows by direct testing that for the case of many Riemann tensors with random contractions, Algorithm~\ref{alg:butler-portugal} appears to have an \emph{average-case} time complexity of $O(n^5)$ for $n$ the number of slots (in fact, our measurements in Section~\ref{sec:testing} show that the particular implementation in \software{Mathematica} runs slightly faster then $O(n^4)$).  While the Riemann tensor is presented as having one of the more complicated slot symmetry groups that is typically encountered in computations, in fact it represents one of the easier (non-trivial) cases for the algorithm to resolve.  The worst-case behavior actually comes from \emph{fully symmetric} tensors (or fully antisymmetric) when they are fully contracted, for example:
\begin{equation}
\label{complexity example}
T_{bdcfae} S^{ebfdac}, \qquad \text{where} \quad T_{abcdef} = T_{(abcdef)}.
\end{equation}
Such tensor monomials suffer from a \emph{re-labelling ambiguity} which maximally frustrates the decision-making of the algorithm, rendering it completely unable to prune the search tree.  Let us follow this example through a few iterations of the algorithm to see why:

First, we examine slot 1.  There are 6 possible ways to put the least label $a$ into slot 1, which correspond to the following slot and label symmetries:
\begin{equation}
	\begin{array}{r|cccccc}
		~			& 1		& 2		& 3		& 4		& 5		& 6 \\
		\hline
		s \in S		& \id	& (1,2)	& (1,3)	& (1,4)	& (1,5)	& (1,6) \\ 
		\ell \in L	& (a,b)	& (a,d)	& (a,c)	& (a,f)	& \id	& (a,e)
	\end{array}\, ,
\end{equation}
which in turn give the configurations
\begin{align}
&T_{adcfbe} S^{eafdbc}, \qquad T_{abcfde} S^{ebfadc}, \qquad T_{adbfce} S^{ebfdca}, \\
&T_{adcbfe} S^{ebadfc}, \qquad T_{adcfbe} S^{ebfdac}, \qquad T_{adcfeb} S^{abfdec}.
\end{align}
Now we must examine slot 2 on \emph{each} of these configurations.  For each configuration, we will find that there are 5 distinct ways to move label $b$ into slot 2, and thus the set of configurations will grow to $6 \cdot 5 = 30$ entries.  At the next step, on each of these entries we will find 4 ways to move label $c$ into slot 3, and so on, so that the largest intermediate result contains $6! = 720$ entries, and the total time examining all of the intermediate results is $1 + 6 + 6 \cdot 5 + 6 \cdot 5 \cdot 4 + \ldots + 6! = 1957$, multiplied of course by whatever time it takes to execute each iteration of the internal loops.  Clearly in such cases the Butler-Portugal algorithm will experience combinatorial explosion in both time and space.  In a sense this behavior is especially frustrating, because to the end-user with human eyes, the canonical result is ``obviously'' the one with all of the indices sorted:
\begin{equation}
T_{abcdef} S^{abcdef},
\end{equation}
and yet the algorithm will take quite a while to work this out.  Increase the number of indices to 12 and the algorithm will use 24 GB of RAM and take a full day.  Increase the number of indices to 14 and the algorithm (if allowed to write intermediate results to disk) will completely fill a 4 TB hard drive over the course of 30 years.\footnotemark{}

\footnotetext{The space requirements here are calculated assuming each configuration is an array of $(n+2)$ 32-bit integers.  The timings are projected from a best-fit curve of the test in Figure~\ref{fig:total-sym-dummies} involving up to 10 indices, which thankfully run in only a minute or two per test iteration (on a machine with a 2.40 GHz processor and 8 GB RAM).  Note that it is difficult to get a reliable fit to these curves due to the paucity of data points, and thus the projected timings may be wildly inaccurate.}

This problem is noted by the authors of \cite{citeulike:13127953}, and they attempt to mitigate it by first sorting any tensor products (such as $TS$ here) by \emph{increasing group order} of their slot symmetry groups.  Thus $S$, since it has no symmetries, will be put first, and then the algorithm becomes \emph{linear} as there is no longer any ambiguity of intermediate results.  This is a dramatic improvement; however, it is still thwarted when \emph{both} tensor factors have a high degree of symmetry, as in
\begin{equation}
T_{bdcfae} T^{ebfdac}.
\end{equation}
Products of tensors with many symmetric or antisymmetric slots show up in several contexts, such as in higher-spin theory \cite{Vasiliev:1995dn,Vasiliev:2004qz}, supergravity \cite{Cremmer:1978km,Freedman:2012zz}, representation theory \cite{Feger:2012bs,fulton1997young,fulton2013representation}, and the construction of spherical harmonics with high quantum numbers and/or in higher dimensions based on homogeneous polynomials \cite{Frye:2012jj}.  Therefore it is worth resolving this problem more thoroughly.

%%%%%%%%%%%%%%
\section{The improved algorithm}
\label{sec:improved}
%%%%%%%%%%%%%%

If one is doing the sort of computation where one expects to see a lot of symmetric/antisymmetric tensors, even of with a moderate number of indices, the behavior of Section~\ref{sec:bp complexity} is troubling.  One needs an efficient solution.  One na\"ive strategy is to treat fully (anti)symmetric tensors in a special way, perhaps by simply sorting their indices (and inserting the appropriate sign for antisymmetric tensors).  But we can quickly see that this will not be sufficient, for two reasons:  First, such tensors can appear in tensor products with tensors with \emph{other} symmetries, and thus to be fully general, an algorithm must be able to deal with arbitrary (anti)symmetric \emph{subsets} of indices, rather than having special code only for the case of full (anti)symmetry.  Second, when including dummy contractions in such tensor products, it is not enough merely to \emph{sort} the (anti)symmetric subsets of indices, because dummy labels may be exchanged while canonicalizing neighboring tensors, thus causing the notion of ``sorted'' to change as well.  A more intelligent approach is needed.

%%%%%%%%%%%%%%
\subsection{Insights from Penrose graphical notation}
\label{sec:penrose}
%%%%%%%%%%%%%%

It turns out that a useful way to approach the problem is to employ the \emph{Penrose graphical notation} for tensor contractions, first published in \cite{penrose1971applications} (see also \cite{Cvitanovic:2008zz} for an alternative version).  In this notation, tensors are represented by various (arbitrary) shapes, and their indices are represented by lines emanating from these shapes.  For example, one might have tensors $\theta$ and $\chi$ given by:
\begin{equation}
\tikzsetnextfilename{theta-abc}
\input{tikz/theta-abc.tikz} = \theta^{ab}{}_c,
\qquad
\tikzsetnextfilename{chi-abcd}
\input{tikz/chi-abcd.tikz} = \chi^a{}_{bcd}.
\end{equation}
Lines going to the top of the diagram represent upper indices, while lines going to the bottom of the diagram represent lower indices.  To represent contracted indices, we merely connect the lines:
\begin{equation}
\tikzsetnextfilename{chi-theta-contractions}
\input{tikz/chi-theta-contraction.tikz} = \theta^{af}{}_c \chi^b{}_{fde}.
\end{equation}
Of course, the label $f$ is meaningless; the important information is the line connecting $\theta$ and $\chi$ which shows the contraction.  One also has special symbols for the metric, inverse metric, and Kronecker delta,
\begin{equation}
\tikzsetnextfilename{metric}
\input{tikz/metric.tikz} = g_{ab},
\qquad
\tikzsetnextfilename{inverse-metric}
\input{tikz/inverse-metric.tikz} = g^{ab},
\qquad
\tikzsetnextfilename{kronecker-delta}
\input{tikz/kronecker-delta.tikz} = \delta_a^b ,
\end{equation}
which are made only of connecting lines, and which obey the obvious graphical relation
\begin{equation}
g_{ac} g^{cb} =
\tikzsetnextfilename{metric-inverse-metric}
\input{tikz/metric-inverse-metric.tikz} =
\tikzsetnextfilename{kronecker-delta}
\input{tikz/kronecker-delta.tikz} = \delta_a^b.
\end{equation}
Combining these with the shapes for tensors like $\theta$ and $\chi$, one can create diagrams for tensor contractions of arbitrary complexity.  There are some additional features of the notation which we will not need here; we refer the reader to \cite{penrose1971applications} for more detail.

%%%%%%%%%%%%%%
\subsubsection{Dummy pairs as links for symmetry propagation}
\label{sec:dummy links}
%%%%%%%%%%%%%%

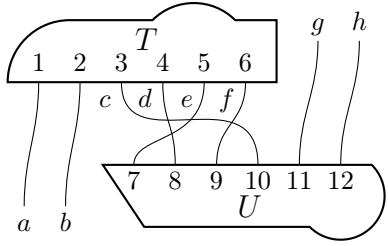
\begin{figure}[t]
	\centering
	\tikzsetnextfilename{TU-contraction}
	\input{tikz/TU-contraction.tikz}
	\caption{Penrose graphical notation for the tensor contraction $T_{abcdef} U^{edfcgh}$.  The slots are numbered 1-12 in positional order.}
	\label{fig:TU contraction}	
\end{figure}

For our purposes, it is useful to enlarge the diagrams a bit and write in the \emph{slot number} to which a line is attached.  For example, the tensor contraction
\begin{equation} \label{TU contraction}
T_{abcdef} U^{edfcgh}
\end{equation}
can be represented by the diagram in Figure~\ref{fig:TU contraction}, where the slots are numbered 1-12 in the order they appear in \eqref{TU contraction}.

\begin{figure}
	\centering
	\begin{subfigure}{0.28\textwidth}
		\tikzsetnextfilename{TU-contraction-symmetry}
		\input{tikz/TU-contraction-symmetry.tikz}
		\caption{Initial configuration.}
		\label{fig:TU initial}
	\end{subfigure}
	\hspace{\fill}
	\begin{subfigure}{0.28\textwidth}
		\tikzsetnextfilename{TU-contraction-unwind}
		\input{tikz/TU-contraction-unwind.tikz}
		\caption{Rearrange $\set{3,4,5,6}$.}
		\label{fig:TU slots rearranged}
	\end{subfigure}
	\hspace{\fill}
	\begin{subfigure}{0.28\textwidth}
		\tikzsetnextfilename{TU-contraction-rename}
		\input{tikz/TU-contraction-rename.tikz}
		\caption{Relabel dummies.}
		\label{fig:TU dummies renamed}
	\end{subfigure}
	\caption{We use the total symmetry group on slots $\set{3,4,5,6}$ to ``disentangle'' the contractions.  In \subref*{fig:TU initial}, we have the initial configuration $T_{abcdef} U^{edfcgh}$.  In \subref*{fig:TU slots rearranged}, we act with symmetry to re-arrange the ends attached to $\set{3,4,5,6}$, giving us $T_{abedfc} U^{edfcgh}$.  Finally in \subref*{fig:TU dummies renamed}, we re-name the dummy labels, giving $T_{abcdef} U^{cdefgh}$.}
\end{figure}
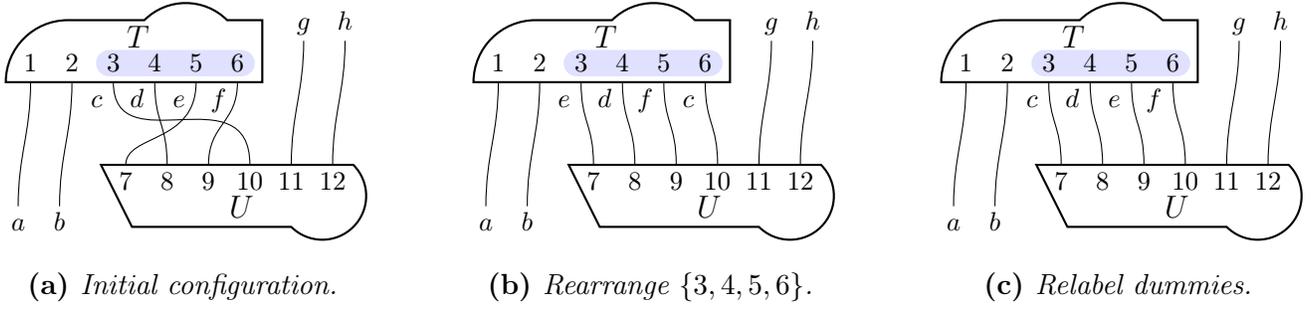

It is instructive to start from the arrangement in Figure~\ref{fig:TU contraction} and walk through the steps a \emph{human} might take to canonicalize it.  Suppose, for example, that the tensor $T$ is totally symmetric over the slots $\set{3,4,5,6}$.  This situation is depicted in Figure~\ref{fig:TU initial}.  The dummy labels $c,d,e,f$ are not meaningful, but the objective is to find a combination of slot- and label-rearrangements which minimizes \eqref{TU contraction} with respect to lexicographical ordering.  The first step is to ``disentangle'' the contracted edges in Figure~\ref{fig:TU initial} by moving the endpoints which are attached to the slots $\set{3,4,5,6}$ where the slot symmetry acts.  This results in
\begin{equation}
T_{abedfc} U^{edfcgh},
\end{equation}
as shown in Figure~\ref{fig:TU slots rearranged}.  The final step is to re-name all of the dummy labels in order of appearance, which produces
\begin{equation}
T_{abcdef} U^{cdefgh},
\end{equation}
as shown in Figure~\ref{fig:TU dummies renamed}.  Assuming the tensor $U$ has no slot symmetries, then we have achieved the least possible lexical order, with the dummies $\seq{c,d,e,f,c,d,e,f}$ repeated in order.

The above process is a strict application of the available symmetry groups: first using slot symmetries to make the two sets of dummies match in order $\seq{e,d,f,c,e,d,f,c}$, and then using label symmetries to rename these into the correct order.  It is easy to describe this process in words, but to put it in an algorithm which visits each slot only once, scanning from left to right, is a bit non-trivial.  Since the Butler-Portugal algorithm does not backtrack once it has made a decision to place a label in a slot, it must instead retain enough information that it can make the decision later.  In this case, it will see $4!$ ways to put $\seq{c,d,e,f}$ into slots $\seq{3,4,5,6}$, and only begin to resolve which of those possibilities to retain when it reaches slot 7.  The decisions needed to reduce $4!$ possibilities down to the 1 correct one are not fully made until it visits slot 10.

Of course, we do not wish to introduce backtracking into the algorithm.  Instead, we will slightly \emph{redefine} the problem in a way that makes it amenable to a strict left-to-right pass without having to retain (and iterate over!) a factorial number of intermediate states.  Our main observation is that the \emph{symmetries at one end of an edge in a Penrose tensor diagram induce symmetries of the opposite end}.  Thus the slot symmetries of tensor $T$ can ``propagate'' along the connecting lines to tensor $U$, and induce new symmetries which appear to act on the slots of $U$ (as we shall see, however, they actually act on the labels!).  This fact is somewhat obvious from the Penrose graphical notation, but is obscured in the standard index notation because the ``propagation'' of symmetries along contractions may be effected only by executing slot and label symmetries at the same time.

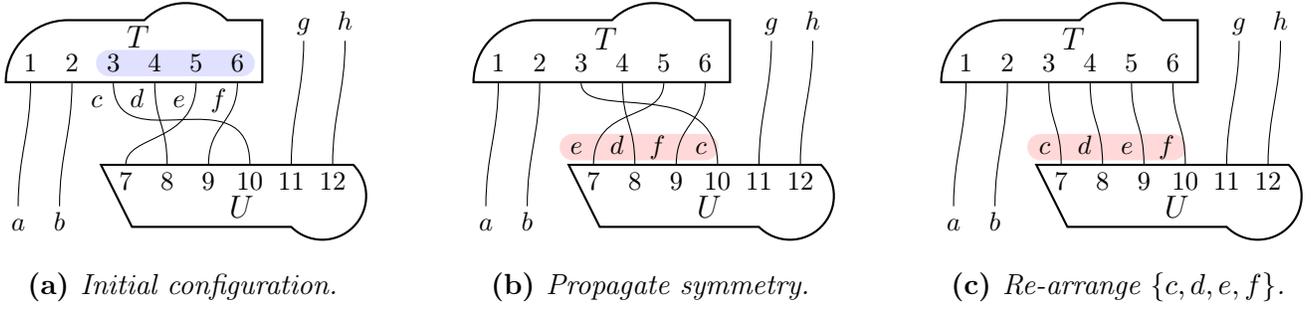
\begin{figure}
	\centering
	\begin{subfigure}{0.28\textwidth}
		\tikzsetnextfilename{TU-contraction-symmetry}
		\input{tikz/TU-contraction-symmetry.tikz}
		\caption{Initial configuration.}
		\label{fig:TU initial 2}
	\end{subfigure}
	\hspace{\fill}
	\begin{subfigure}{0.28\textwidth}
		\tikzsetnextfilename{TU-contraction-propagate}
		\input{tikz/TU-contraction-propagate.tikz}
		\caption{Propagate symmetry.}
		\label{fig:TU propagate}
	\end{subfigure}
	\hspace{\fill}
	\begin{subfigure}{0.28\textwidth}
		\tikzsetnextfilename{TU-contraction-unwind2}
		\input{tikz/TU-contraction-unwind2.tikz}
		\caption{Re-arrange $\set{c,d,e,f}$.}
		\label{fig:TU labels rearranged}
	\end{subfigure}
	\caption{Alternate way to view the process.  In \subref*{fig:TU initial 2}, we have the initial configuration $T_{abcdef} U^{edfcgh}$.  In \subref*{fig:TU propagate}, we freeze slots $\set{3,4,5,6}$ in place.  The configuration remains $T_{abcdef} U^{edfcgh}$, but we \emph{propagate the symmetry} along the edges, so that the original $\set{3,4,5,6}$ slot symmetry of $T$ becomes a \emph{label symmetry} of the labels attached to $\set{7,8,9,10}$ of $U$.  Finally, in \subref*{fig:TU labels rearranged} we apply this label symmetry  to re-arrange the $U$-attached endpoints (remembering that the $T$-attached ends do not participate!), giving the final configuration $T_{abcdef} U^{cdefgh}$.}
\end{figure}

To illustrate, we walk through the same example.  We begin with \eqref{TU contraction}, as shown in Figure~\ref{fig:TU initial 2}.  Now, instead of rearranging the endpoints attached to $\set{3,4,5,6}$, let us actually freeze them in place; after all, the labels there are already $\seq{c,d,e,f}$, which is what we want.  But there is a symmetry among the slots $\set{3,4,5,6}$ which we have not used, and which is crucial to obtain the correct result.  Let us ``propagate'' this symmetry forward, along the legs attached at $\set{3,4,5,6}$, to the slots $\set{7,8,9,10}$ on tensor $U$.  In fact, for reasons we will soon explain, we will consider this symmetry to be associated to the \emph{labels} $\set{e,d,f,c}$ which are present at slots $\set{7,8,9,10}$, as depicted in Figure~\ref{fig:TU propagate}.

The tensor $U$ does not actually have a symmetry in the slots $\set{7,8,9,10}$, but by \emph{simultaneously} applying the slot symmetries on $\set{3,4,5,6}$ and the standard label symmetries, one can give the \emph{appearance} that $U$ has such a symmetry:
\begin{equation}
T_{abcdef} U^{edfcgh} = T_{abcdef} U^{fedcgh} = T_{abcdef} U^{cdefgh} = \text{etc.}
\end{equation}
This is what we mean by a ``propagated'' symmetry.  Thus we may freeze the $T$-attached endpoints of the contractions in place, and treat the $U$-attached endpoints as though they are free to move.  This allows us to rearrange the labels $\set{e,d,f,c}$ on tensor $U$ \emph{independently} of those on tensor $T$, and thus reach the canonical configuration as shown in Figure~\ref{fig:TU labels rearranged}.

With this notion of symmetry propagation along dummy contractions, we can effectively implement the necessary decision process in a single left-to-right pass without storing many intermediate configurations.  Instead we must store information about the symmetries to be propagated forward.  For arbitrary symmetry groups, this might become unwieldy; but if this method is used only for \emph{totally (anti)symmetric} subgroups, then it is simple to implement.  One needs only a single array of length $n$ to store \emph{all} of the information regarding propagated symmetries (as will be described in Section~\ref{sec:data structures}).  One can then take advantage of this information to make a \emph{single} definite decision at each new visited slot, completely eliminating the $O(n!)$ intermediate steps needed in the Butler-Portugal approach.

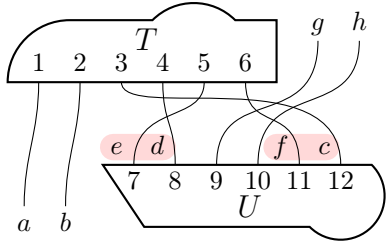
\begin{figure}
	\centering
	\tikzsetnextfilename{TU-contraction-propagate-shift}
	\input{tikz/TU-contraction-propagate-shift.tikz}
	\caption{Symmetry propagation with additional slot symmetries.  If $U$ also has a slot symmetry, for example $(9,11)(10,12)$, then the propagated symmetry (over the labels $\set{c,d,e,f}$) must move to the new slots.  Hence propagated symmetries are \emph{label} symmetries; not slot symmetries.}
	\label{fig:TU propagate shift}
\end{figure}

We have stated that slot symmetries which have been propagated along dummy contractions become \emph{label} symmetries, and this requires some explanation.  Up until now, the tensor $U$ had no symmetries of its own.  But suppose instead it had a single symmetry $(9,11)(10,12)$ which exchanges the last two pairs of slots.  Our lexicographical ordering prioritizes free indices ahead of dummies, so we should use this symmetry to move $\seq{g,h}$ ahead of $\seq{f,c}$.  But the propagated symmetries, which originated from total symmetry of slots $\set{3,4,5,6}$ of $T$, must remain associated to the contraction edges which attach to slots $\set{3,4,5,6}$.  Therefore, the propagated symmetries will now be associated with slots $\set{7,8,11,12}$ as shown in Figure~\ref{fig:TU propagate shift}.  It is clear that we should really associate the propagated symmetry with the \emph{labels} $\set{e,d,f,c}$, and thus the generators
\begin{equation}
\set{ (c_2, d_2), \; (d_2, e_2), \; (e_2, f_2) }
\end{equation}
should be appended to the label group $L$ (here, as before, the subscript $_2$ refers to the appearance of the label as an \emph{upper} index; namely on the tensor $U$).

%%%%%%%%%%%%%%
\subsection{Data structures}
\label{sec:data structures}
%%%%%%%%%%%%%%

In this section we give several data structures which are needed to implement the symmetry-propagation mechanism described in Section~\ref{sec:dummy links}.  These data structures are also useful for improving the basic Butler-Portugal algorithm as they can reduce most of the label-group computations of Algorithm~\ref{alg:butler-portugal} to mere array lookups.  These data structures we will refer to as the \variable{Values} array, the \variable{Label-Groups} array, the \variable{Symmetric-Subsets} array, and the \variable{Propagated-Symmetries} array.  All of these are simple length-$n$ arrays, and thus using these data structures incurs an additional $O(n)$ memory cost only.

We will also describe \variable{Config-Slots-of-Least-Value}, which is a slightly different way of tracking the current level of the search tree, and the notion of a \variable{Least-Value-Set}, which is just a useful way to group the entries of \variable{Config-Slots-of-Least-Value}.

%%%%%%%%%%%%%%
\subsubsection{The \variable{Values} array}
\label{sec:values array}
%%%%%%%%%%%%%%

Lines \ref{leastloop start}--\ref{leastloop end} of Algorithm~\ref{alg:butler-portugal} are occupied with finding the least available label which can be moved into the current slot.  The costliest parts of this code are line~\ref{leastloop start} and line~\ref{label orbit}, where the orbits $\Delta_i^S$ and $\Delta^{L}_{g\access{j}}$ are computed.  In fact, the slot orbit $\Delta_i^S$ can be computed once-and-for-all at the time a tensor and its symmetries are declared, and stored in a type of Schreier tree structure, which we will not detail here (see, for example, \cite{brown1989new,Cooperman:1992:FCB:143242.143316,JERRUM198660,Knuth1991}).  However, the label orbit $\Delta^{L}_{g\access{j}}$ depends on which labels have been placed in a tensor's slots, which comes from the context of a specific tensor expression.  It would appear that this must be computed on-the-fly.

However, we can introduce the concept of the \emph{value} of a label, which can achieve two useful things simultaneously:  First, it can reduce lines \ref{label orbit} and \ref{least in orbit} of Algorithm~\ref{alg:butler-portugal} to a simple array lookup, thus skipping the computation of $\Delta^{L}_{g\access{j}}$ altogether; second, it can concisely encode \emph{all} the data as to which indices are free, or belong to a given dummy set or repeated set, into a single array of length $n$.  The concept is simple, and satisfies the properties of Definition~\ref{def:value}:

\begin{definition}
	\caption{Properties of the \variable{Values} array.}
	\label{def:value}
	\begin{enumerate}[label=\textbf{V\arabic*.},ref=V\arabic*]
		\item The \emph{value} of a label is equal to the (number encoding) the least label in the group to which a label belongs;
		\item Each label in a group of interchangeable labels has the same value;
		\item Every free index belongs to its own group;
		\item A group of interchangeable labels is \emph{usually} contiguous in label space.\footnotemark{}
	\end{enumerate}
\end{definition}

\footnotetext{The exception is when a label belongs to a dummy set for a vector bundle on which there is no defined metric.  Then upstairs labels cannot mix with downstairs ones.  The mechanics of this case will be explained in Section~\ref{sec:label groups array}.}

Take, for example, the tensor expression
\begin{equation}
T_{11ab}{}^{bc} = T_{1_1 1_2 a b_1}{}^{b_2 c},
\end{equation}
where as before, we insert the subscripts $_1, \, _2$ to distinguish identical labels.  The labels have a natural lexicographical ordering $\seq{a,c,1_1,1_2,b_1,b_2}$ which we use as the index into the array.  Then the \variable{Values} array can be written
\begin{equation}
	\begin{array}{r|cccccc}
		\text{Label}	& a		& c		& 1_1	& 1_2	& b_1	& b_2	\\
		\hline
		\text{Value}	& 1		& 2		& 3		& 3		& 5		& 5
	\end{array}\, .
\end{equation}
As a more elaborate example, consider again the contraction
\begin{equation} \label{TU contraction subscripts}
T_{abcdef} U^{edfcgh} = T_{abc_1d_1e_1f_1} U^{e_2d_2f_2c_2gh}.
\end{equation}
The \variable{Values} array corresponding to this expression is given by
\begin{equation} \label{values 12 slot}
	\begin{array}{r|cccccccccccc}
		\text{Label}	& a		& b		& g		& h		& c_1	& c_2	& d_1	& d_2	& e_1	& e_2	& f_1	& f_2 \\
		\hline
		\text{Value}	& 1		& 2		& 3		& 4		& 5		& 5		& 5		& 5		& 5		& 5		& 5		& 5
	\end{array}\, .
\end{equation}
Notice in each case that all labels which can be interchanged by label symmetries have the same value; moreover, that value is equal to the (numerical) index of the least label itself.  Then, for example, if one needs to know the least label that can be moved into slot 7 of \eqref{TU contraction subscripts}, the process is simple:
\begin{enumerate}
	\item Look up the label in slot 7 of \eqref{TU contraction subscripts}: $e_2$.
	\item In the \variable{Values} array \eqref{values 12 slot}, look up the value of $e_2$: 5.
	\item In the list of sorted labels $\seq{a,b,g,h,c_1,c_2,d_1,d_2,e_1,e_2,f_1,f_2}$, take the 5th entry: $c_1$.
\end{enumerate}
and thus we see that $c_1$ is the least label that can be moved into slot 7, without having to explicitly enumerate the orbit $\Delta^{L}_{g\access{7}}$.

Of course, since the algorithm visits the slots in order, we should have already placed $a$, $b$, and four of the dummy labels by the time we reach slot 7.  Once a label has been consumed by placing it in a slot, it is no longer interchangeable with the unused labels, and the \variable{Values} array must reflect that fact for the correct functioning of the algorithm.  As we shall detail in Section~\ref{sec:full algorithm}, the \variable{Values} array will be updated at the end of every step to ensure that the properties of Definition~\ref{def:value} continue to hold.

%%%%%%%%%%%%%%
\subsubsection{The \variable{Label-Groups} array}
\label{sec:label groups array}
%%%%%%%%%%%%%%

The \variable{Values} array can efficiently keep track of which subsets of labels are interchangeable; however, it does not contain information about \emph{how} they can be interchanged.  In Algorithm~\ref{alg:butler-portugal}, this information is contained in the label symmetry group $L$, which must be continually updated in line \ref{reorder L} via the procedure \textproc{Re-order-Base} whenever any two labels are swapped.  The main insight of \cite{2002IJMPC..13..859M} is that \textproc{Re-order-Base} can be implemented by conjugating the members of $L$, which is much faster than a generic base-change algorithm.

However, even this can be improved upon.  To this end, we introduce another data structure of length $n$, the \variable{Label-Groups} array.  Instead of literally storing a base-and-strong-generating-set for the label group $L$, we will store an array of \emph{group codes} which indicate how labels may be exchanged among a given equal-value set.  Like the \variable{Values} array, the \variable{Label-Groups} array is indexed by the labels, in lexicographic order, and thus should be thought of as living ``in label space''.  Each entry in the \variable{Label-Groups} array shall be one of the \emph{group codes} as described in Table~\ref{tab:group codes} (these can be implemented, for example, by an \code{enum} type or similar).

\begin{table}
	\centering
	\caption{Group codes for \variable{Label-Groups} array.}
	\label{tab:group codes}
	\begin{tabular}{c|l}
		Code			& Meaning \\
		\hline
		$\free$			& No label exchange \\
		$\component$	& Component labels such as $\seq{1_1, 1_2, 1_3, 1_4}$ \\
		$\sdummy$		& Dummy labels with symmetric metric such as $g_{ab}$ \\
		$\adummy$		& Dummy labels with antisymmetric metric such as $\varepsilon_{\alpha \beta}$ \\
		$\ldummy$		& \emph{Lower} dummy labels with no metric \\
		$\udummy$		& \emph{Upper} dummy labels with no metric
	\end{tabular}
\end{table}

It is useful to display the \variable{Values} array along with the \variable{Label-Groups} array, as together these constitute all the information about the label symmetry group $L$.  To revisit our earlier examples, the expression
\begin{equation}
T_{11ab}{}^{bc} = T_{1_1 1_2 a b_1}{}^{b_2 c}
\end{equation}
corresponds to the \variable{Values} and \variable{Label-Groups}
\begin{equation} \label{T comp}
\begin{array}{r|cccccc}
\text{Label}	& a		& c		& 1_1			& 1_2			& b_1		& b_2	\\
\hline
\text{Value}	& 1		& 2		& 3				& 3				& 5			& 5		\\
\text{Group}	& \free	& \free	& \component	& \component	& \sdummy	& \sdummy
\end{array}\, ,
\end{equation}
and the expression of our earlier example,
\begin{equation}
	T_{abcdef} U^{edfcgh} = T_{abc_1d_1e_1f_1} U^{e_2d_2f_2c_2gh},
\end{equation}
corresponds to
\begin{equation} \label{TU sym metric}
	\begin{array}{r|cccccccccccc}
		\text{Label}	& a		& b		& g		& h		& c_1		& c_2		& d_1		& d_2		& e_1		& e_2		& f_1		& f_2 \\
		\hline
		\text{Value}	& 1		& 2		& 3		& 4		& 5			& 5			& 5			& 5			& 5			& 5			& 5			& 5		\\
		\text{Group}	& \free	& \free	& \free	& \free	& \sdummy	& \sdummy	& \sdummy	& \sdummy	& \sdummy	& \sdummy	& \sdummy	& \sdummy
	\end{array}\, ,
\end{equation}
if we assume that there is a metric which can exchange upper and lower indices.  If no metric exists, then we must remember that \emph{only labels which can be exchanged} can have the same value; thus the \variable{Values} and \variable{Label-Groups} arrays will read
\begin{equation} \label{TU no metric}
\begin{array}{r|cccccccccccc}
\text{Label}	& a		& b		& g		& h		& c_1		& c_2		& d_1		& d_2		& e_1		& e_2		& f_1		& f_2 \\
\hline
\text{Value}	& 1		& 2		& 3		& 4		& 5			& 6			& 5			& 6			& 5			& 6			& 5			& 6		\\
\text{Group}	& \free	& \free	& \free	& \free	& \ldummy	& \udummy	& \ldummy	& \udummy	& \ldummy	& \udummy	& \ldummy	& \udummy
\end{array}\, .
\end{equation}
In this case, the set of labels with value 5 is not contiguous (likewise the set with value 6), and thus our algorithm cannot strictly rely on the \variable{Values} array being a non-decreasing sequence.  However, the entries in the \variable{Label-Groups} array give enough hinting to deduce that the \variable{Values} array must have a contiguous sequence of $\seq{5,6}$ blocks, and \emph{this} fact we can take advantage of.

In any of these cases, the \variable{Values} and \variable{Label-Groups} arrays in \eqref{T comp}, \eqref{TU sym metric}, and \eqref{TU no metric} contain sufficient information to reconstruct any necessary label exchange to bring the least label into a slot containing a given label.  Suppose, for example, we are given the label $e_2$ in the presence of a symmetric metric, thus corresponding to \eqref{TU sym metric}.  We construct the label exchange as follows:
\begin{enumerate}
	\item The value of $e_2$ in \eqref{TU sym metric} is 5, which means it can be swapped with the label in the 5th position, $c_1$.
	\item The group of $e_2$ is $\sdummy$, which means that each of the swaps $(c_1,e_1)(c_2,e_2)$ and $(e_1,e_2)$ are allowed.
	\item The position of $e_2$ is 10; since $(10-5)$ is odd, we determine that \emph{both} exchanges must be made in order to map $e_2 \mapsto c_1$.  This can be achieved by
	\begin{equation}
	(e_1,e_2) \cdot (c_1,e_1)(c_2,e_2) = (c_1,e_1,c_2,e_2).
	\end{equation}
\end{enumerate}
By contrast, suppose instead that we have no metric, as in \eqref{TU no metric}.  In this case, the label exchange can be constructed as follows:
\begin{enumerate}
	\item The value of $e_2$ in \eqref{TU no metric} is 6, which means it can be swapped with the label in the 6th position, $c_2$.
	\item The group of $e_2$ is $\udummy$, which means it is an upper index.  In our conventions, the lower indices are ordered before each corresponding upper one; therefore we conclude that each of $c_2, \, e_2$ should be paired with the label immediately to its \emph{left} in the array.
	\item With this information, we can construct the necessary label exchange: $(c_1, e_1)(c_2, e_2)$.
\end{enumerate}

Thus we can see how the \variable{Values} and \variable{Label-Groups} arrays can be used to quickly circumvent many of the more expensive steps in the inner loops of Algorithm~\ref{alg:butler-portugal}.  It should also be straightforward how one might construct these arrays in the first place; the \variable{Values} are merely a means of encoding the various sets of indices that appear in an expression, and the \variable{Label-Groups} are fixed entirely by the combination of label types and type of metric tensor currently defined (or not defined) on a given vector bundle.

%%%%%%%%%%%%%%
\subsubsection{The \variable{Symmetric-Subsets} array}
\label{sec:symmetric subsets array}
%%%%%%%%%%%%%%

The previous data structures can be used to improve the performance of the standard Butler-Portugal algorithm even without considering the new concepts introduced in Section~\ref{sec:dummy links}.  In this section and the next, we will introduce further data structures for implementing the symmetry-propagation ideas which allow fast canonicalization of totally (anti)symmetric subsets of indices.

The first of these is the \variable{Symmetric-Subsets} array, which is an extra \emph{input} to the algorithm, of length $n$, which stores condensed information on where to find (anti)symmetric subsets in the slot symmetry group $S$.  The \variable{Symmetric-Subsets} array can be generated by $O(n^2)$ group-membership tests (slightly modified for \emph{signed} permutations) to find the elementary length-2 cycles which generate each (anti)symmetric subgroup (see \cite{DBLP:Butler91,holt2005handbook,seress2003permutation}).  Since the information contained in \variable{Symmetric-Subsets} is derived only from the slot symmetry group $S$, it can be obtained ahead of time at tensor declaration time and stored at $O(n)$ cost, which is trivial compared to the cost of the Schreier vector structure needed for $S$ itself.\footnotemark{}

\footnotetext{We point out that the \variable{Symmetric-Subsets} array in fact describes \emph{redundant} information which is already strictly present in the Schreier vector structure of $S$.  However, this information is in a special format which allows quick manipulation by taking advantage of the simple structure of totally (anti)symmetric groups.  One could set \variable{Symmetric-Subsets} to the zero array and the algorithm will still produce correct results, but more slowly.}

\begin{definition}
	\caption{Properties of the \variable{Symmetric-Subsets} array.}
	\label{def:symmetric-subsets}
	\begin{enumerate}[label=\textbf{S\arabic*.},ref=S\arabic*]
		\item A zero entry indicates that a slot does not participate in any (anti)symmetric subsets.
		\item Positive integers represent a symmetric subset of slots.  All slots with the same positive integer can be exchanged with each other.
		\item Negative integers represent antisymmetric subsests.  All slots with the same negative integer can be exchanged at the cost of a minus sign.
		\item In addition, we shall require that the integers be chosen in a sequence of increasing absolute value, and that each absolute value is used for only one subset (thus, 1 and $-1$ should not both appear, but rather 1 and $-2$, etc.).
	\end{enumerate}
\end{definition}

The structure of the \variable{Symmetric-Subsets} array is simple:  The positions in the array represent \emph{slots} (and thus \variable{Symmetric-Subsets} lives in slot space, in contrast to \variable{Values} and \variable{Label-Groups} which live in label space).  Each entry is an integer, with meanings as given in Definition~\ref{def:symmetric-subsets}.  For example, suppose we have a tensor $T_{abcdef}$ which is symmetric on the last 4 slots; thus the slot symmetry group $S$ is given by the strong generating set
\begin{equation}
S = \set{ +(3,4), \; +(4,5), \; +(5,6) }.
\end{equation}
The corresponding \variable{Symmetric-Subsets} array is then given by
\begin{equation}
\begin{array}{r|cccccc}
\text{Slot}		& 1		& 2		& 3		& 4		& 5		& 6	\\
\hline
\text{Subset}	& 0		& 0		& 1		& 1		& 1		& 1
\end{array}\, .
\end{equation}
Similarly, the Riemann tensor $R_{abcd}$ has the slot symmetry group given by the strong generating set
\begin{equation}
S = \set{ -(1,2), \; +(1,3)(2,4), \; -(3,4) },
\end{equation}
and the corresponding \variable{Symmetric-Subsets} array is
\begin{equation}
\begin{array}{r|rrrr}
\text{Slot}		& 1		& 2		& 3		& 4	\\
\hline
\text{Subset}	& -1	& -1	& -2	& -2
\end{array}\, .
\end{equation}
Finally, consider the contraction $T_{abcdef} R^{cd}{}_{gh}$.  This has the slot symmetry group given by
\begin{equation}
S = \set{ +(3,4), \; +(4,5), \; +(5,6), \; -(7,8), \; +(7,9)(8,10), \; -(9,10) },
\end{equation}
and the \variable{Symmetric-Subsets} array will read
\begin{equation} \label{TR-sym-subsets}
\begin{array}{r|rrrrrrrrrr}
\text{Slot}		& 1		& 2		& 3		& 4		& 5		& 6		& 7		& 8		& 9		& 10	\\
\hline
\text{Subset}	& 0		& 0		& 1		& 1		& 1		& 1		& -2	& -2	& -3	& -3
\end{array}\, .
\end{equation}
Of course, we know that the expression $T_{abcdef} R^{cd}{}_{gh} = 0$ because the contraction of the indices $cd$ will bring the symmetry of slots $\seq{3,4}$ in conflict with the antisymmetry of slots $\seq{7,8}$.  Thus we can appreciate the usefulness of having pre-computed the structure in \eqref{TR-sym-subsets}; in cases like this one, this will allow our algorithm to detect zero contractions much earlier than Butler-Portugal, as will be shown in Section~\ref{sec:testing}.

In addition we note that \variable{Symmetric-Subsets}, being a container for \emph{slot-symmetry} information, does not take into account that $cd$ are dummy indices while $ef$ are free; the slots $\set{3,4,5,6}$ participate in the same slot symmetry regardless of which indices they contain.

%%%%%%%%%%%%%%
\subsubsection{The \variable{Propagated-Symmetries} array}
\label{sec:propagated symmetries array}
%%%%%%%%%%%%%%

In this section we will discuss the \variable{Propagated-Symmetries} array, whose purpose is to track the information needed to implement the symmetry-propagation mechanism of Section~\ref{sec:dummy links}.  The types of information that need to be stored are twofold:  First, one needs to know which portions of the \variable{Symmetric-Subsets} are attached to dummy indices; second, one needs to know where the other ``ends'' of those dummy contractions reside, in the sense of the Penrose graphical notation introduced in Section~\ref{sec:penrose}.  Both of these needs can be served using a single array of length $n$, whose structure is similar to that of \variable{Symmetric-Subsets}, but with an additional distinction between odd- and even-numbered entries, which will be explained shortly.

Before defining the entries in \variable{Propagated-Symmetries}, however, we must decide whether it should be indexed by slots or labels.  We first note that the symmetries described in \variable{Propagated-Symmetries} are \emph{label} symmetries, as suggested by Figure~\ref{fig:TU propagate shift}, and thus it seems sensible for \variable{Propagated-Symmetries} to live in label space.  However, as will be explained in Section~\ref{sec:propagated symmetries array}, the \variable{Propagated-Symmetries} array can never be accessed \emph{directly}, because each configuration $g \in \variable{Configs}$ which we must iterate over (as in line~\ref{configs loop start} of Algorithm~\ref{alg:butler-portugal}) may have been reached by a \emph{different} combination of slot and label symmetries.  That is, the labels may have been renamed, and/or the slots shifted.  The total configuration is given by $g = \ell\access{g_\text{init}\access{s}}$ where $(\ell, s)$ are the independent label- and slot-actions which have been accumulated in the course of the algorithm.\footnotemark{}  The \variable{Propagated-Symmetries} array is a \emph{global} variable whose information is needed in every configuration we iterate over.  Thus if \variable{Propagated-Symmetries} is label-indexed, we must know $\ell$ in order to access it properly; alternatively, if it is slot-indexed, we must know $s$.  Neither convention seems to offer a computational advantage.  We will choose \variable{Propagated-Symmetries} to be \emph{slot}-indexed, despite that fact that its entries represent label symmetries, because we find it easier to relate to the diagrams of Section~\ref{sec:penrose} this way.

\footnotetext{In our version of the Butler-Portugal algorithm in Algorithm~\ref{alg:butler-portugal}, we store only the total configuration $g$, whereas in the original version in \cite{2002IJMPC..13..859M}, the label- and slot-actions $(\ell, s)$ are stored separately.}

\begin{definition}
	\caption{Properties of the \variable{Propagated-Symmetries} array.}
	\label{def:prop-sym}
	\begin{enumerate}[label=\textbf{P\arabic*.},ref=P\arabic*]
		\item A zero entry indicates that the label at a given slot does not participate in any (anti)symmetric subsets.
		\item An \emph{odd} integer (positive or negative) indicates that the label at a given slot has a symmetry or antisymmetry, according to the sign of the integer.  All labels at slots with the same integer participate in the same symmetry.  Odd integers may be entered for both dummy labels and component labels.
		\item An \emph{even} integer (positive or negative) indicates that the dummy label at a given slot has a symmetry/antisymmetry which has been \emph{propagated} from its corresponding partner with odd-numbered entry.  If the even-numbered entry is $k$, then its partner symmetry is the odd number of the same sign, and one less in absolute value, $(\sign k) \cdot (\abs{k} - 1)$.
		\item \label{prop-sym-precedence} The process which creates the entries shall work left-to-right in slot order, alternatively enumerating the odd-numbered (anti)symmetric sets and then propagating to their corresponding even-numbered partners.  If at any time this process would cause a new (nonzero) entry to overwrite an old (nonzero) entry, the entry with \emph{lower absolute value} shall take precedence (such conflicts can happen when both ends of a contraction are attached to the \emph{same} set of symmetric slots, or when the ``far'' end of a contraction is attached to another symmetric subset, distinct from the subset to which the ``near'' end is attached).
		\item \label{diff-values} If there are labels whose slots have the same \variable{Symmetric-Subsets} entry, but whose \variable{Values} entries are different (thus, they belong to different component-label or dummy sets), then they will be given \emph{different} odd integers; likewise, any of their propagated partners must receive different (and corresponding) even integers.  Thus the entries in \variable{Propagated-Symmetries} must always represent valid label symmetries.
		\item \label{no-singletons} Nonzero entries of either type will be entered only if they are shared by at least two dummy labels (\emph{after} applying Rule~\ref{prop-sym-precedence}); ``symmetric subsets of length 1'' are equivalent to non-symmetric labels, and should have the entry 0.
		\item As with the \variable{Symmetric-Subsets} array, odd integers will be chosen in a sequence of increasing absolute value, and each absolute value will be used for only one subset.  Even integers are always entered as partners of odd ones, and hence also form such a sequence.  The even-numbered sequence may have gaps, either because component labels do not have partners, or as a result of Rule~\ref{prop-sym-precedence}.
	\end{enumerate}
\end{definition}

The properties satisfied by \variable{Propagated-Symmetries} are given in Definition~\ref{def:prop-sym}.  Of all of the data structures we will introduce, this one has the most complex system of rules, and some examples will help clarify how it is to be filled out.  First, suppose we have a tensor $T_{abcdef}$ which is symmetric in its last 4 slots, and thus has the \variable{Symmetric-Subsets} array
\begin{equation}
\begin{array}{r|cccccc}
\text{Slot}		& 1		& 2		& 3		& 4		& 5		& 6	\\
\hline
\text{Subset}	& 0		& 0		& 1		& 1		& 1		& 1
\end{array}\, .
\end{equation}
We have previously considered the expression
\begin{equation}
T_{11ab}{}^{bc} = T_{1_1 1_2 a b_1}{}^{b_2 c},
\end{equation}
whose \variable{Values} array is
\begin{equation}
\begin{array}{r|cccccc}
\text{Label}	& a		& c		& 1_1	& 1_2	& b_1	& b_2	\\
\hline
\text{Value}	& 1		& 2		& 3		& 3		& 5		& 5
\end{array}\, .
\end{equation}
Now consider how to fill out its \variable{Propagated-Symmetries} array according to the rules in Definition~\ref{def:prop-sym}.  It has one subset of equivalent component labels $\set{1_1, 1_2}$, but they are in slots $\seq{1,2}$ which do not have a slot symmetry.  Therefore the $\seq{1,2}$ entries of \variable{Propagated-Symmetries} should be 0.  Next, there is one dummy pair, $\set{b_1, b_2}$ in slots $\seq{4,5}$.  These slots are in a symmetric subset with each other, and thus should receive the odd-number entry 1.  Because of the precedence established in Rule~\ref{prop-sym-precedence}, there will be no entries for propagated symmetries, because both ends of the $b_1 \smile b_2$ contraction are already attached to the same symmetric subset.  The \variable{Propagated-Symmetries} array is therefore
\begin{equation}
\begin{array}{r|cccccc}
\text{Slot}		& 1		& 2		& 3		& 4		& 5		& 6	\\
\hline
\text{Symmetry}	& 0		& 0		& 0		& 1		& 1		& 0
\end{array}\, .
\end{equation}

Next, consider a slightly more elaborate example: the expression $T_{abcdef} R^{cd}{}_{gh}$. Again we take $T$ to be symmetric on its last 4 slots, and $R$ is the Riemann tensor.  We first write out the \variable{Values} array (remembering that free indices are always sorted before dummies):
\begin{equation}
\begin{array}{r|cccccccccc}
\text{Label}	& a		& b		& e		& f		& g		& h		& c_1	& c_2	& d_1	& d_2 \\
\hline
\text{Value}	& 1		& 2		& 3		& 4		& 5		& 6		& 7		& 7		& 7		& 7
\end{array}\, .
\end{equation}
Next, the \variable{Symmetric-Subsets} array is as in \eqref{TR-sym-subsets}:
\begin{equation}
\begin{array}{r|rrrrrrrrrr}
\text{Slot}		& 1		& 2		& 3		& 4		& 5		& 6		& 7		& 8		& 9		& 10	\\
\hline
\text{Subset}	& 0		& 0		& 1		& 1		& 1		& 1		& -2	& -2	& -3	& -3
\end{array}\, .
\end{equation}
Finally, to construct the \variable{Propagated-Symmetries} array, we note that the only non-free labels are in slots $\set{3,4,7,8}$, and by the precedence given in Rule~\ref{prop-sym-precedence}, we obtain
\begin{equation}
\begin{array}{r|cccccccccc}
\text{Slot}		& 1		& 2		& 3		& 4		& 5		& 6		& 7		& 8		& 9		& 10	\\
\hline
\text{Symmetry}	& 0		& 0		& 1		& 1		& 0		& 0		& 2		& 2		& 0		& 0
\end{array}\, .
\end{equation}
In particular, we do \emph{not} get $-3,-3$ in slots $\seq{7,8}$.  This we see that \variable{Propagated-Symmetries} allows us to ``look ahead'' and see that this contraction must be zero, because the entries in slots $\seq{7,8}$ of \variable{Propagated-Symmetries} disagree in sign from those in $\seq{7,8}$ of \variable{Symmetric-Subsets}.

As a final example, consider the expression $T_{ab11cd} R^c{}_e{}^d{}_f$, again with the same tensors.  This time, the \variable{Values} array is
\begin{equation}
\begin{array}{r|cccccccccc}
\text{Label}	& a		& b		& e		& f		& 1_1	& 1_2	& c_1	& c_2	& d_1	& d_2 \\
\hline
\text{Value}	& 1		& 2		& 3		& 4		& 5		& 5		& 7		& 7		& 7		& 7
\end{array}\, ,
\end{equation}
and the \variable{Symmetric-Subsets} array remains the same as in \eqref{TR-sym-subsets}:
\begin{equation}
\begin{array}{r|rrrrrrrrrr}
\text{Slot}		& 1		& 2		& 3		& 4		& 5		& 6		& 7		& 8		& 9		& 10	\\
\hline
\text{Subset}	& 0		& 0		& 1		& 1		& 1		& 1		& -2	& -2	& -3	& -3
\end{array}\, .
\end{equation}
However, to construct the \variable{Propagated-Symmetries} array, we must apply Rule~\ref{diff-values}, since there are both component labels and dummy labels appearing in the symmetric subset of slots $\set{3,4,5,6}$, and these label sets have different entries in the \variable{Values} array.  The symmetric subset ``splits'' into two pieces, giving the \variable{Propagated-Symmetries} array
\begin{equation}
\begin{array}{r|cccccccccc}
\text{Slot}		& 1		& 2		& 3		& 4		& 5		& 6		& 7		& 8		& 9		& 10	\\
\hline
\text{Symmetry}	& 0		& 0		& 1		& 1		& 3		& 3		& 4		& 0		& 4		& 0
\end{array}\, .
\end{equation}
Notice that slots $\set{7,9}$ must be marked $4,4$, which shows that they have an induced symmetry propagated from slots $\set{5,6}$.  There are no entries with the number 2, because the labels in slots $\set{3,4}$ are component labels and do not have partners.

Finally, note that because we have chosen \variable{Propagated-Symmetries} to be \emph{slot}-indexed, we must store the slot symmetry $s$ which was used to reach the current configuration $g$ from the initial one $g_\text{init}$.  In Algorithm~\ref{alg:improved}, we will re-define \variable{Configs} to be the set of ordered pairs $(g,s)$.  Alternatively, one could store $(\ell, s)$ as was done in the original implementation of Butler-Portugal \cite{2002IJMPC..13..859M} and use the relation $g = \ell\access{g_\text{init}\access{s}}$; however, we find it more useful to store $(g,s)$ as $\ell$ alone is not typically needed.\footnotemark{}

\footnotetext{If we had chosen \variable{Propagated-Symmetries} to be label-indexed, we would instead need to store $(g, \ell)$.}

%%%%%%%%%%%%%%
\subsubsection{The \variable{Config-Slots-of-Least-Value} array and its \variable{Least-Value-Set} entries}
\label{sec:config-slots-of-least-value}
%%%%%%%%%%%%%%

One final data structure we will need is the \variable{Config-Slots-of-Least-Value} array, which replaces the \variable{Config-Slots-of-Least-Label} array appearing in Algorithm~\ref{alg:butler-portugal}.  The overall purpose of \variable{Config-Slots-of-Least-Value} will be the same: to store the configurations and slot numbers at which one can find a label with least value.  However, because the properties of the label symmetry group are now distributed between three data structures \variable{Values}, \variable{Label-Groups}, and \variable{Propagated-Symmetries}, some slight changes will be necessary.

First, as mentioned in Section~\ref{sec:propagated symmetries array}, the \variable{Configs} array will now have to store ordered pairs $\set{(g,s)}$ where $g$ is the total slot-to-label map of the configuration, and $s$ is the total action of the \emph{slot} symmetry group which brought us to $g = \ell\access{g_\text{init}\access{s}}$.  Therefore each entry in the \variable{Config-Slots-of-Least-Value} array must contain a reference to the ordered pair $(g,s) \in \variable{Configs}$ from which the current least-value slot numbers originate (that is, $(g,s)$ is a reference to the parent node in the search tree).

In addition, each entry of \variable{Config-Slots-of-Least-Value} must contain an ordered \emph{pair} slot numbers $(p,q)$ rather than just a single one.  The reason for this is because label symmetries can come from two different sources: the original label symmetry group $L$ which is represented by the combination of \variable{Values} and \variable{Label-Groups}; and the new label symmetries which come from propagation of slot symmetries, as represented in \variable{Propagated-Symmetries}.\footnotemark{}

\footnotetext{One could alternatively use a Schreier-tree structure to store the entire label group as was done in Algorithm~\ref{alg:butler-portugal}, but this would sacrifice quite a bit of speed, as one would have to add the propagated symmetries by ``sifting'' them into the Schreier tree, and then re-compute orbits, etc.}

Thus the \variable{Config-Slots-of-Least-Value} array will have the following structure:
\begin{equation}
\begin{array}{r|cccc}
\text{Configuration}	& (g_1, s_1)	& (g_2, s_2)	& \dotsc	& (g_m, s_m) \\
\hline
\text{Slot pair}		& (p_1, q_1)	& (p_2, q_2)	& \dotsc	& (p_m, q_m)
\end{array}\, ,
\end{equation}
where one can think of each $(g_k,s_k)$ as a reference into the \variable{Configs} array.  We do not expect each of the $(g_k, s_k)$ to be distinct, as it is possible that a given configuration contain multiple labels of (equal) least value in the orbit of the slot currently under consideration.  We will find it useful to collect together all of the \variable{Config-Slots-of-Least-Value} coming from the \emph{same} configuration $(g,s)$ into a \variable{Least-Value-Set}; thus, the \variable{Least-Value-Set} corresponding to a given $(g,s)$ is precisely the set of child nodes of the search tree under the parent node $(g,s)$.\footnotemark{}  The meanings of the slot pairs $(p_k, q_k)$ are as follows:
\begin{enumerate}
	\item The $p_k$ are the slot numbers in the slot-symmetry orbit $\Delta^S_i$ where a least-value label can be found directly.
	\item The $q_k$ are the slot numbers where a least-value label can be found by application of \variable{Propagated-Symmetries}, even if that slot is \emph{outside} the orbit $\Delta^S_i$.
\end{enumerate}
Usually, one has $p_k = q_k$ for each $k$, \emph{except} in the case where the \variable{Propagated-Symmetries} have been used to step outside the usual slot-symmetry orbit to find the least-value label.  Then $p_k$ gives the point in the slot-symmetry orbit we jumped away from, and $q_k$ gives the place where the least-value label was actually found.  These two slot numbers together give us implicit information about how to reconstruct the necessary slot- and label-symmetries to put the least label into our current position.

\footnotetext{Thus \variable{Config-Slots-of-Least-Value} could be implemented as a multi-dimensional array containing \variable{Least-Value-Set}'s, although one must take into account that each \variable{Least-Value-Set} may be of different length.  For the timing data in Section~\ref{sec:testing}, we have chosen a flat implementation.}

%%%%%%%%%%%%%%
\subsection{The main algorithm}
\label{sec:full algorithm}
%%%%%%%%%%%%%%

Having developed the necessary preliminary concepts, we now present the improved algorithm.  We will first give an overview in Algorithm~\ref{alg:improved} of the main loop which iterates once over the slots of the initial configuration $g_\text{init}$.  There are several subprocedures called by the main loop, most of whose code we relegate to Appendix~\ref{app:subprocedures}, with the exception of \textproc{Append-Non-Redundant-Instances} which contains the essential logic that prevents iteration over redundant branches of the search tree.

%%%%%%%%%%%%%%%
% Improved algorithm
%%%%%%%%%%%%%%%

% the algorithm itself
\begin{algorithm}
	\caption{A faster algorithm for tensor canonicalization}
	\label{alg:improved}
	\vspace{0.5ex}
	\begin{algorithmic}
		\Input Configuration $g_\text{init}$;~ slot symmetry group $S$;~ \variable{Values}, \variable{Label-Groups}, and \variable{Symmetric-Subsets}
		\Output Canonical configuration $g_\text{can}$ or zero if tensor vanishes
	\end{algorithmic}
	\vspace{0.5ex} \hrule \vspace{0.5ex}
	\begin{algorithmic}[1]\raggedright
		\Procedure{Canonicalize}{$g_\text{init}, S, \variable{Values}, \variable{Label-Groups}, \variable{Symmetric-Subsets}$}
			\State $\variable{Configs} \gets \set{ (g_\text{init}, \id) }$
			\State $\variable{Propagated-Symmetries} \gets \seq{0,0,\dotsc,0}_n$
			\For{$i \gets 1$ to $n$}
			\label{main slot loop start}
				\State $\Delta^S_i \gets$ \set{orbit of slot $i$ under $S^{(i)}$}
				\label{get orbit}
				\State $\set{\variable{least-value}, \variable{Config-Slots-of-Least-Value}} \gets \function{Get-Least-Value-Instances}{i, \Delta^S_i,  \variable{Configs}, \variable{Values}, \variable{Propagated-Symmetries}}$
				\label{get config slots}
				\State $\variable{Next-Configs} \gets \set{}$
				\ForEach{$(g,s) \in \variable{Configs}$}
				\label{config loop start}
					\State $\variable{Least-Value-Set} \gets$ \set{all entries of \variable{Config-Slots-of-Least-Value} coming from $(g,s)$}
					\State $\variable{Propagated-Symmetries} \gets \linebreak \function{Update-Propagated-Symmetries}{\variable{Least-Value-Set}, (g,s), \variable{least-value}, \linebreak \hspace*\continueindent \variable{Label-Groups}, \variable{Symmetric-Subsets}, \variable{Propagated-Symmetries}}$
					\If{\big($\function{Zero-Due-to-Propagated-Symmetries}{(g,s), \variable{least-value}, \linebreak \hspace*\continueindent \variable{Label-Groups}, \variable{Symmetric-Subsets}, \variable{Propagated-Symmetries}}$\big)}
						\State \Return 0
					\EndIf
					\State $\variable{Next-Configs} \gets \function{Append-Non-Redundant-Instances}{\variable{Next-Configs}, \variable{Least-Value-Set}, (g,s), \linebreak \hspace*\continueindent \variable{least-value}, S, i, \variable{Label-Groups}, \variable{Symmetric-Subsets}, \variable{Propagated-Symmetries}}$
				\EndFor
				\label{config loop end}
				\State $\set{\variable{Values}, \variable{Label-Groups}} \gets \function{Update-Values-and-Label-Groups}{\variable{Values}, \variable{Label-Groups}, \variable{least-value}}$
				\label{last bit start}
				\State $\variable{Configs} \gets \function{Remove-Duplicates}{\function{Sort}{\variable{Next-Configs}}}$
				\If{\big(\variable{Configs} contains equal configurations $+g$ and $-g$ of opposite sign\big)}
					\State \Return 0
				\EndIf
				\label{last bit end}
			\EndFor
			\label{main slot loop end}
			\State \Return $g_\text{can} = $ first element of \variable{Configs}
		\EndProcedure
	\end{algorithmic}
\end{algorithm}

% Need to construct manual reference to this algorithm for externalized PGFplots
\edef\theimprovedalgorithm{\thealgorithm}

The general strategy of Algorithm~\ref{alg:improved} is the same as Algorithm~\ref{alg:butler-portugal}: namely, to visit each slot one at a time from left to right, determine the least label which can be moved into that slot, place that label, and then move on to the next slot.  As before, if there are multiple ways in which the (same) least label can be moved into the current slot, then the search tree bifurcates, and the next iteration must look at all of the resulting possibilities.  The key difference is that in Algorithm~\ref{alg:improved}, we make an effort to prune these extra branches whenever they would be redundant.  Not all types of redundancy are detected, but we do handle the most common case of (anti)symmetric subgroups filled with dummy labels.  We apply the symmetry propagation concepts of Section~\ref{sec:dummy links} in order to make an early decision about placing such dummy labels into slots, while preserving the symmetry information needed to ensure that later slots can be canonicalized.  The same information can be used for early detection of zero results.

The main loop over the slots of $g_\text{init}$ runs from lines \ref{main slot loop start} through \ref{main slot loop end} of Algorithm~\ref{alg:improved}.  In line~\ref{get orbit}, we obtain the orbit $\Delta^S_i$ of the current slot $i$ under the slot symmetry group $S$ (an operation which can be made quite fast if we have stored $S$ in the form of a Schreier tree structure ahead of time).  In line~\ref{get config slots}, we populate the \variable{Config-Slots-of-Least-Value} array described in Section~\ref{sec:config-slots-of-least-value}, as well as record the \variable{least-value} label which can be moved into this slot (either via the slot symmetry group $S$, or via the extra label symmetries in \variable{Propagated-Symmetries}).

Next in lines \ref{config loop start} through \ref{config loop end}, we iterate over the next level of nodes of the search tree, grouped by their parent node $(g,s) \in \variable{Configs}$ in the form of a \variable{Least-Value-Set}.  On each \variable{Least-Value-Set} we make \emph{two} passes; first to iteratively add information to the \variable{Propagated-Symmetries} array, and then to use the information to make decisions about which leaf nodes to append to \variable{Next-Configs}.  Note that the data in \variable{Propagated-Symmetries} is filled out according to the rules in Definition~\ref{def:prop-sym}, but it is not done all at once; rather, only the \variable{Symmetric-Subsets} which are encountered within the current \variable{Least-Value-Set} are entered.  Thus at any given time before the last iteration of the algorithm, the \variable{Propagated-Symmetries} array is incomplete; however, it always contains just enough information to be used in \textproc{Append-Non-Redundant-Instances}.\footnotemark{}

\footnotetext{Propagated symmetries always come from slot symmetries, and if there is a symmetric subgroup to be propagated, then it is always true that all of its slot numbers should be listed in the orbit $\Delta^S_i$, and thus appear in the current \variable{Least-Value-Set}.  Thus at the very latest, each symmetry which can be propagated is entered into \variable{Propagated-Symmetries} immediately before it is needed for the decision-making phase.  One \emph{could} populate the entire \variable{Propagated-Symmetries} ahead of time, but doing so iteratively as we do here prevents us having to do any more work than necessary (in the event, for example, that the entire algorithm is short-circuited by a zero detection).}

In lines \ref{last bit start} through \ref{last bit end}, we do some final steps before moving on to the next slot iteration.  First we must update the \variable{Values} and \variable{Label-Groups} arrays to maintain the properties in Definition~\ref{def:value} and Table~\ref{tab:group codes}.  This effectively implements ``re-ordering the base'' for the label symmetry group, as the lowest-value label will be marked as used and its exchange symmetry with other labels will be erased.  Finally we sort and remove duplicates from \variable{Next-Configs}, copy the result back into \variable{Configs}, and check for zero.

The subprocedures \textproc{Get-Least-Value-Instances}, \textproc{Update-Propagated-Symmetries}, and \textproc{Update-Values-and-Label-Groups} are intended to populate the various data structures defined in Section~\ref{sec:data structures} and maintain the conditions of their definitions for each loop iteration.  The conditional \textproc{Zero-Due-to-Propagated-Symmetries} constitutes the early check for zero using the extra information assembled in the data structures of Section~\ref{sec:data structures}.  However, we emphasize here that this check is \emph{not} optional; rather, it is \emph{necessary} to take into account all the possibilities for a zero result that might be contained in \variable{Propagated-Symmetries} immediately.  This allows \textproc{Append-Non-Redundant-Instances} to be quite liberal in its rejection of redundant search-tree branches.  A detailed description of these subprocedures can be found in Appendix~\ref{app:subprocedures}.

%%%%%%%%%%%%%%%
% Append-Non-Redundant-Instances
%%%%%%%%%%%%%%%
\begin{algorithm}
	\caption{Append-Non-Redundant-Instances}
	\label{alg:append-non-redundant-instances}
	\vspace{0.5ex}
	\begin{algorithmic}\raggedright
		\Input Current list of \variable{Next-Configs} and \variable{Least-Value-Set};~ configuration $(g,s)$;~ the \variable{least-value} in this configuration;~ slot group $S$;~ slot $i$;~ \variable{Label-Groups}, \variable{Symmetric-Subsets}, \linebreak and \variable{Propagated-Symmetries}
		\Require The possibility of obtaining a zero result merely from the conflict of \variable{Propagated-Symmetries} with \variable{Symmetric-Subsets} has already been taken into account
		\Output New \variable{Next-Configs} with non-redundant items added from \variable{Least-Value-Set}
	\end{algorithmic}
	\vspace{0.5ex} \hrule \vspace{0.5ex}
	\begin{algorithmic}[1]\raggedright
		\Procedure{Append-Non-Redundant-Instances}{$\variable{Next-Configs}, \variable{Least-Value-Set}, (g,s), \variable{least-value}$, \hspace*\continueindent $S, i, \variable{Label-Groups}, \variable{Symmetric-Subsets}, \variable{Propagated-Symmetries}$}
			\State $\variable{Visited-Subsets} \gets \seq{0,0,\dotsc,0}_n$
			\label{initialize visited subsets}
			\ForEach{$(p,q) \in \variable{Least-Value-Set}$}
				\State $\variable{label} \gets g\access{q}$
				\Comment Note that we use $q$ here
				\If{\big($\variable{Symmetric-Subsets}\access{p} \neq 0$\big)}
				\label{ignore redundant start}
					%\State $\variable{subset-index} \gets \abs{\variable{Symmetric-Subsets}\access{p}}$
					\If{\big($\variable{Visited-Subsets}\access{\, \abs{\variable{Symmetric-Subsets}\access{p}} \,} = 0$\big)}
					\label{check visited}
						\State $\variable{Visited-Subsets}\access{\, \abs{\variable{Symmetric-Subsets}\access{p}} \,} \gets 1$
						\label{mark visited}
					\Else
						\State \Skip to next $(p,q)$
					\EndIf
				\EndIf
				\label{ignore redundant end}
				\State $\tilde \ell \gets \id$
				\If{\big($\variable{Propagated-Symmetries}\access{s\access{q}} \neq 0$ and $p \neq q$\big)}
					\State $\varepsilon \gets \function{Sign}{\variable{Propagated-Symmetries}\access{s\access{q}}}$
					\State $\tilde \ell \gets \tilde \ell \access{\text{signed permutation} \; \varepsilon \times (\variable{label}, g\access{p}) \; \text{in array form}}$
				\EndIf
				\State $\tilde \ell \gets \tilde \ell \access{\function{Label-Permutation-from-Group}{\variable{label}, \variable{least-value}, \variable{Label-Groups}\access{\variable{label}}}}$
				\State $\tilde s \gets \function{Coset-Rep}{p,S^{(i)}}$
				\Comment We require $p$ here, not $q$
				\State $g' \gets \tilde \ell \access{g\access{\tilde s}}$
				\State $s' \gets s\access{\tilde s}$
				\State $\variable{Next-Configs} \gets \function{Append}{\variable{Next-Configs}, (g', s')}$
			\EndFor
			\State \Return \variable{Next-Configs}
		\EndProcedure
	\end{algorithmic}
\end{algorithm}

However, we single out \textproc{Append-Non-Redundant-Instances} for discussion here, as this subprocedure contains the key logic which prevents factorial growth of intermediate results.  We present this subprocedure in Algorithm~\ref{alg:append-non-redundant-instances}.  The rejection functionality is achieved with a simple check in lines \ref{ignore redundant start} through \ref{ignore redundant end} which ensures that within the current \variable{Least-Value-Set}, only one representative from each symmetry subset is kept.  This is done via an array \variable{Visited-Subsets} which marks whether a subset has been visited; in lines \ref{check visited} and \ref{mark visited}, we use the absolute value of $\variable{Symmetric-Subsets}\access{p}$ as the index into \variable{Visited-Subsets}.  Note that for any $(p,q)$ within a \variable{Least-Value-Set}, all of entries $\variable{Values}\access{q}$ in the \variable{Values} array must be the same (since they are the least value), and thus any two entries belonging to the same symmetric subset must in fact be exchangeable (even if there may be labels with \emph{other} values in the other slots of this symmetric subset!).  In order to rely on this complete exchange equivalence (i.e., the slots are totally (anti)symmetric, and we likewise treat the labels as totally (anti)symmetric), we must use the symmetry-propagation mechanism of Section~\ref{sec:dummy links}.  And since we will now throw out any other intermediate configurations resulting from these symmetries, we will not be able to detect a future sign conflict, so it must have already been dealt with (via \textproc{Zero-Due-to-Propagated-Symmetries}).

%%%%%%%%%%%%%%
\subsection{Comments on complexity}
\label{sec:complexity}
%%%%%%%%%%%%%%

Rather than perform detailed complexity analysis, we will present data in Section~\ref{sec:testing} which shows that our algorithm, like the original Butler-Portugal algorithm, is polynomial in the most common situations.  Here we will make only a few comments which relate to our improvements to the algorithm.

We have made two main improvements:  First, we have chosen a means to represent the label symmetry group $L$ by distributing this information among several small arrays:  \variable{Values}, \variable{Label-Groups}, and \variable{Propagated-Symmetries} for storing the additional label symmetries which might be added during the course of the algorithm.  Each of these arrays is just a single row of length $n$, and thus we incur an $O(n)$ memory cost, which is dwarfed by the $O(n^3)$ cost of storing the Schreier-tree structure of the slot symmetry group $S$.  While the new data structures require special logic to deal with them, their benefit is to reduce all group-theory computations involving the label symmetry group to either to array lookups which take $O(1)$ time, or single scans through the array which take $O(n)$ time.  In either case, the time taken to compute with the label symmetry group is now insignificant.

Second, the two arrays \variable{Symmetric-Subsets} and \variable{Propagated-Symmetries} allow us to eliminate equivalent choices from the search tree when presented with several interchangeable labels in a set of slots which are (anti)symmetric.  At a cost of $O(n)$ space, we can avoid having to store $O(n!)$ intermediate results, which take $O(n!)$ time to process, a dramatic improvement.

However, we note that $O(n!)$ behavior has not been eliminated entirely.  Our algorithm is only capable of detecting subsets of indices which are totally symmetric or antisymmetric.  But there are other possibilities which lead to combinatorial explosion.  For example, suppose that a tensor has many indices which are \emph{pairwise} symmetric, with a slot symmetry group generated by the pairwise exchanges
\begin{equation}
S = \{ (1,3)(2,4), \; (3,5)(4,6), \; \dotsc, \; (2n-3,2n-1)(2n-2,2n) \}.
\end{equation}
Such a slot symmetry group has order $n!$ and will not be detected because it does not involve the direct exchange between two slots.  One can imagine other factorial-size groups such as alternating groups or groups which symmetrize length-$k$ subsets, etc.  These are not caught by the algorithm, and thus it will exhibit $O(n!)$ behavior in these cases.  Fortunately, however, these cases are rare in actual calculation---in contrast to the (anti)symmetric case which we have addressed!

One can see, in fact, that it is impossible to eliminate all possible sources of $O(n!)$ behavior (as should be expected \cite{DBLP:Butler91}).  Take the simple case of symmetric exchanges of length-$k$ subsets, for example.  It may be possible, with some cleverly-designed data structures, to write an algorithm which---\emph{given data about these exchange symmetries ahead of time}---can resolve canonicalization problems in polynomial time for such groups.  But even if such an algorithm were designed, this is not enough, because that data about the symmetric exchange of length-$k$ subsets must \emph{first} be somehow obtained.  To generate it algorithmically for a \emph{specific} $k$ requires $O(n^{k+1})$ group membership tests (and thus, in our particular case which resolves the question for length-1 exchanges, one can construct the \variable{Symmetric-Subsets} structure using $O(n^2)$ group-membership tests).  To do so for \emph{all} $k$ (thus, up to $k=n/2$) would then require $O(n!)$ group membership tests, and so we gain nothing; removing one head from the hydra only springs forth more.

We point out, however, that there is a case in which data about length-$k$ exchanges \emph{can} be provided ahead of time, in polynomial time:  The case where a tensor monomial is constructed out of the tensor product of several identical factors,
\begin{equation}
T_{abc} T_{def} T_{ghi} \dotsm.
\end{equation}
In such cases, it may be useful to have an algorithm which efficiently handles this type of symmetry.  But even so, it will still be a special case, and $O(n!)$ behavior is still possible by some other means.

%%%%%%%%%%%%%%
\section{Performance testing} 
\label{sec:testing}
%%%%%%%%%%%%%%

To test the effectiveness of Algorithm~\ref{alg:improved}, we have fully implemented it and tested it in several situations against two implementations of the Butler-Portugal algorithm: the one in \software{xPerm} \cite{MartinGarcia2008597} which is used to simplify tensor expressions in \software{xAct} \cite{citeulike:13127953}, and the built-in function \code{TensorReduce} which was introduced in \software{Mathematica} version 9.  For the purpose of these tests, Algorithm~\ref{alg:improved} has been implemented in \software{Mathematica} code, and wrapped within the command \code{Compile[\ldots, CompilationTarget$\phantom{}\to\phantom{}$\textquotedbl C\textquotedbl]}.  This causes \software{Mathematica} to automatically generate \code{C} code (from the enclosed \software{Mathematica} code), which is then compiled to machine language via an external \code{C} compiler.\footnotemark{}  We have used the most up-to-date version of \software{xPerm}, from \software{xAct} version \code{1.1.2}.  The function \code{TensorReduce} is that available in \software{Mathematica} version \code{11.0.0.0}, which was used to run all of these tests.

\footnotetext{It is therefore conceivable that one might achieve slightly better performance by coding directly in \code{C}, which would allow more control over memory resources.}

In each test, we run each of the three implementations on randomized input with certain constraints as will be explained below.  We have attempted to increase the input size until the asymptotic behavior becomes clearly visible, or at least until the total running time begins to exceed a few minutes.  For each input size, we run the test several times (typically 10, except for the Riemann tensor tests which used 30 trials in order to generate enough results for each of the zero and nonzero bins).  Different random input is generated for each trial.  In each graph, the error bars represent the full range of running times over the trials for a given input size.  The plot marks are placed at the geometric mean of the running times (thus, in these log plots, representing the mean of the logs).

Each of the three implementations takes slightly different input, but we have made our best effort to eliminate as much overhead as possible.  Algorithm~\ref{alg:improved} takes a pre-computed Scheier tree structure as input, as well as \variable{Symmetric-Subsets}, and represents the label symmetry group via the inputs \variable{Values} and \variable{Label-Groups}.  For \software{xPerm}, we have given it a pre-computed base-and-strong-generating-set (BSGS) for the slot symmetry group, and checked a flag indicating that this does not need to be re-computed.  Unfortunately with \software{Mathematica}'s built-in function \code{TensorReduce}, there is no way to specify that the slot symmetries are already a BSGS, and so it spends time running the Schreier-Sims algorithm attempting to construct a BSGS which dominates the timing for low numbers of slots.  Therefore we should point out that the data here are not entirely ``fair''; however, the differences in performance are quite significant, in some cases differing by a whole power of $n$ and not just a constant factor.

We also point out that each of these three implementations produces slightly different \emph{output}; that is, each algorithm is using a slightly different notion of what ``canonical'' means.  \software{xPerm} works in two stages, first moving all of the free indices as far forward as possible, then fixing them in place and canonicalizing the remaining indices by lexicographic order.  \software{Mathematica} uses a definition of ``canonical'' based on the \emph{slot numbers} of the contracted pairs, rather than the labels that appear in those slots, and so it may produce a different ordering.  Algorithm~\ref{alg:improved} uses similar criteria for ``canonical'' as does \software{xPerm}, but works in one stage only, moving free and dummy indices as they appear; it is \emph{possible} for them to give different results.  Nevertheless, we have tested Algorithm~\ref{alg:improved} extensively for correctness, before running these timing tests.

The \software{Mathematica} source code used to implement the algorithm and run these tests can be found at \url{https://github.com/bniehoff/tensor-canonicalizer}.

%%%%%%%%%%%%%%%%
\subsection{Preliminaries}
\label{sec:preliminaries}
%%%%%%%%%%%%%%%%

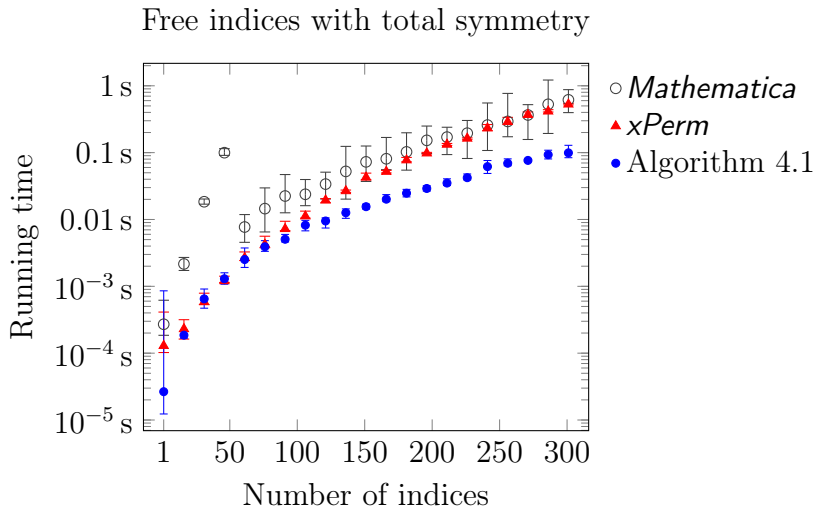
\begin{figure}
	\centering
	\tikzsetnextfilename{sym-frees}
	\input{tikz/sym-frees.tikz}
	\caption{Free indices with total symmetry}
	\label{fig:sym-frees}
\end{figure}

We start with three tests which do not significantly stress any of the algorithms, and give a sort of basis of comparison which may help counteract the imbalance just mentioned.  First, in Figure~\ref{fig:sym-frees}, we run the algorithms with \emph{free} indices, with total slot symmetry.  Thus there is no label renaming ambiguity; the algorithms need only sort the list of indices.  From Figure~\ref{fig:sym-frees} it is clear that \software{Mathematica} is doing extra work by running Schreier-Sims every time; furthermore it is clear that at around 50 indices, \software{Mathematica} changes strategy to use a randomized Schreier-Sims algorithm.  To obtain an estimate on the asymptotic complexity, we attempt to fit a power law to the data points with at least 100 indices (thus ignoring the transient effects for low numbers of indices).  We obtain the following power law growth:
\begin{equation}
\software{Mathematica} \sim O(n^{2.91}), \qquad \software{xPerm} \sim O(n^{3.65}), \qquad \text{Algorithm~\ref*{alg:improved}} \sim O(n^{2.53}).
\end{equation}
The slightly smaller power in the growth rate for Algorithm~\ref{alg:improved} compared to \software{Mathematica} is likely due to the fact that Algorithm~\ref{alg:improved} starts from slot-symmetry data where the orbits $\Delta^S_i$ are pre-computed (we point out that in a system where tensors are declared in advance, it is always possible to do this).  Another possible explanation is that the original version of Butler-Portugal canonicalizes the free indices separately from the dummy ones using a slightly different algorithm \cite{MartinGarcia2008597,2001math.ph...7031P}.

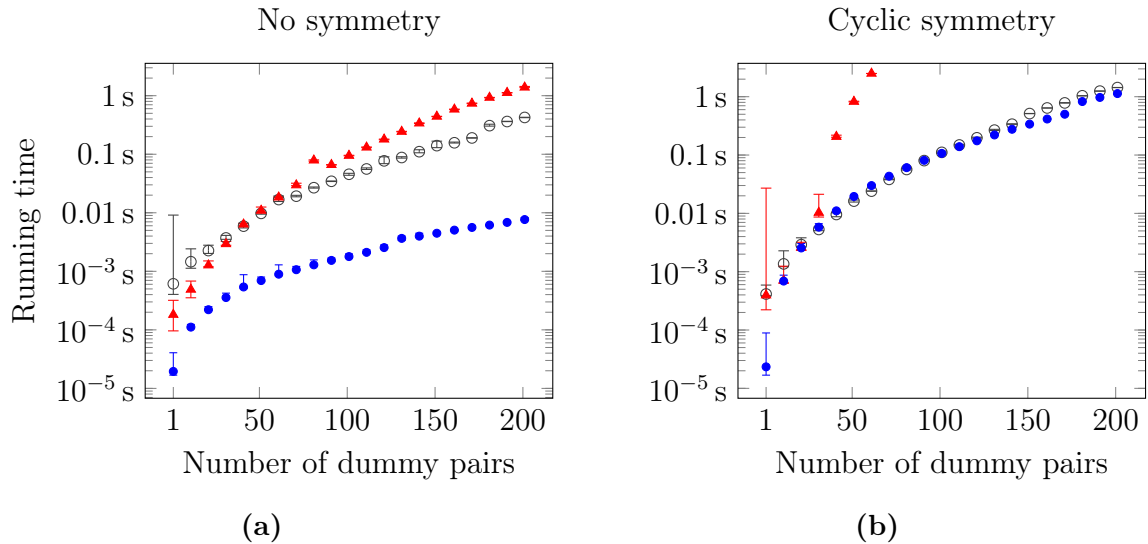
\begin{figure}
	\centering
	\begin{subfigure}{0.4\textwidth}
		\tikzsetnextfilename{no-sym-dummies}
		\input{tikz/no-sym-dummies.tikz}
		\vspace{-\baselineskip}
		\subcaption{}
		\label{fig:no-sym-dummies}
	\end{subfigure}
	\hspace{1cm}
	\begin{subfigure}{0.4\textwidth}
		\tikzsetnextfilename{cyclic-dummies}
		\input{tikz/cyclic-dummies.tikz}
		\vspace{-\baselineskip}
		\subcaption{}
		\label{fig:cyclic-dummies}
	\end{subfigure}
	\caption{Dummy indices without symmetry, and with cyclic symmetry.}
	\label{fig:dummy-no-sym-and-cyclic}
	\tikzsetnextfilename{legend}
	\input{tikz/legend.tikz}
\end{figure}

Next we run two basic tests with dummy indices in Figure~\ref{fig:dummy-no-sym-and-cyclic}.  In the first test there are \emph{no} slot symmetries, but instead we give a random sequence of dummy indices.  Thus, all the dummies need to be renamed in the optimal order, but since there are no slot symmetries, there should be no ambiguity in the search tree.  In Figure~\ref{fig:no-sym-dummies}, one can clearly see the advantage of our data structures \variable{Values} and \variable{Label-Groups} for representing the label group $L$, which allow us to reduce the computation of the label-group orbit $\Delta^{L}_{g\access{j}}$ to a simple array lookup.  Using a similar method of best-fit power law while throwing out the points with a small number of indices, we obtain the time complexities
\begin{equation}
\software{Mathematica} \sim O(n^{2.67}), \qquad \software{xPerm} \sim O(n^{3.57}), \qquad \text{Algorithm~\ref*{alg:improved}} \sim O(n^{1.84}).
\end{equation}

In Figure~\ref{fig:cyclic-dummies} we run a test which eliminates both the advantages of Algorithm~\ref{alg:improved} \emph{and} the drawbacks of the original Butler-Portugal algorithm.  The algorithms are given the task to canonicalize the expression
\begin{equation}
T_{\pi_1(a_1 a_2 \dotsc a_n)} U^{\pi_2(a_1 a_2 \dotsc a_n)}, \qquad T_{a_1 a_2 \dotsc a_n} = T_{a_2 \dotsc a_n a_1}, \qquad U^{a_1 a_2 \dotsc a_n} = U^{a_2 \dotsc a_n a_1},
\end{equation}
where the tensors $T, U$ have a \emph{cyclic} symmetry over all of their indices.  In the \software{Mathematica} algorithm, since the strong-generating-set of a cyclic group only has one generator, the Schreier-Sims step finishes quickly without significantly impacting performance.  In Algorithm~\ref{alg:improved}, one gets no benefit from the symmetry-propagation mechanism, because there are no (anti)symmetric subsets of slots.  Nevertheless, the cyclic symmetry, combined with dummy renaming, provide enough computational challenge that it dominates in both cases, and Figure~\ref{fig:cyclic-dummies} shows nearly identical performance.\footnotemark{}  The best-fit power law complexity is around $O(n^3)$.

\footnotetext{The behavior of \software{xPerm} in Figure~\ref{fig:cyclic-dummies} is somewhat strange in that there is a sharp corner around 35 dummy pairs where it suddenly jumps to a different power-law behavior.  We have discussed this with the author of \cite{MartinGarcia2008597} but cannot ascertain the cause.  \software{xPerm} does not show any strange behavior in any other tests.}

%%%%%%%%%%%%%%%%
\subsection{Random Riemann invariants}
\label{sec:random-riemann-invariants}
%%%%%%%%%%%%%%%%

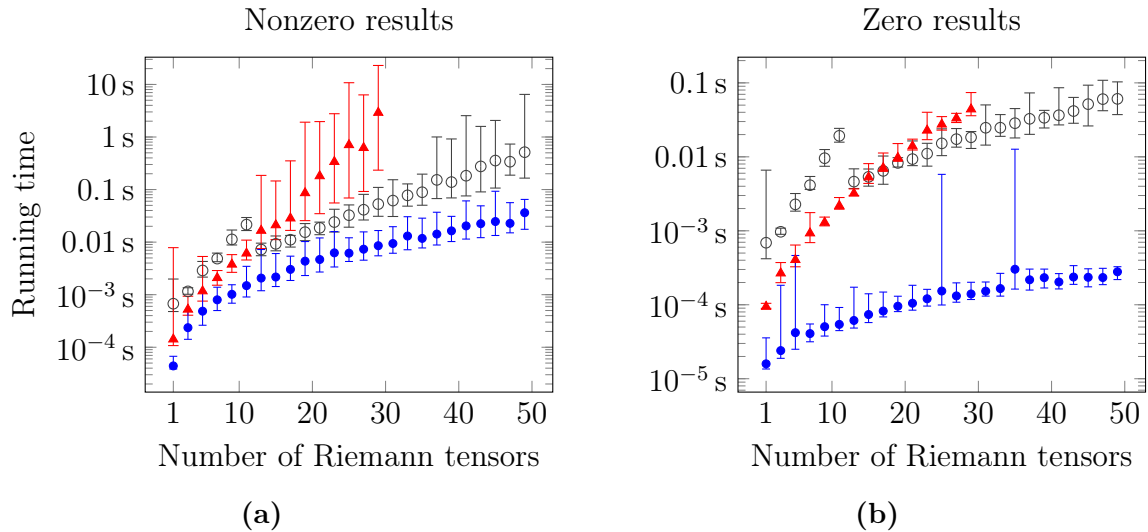
\begin{figure}
	\centering
	\begin{subfigure}{0.4\textwidth}
		\tikzsetnextfilename{nonzero-riemanns}
		\input{tikz/nonzero-riemanns.tikz}
		\vspace{-\baselineskip}
		\subcaption{}
		\label{fig:nonzero-riemanns}
	\end{subfigure}
	\hspace{1cm}
	\begin{subfigure}{0.4\textwidth}
		\tikzsetnextfilename{zero-riemanns}
		\input{tikz/zero-riemanns.tikz}
		\vspace{-\baselineskip}
		\subcaption{}
		\label{fig:zero-riemanns}
	\end{subfigure}
	\caption{Randomized Riemann contractions, nonzero and zero results.}
	\label{fig:riemanns-zero-nonzero}
	\tikzsetnextfilename{legend}
	\input{tikz/legend.tikz}
\end{figure}

The next test is to repeat the benchmarking done in \cite{2002IJMPC..13..859M,MartinGarcia2008597,MartinGarcia:2008qz} for randomly-contracted Riemann monomials of the form
\begin{equation}
R^\circ{}_{\circ \circ \circ} R_{\circ \circ}{}^{\circ \circ} R^\circ{}_\circ{}^\circ{}_\circ \dotsm,
\end{equation}
where each $\circ$ should be filled with a random dummy index.\footnotemark{}  Since the Riemann tensor has antisymmetries, these invariants are occasionally zero.  The cases which turn out to be zero tend to have vastly shorter timing, and so we separate results into two bins, as shown in Figure~\ref{fig:riemanns-zero-nonzero}.  In Figure~\ref{fig:nonzero-riemanns}, for nonzero results we obtain the best-fit complexities
\begin{equation}
\software{Mathematica} \sim O(n^{3.86}), \qquad \software{xPerm} \sim O(e^{0.322 \, n}), \qquad \text{Algorithm~\ref*{alg:improved}} \sim O(n^{2.14}).
\end{equation}
We observe that not only is Algorithm~\ref{alg:improved} faster than \software{Mathematica} by at least an order of magnitude, it is also faster by about $n^{1.5}$ in time complexity.  It appears \software{xPerm} is not well-fit by a power law in this case, although perhaps we have only managed to capture its transient behavior.

\footnotetext{Unlike \cite{MartinGarcia:2008qz}, we do not include derivatives of the Riemann tensor, but only algebraic invariants.  We also do not impose any dimension-dependent identities, but use only the Riemann tensor symmetries.}

For invariants which turn out to be zero, Figure~\ref{fig:zero-riemanns} shows an even more dramatic improvement of up to 3 orders of magnitude and the following best-fit complexities:
\begin{equation}
\software{Mathematica} \sim O(n^{2.23}), \qquad \software{xPerm} \sim O(n^{3.08}), \qquad \text{Algorithm~\ref*{alg:improved}} \sim O(n^{1.13}).
\end{equation}
All of the algorithms are faster; again we see that Algorithm~\ref*{alg:improved} is faster than \software{Mathematica} by about one power of $n$ and a large constant factor.  Here we see the benefits of the symmetry-propagation mechanism for early zero detection.

%%%%%%%%%%%%%%%%
\subsection{Total symmetry}
\label{sec:total-symmetry}
%%%%%%%%%%%%%%%%

\begin{figure}
	\centering
	\begin{subfigure}{0.4\textwidth}
		\tikzsetnextfilename{total-sym-dummies}
		\input{tikz/total-sym-dummies.tikz}
		\vspace{-\baselineskip}
		\subcaption{}
		\label{fig:total-sym-dummies}
	\end{subfigure}
	\hspace{1cm}
	\begin{subfigure}{0.4\textwidth}
		\tikzsetnextfilename{random-total-sym-dummies}
		\input{tikz/random-total-sym-dummies.tikz}
		\vspace{-\baselineskip}
		\subcaption{}
		\label{fig:random-total-sym-dummies}
	\end{subfigure}
	\caption{Total symmetry with contracted dummies, both frustrated and randomized.}
	\label{fig:dummies-sym-and-random}
	\tikzsetnextfilename{legend}
	\input{tikz/legend.tikz}
\end{figure}
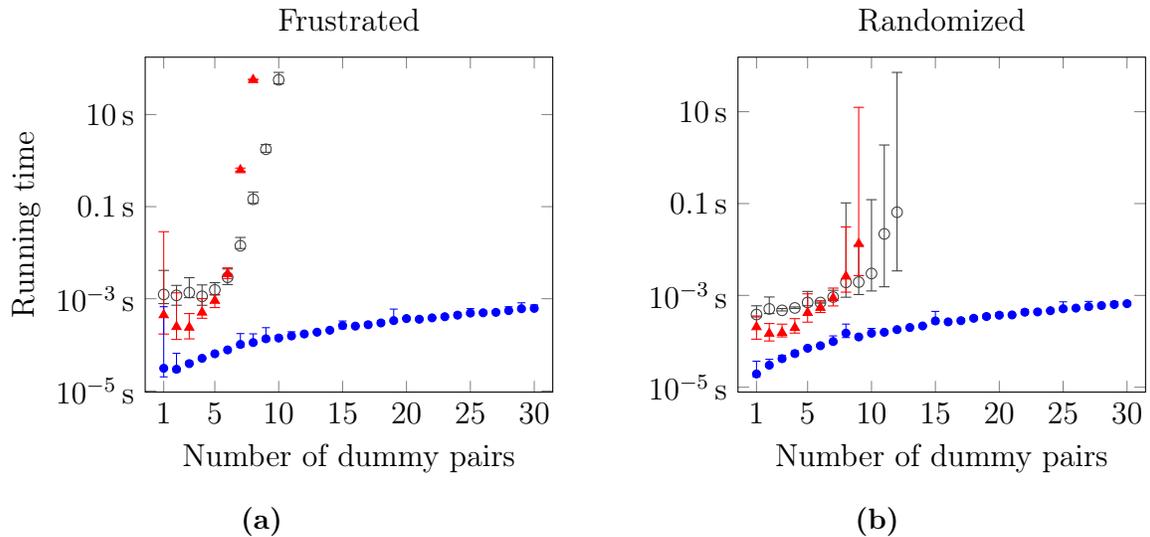

Next we run a test which exhibits the condition Algorithm~\ref{alg:improved} was specifically designed for:  namely, contractions between \emph{totally symmetric} tensors $T$ and $U$:
\begin{equation}
T_{a_1 a_2 \dotsc a_n} = T_{(a_1 a_2 \dotsc a_n)}, \qquad U^{a_1 a_2 \dotsc a_n} = U^{(a_1 a_2 \dotsc a_n)},
\end{equation}
as shown in Figure~\ref{fig:dummies-sym-and-random}.  We run two different versions of the test.  First in Figure~\ref{fig:total-sym-dummies} is the ``maximally frustrated'' situation, where the dummy pairs are distributed so that each pair has one leg on $T$ and one leg on $U$:
\begin{equation}
T_{\pi_1(a_1 a_2 \dotsc a_n)} U^{\pi_2(a_1 a_2 \dotsc a_n)}.
\end{equation}
This situation represents the absolute worst-case scenario for the Butler-Portugal algorithm.  In Figure~\ref{fig:random-total-sym-dummies}, we do instead a fully randomized test, where the indices are distributed across both tensor factors randomly.  This is a slightly better situation for \software{Mathematica} and \software{xPerm}, but the results are not much different.  As expected, Algorithm~\ref{alg:improved} clearly exhibits better performance in this test, showing polynomial growth whereas \software{Mathematica} and \software{xPerm} grow as $O(n!)$.  The data points for Algorithm~\ref{alg:improved} in the frustrated case of Figure~\ref{fig:total-sym-dummies} are well-fit by the power law
\begin{equation}
\text{Algorithm~\ref*{alg:improved}} \sim O(n^{1.36}).
\end{equation}
While \software{Mathematica} and \software{xPerm} are clearly factorial, there are too few data points to get a good fit.

%%%%%%%%%%%%%%%%
\subsection{Pairwise symmetry}
\label{sec:pairwise-symmetry}
%%%%%%%%%%%%%%%%

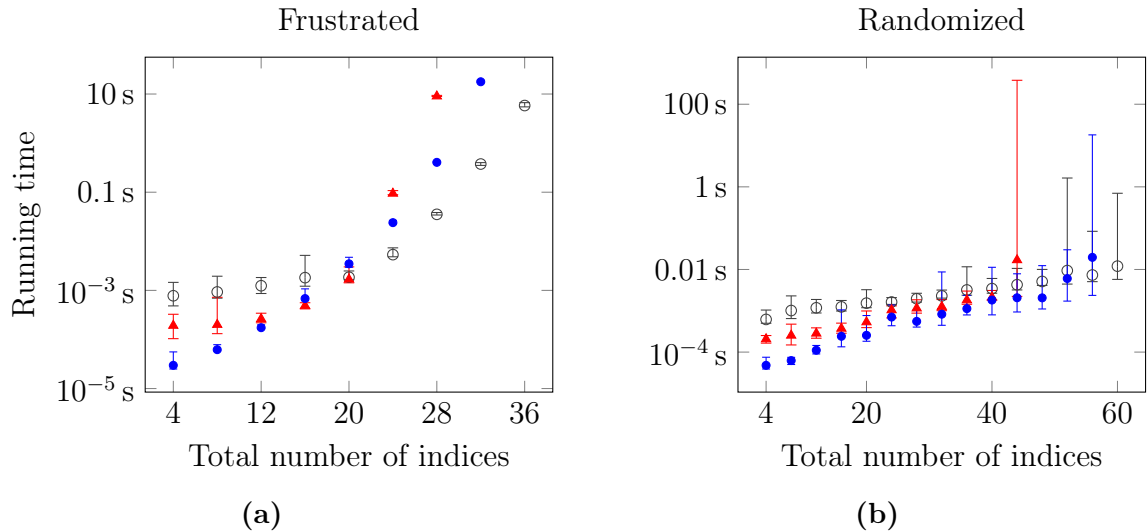
\begin{figure}
	\centering
	\begin{subfigure}{0.4\textwidth}
		\tikzsetnextfilename{pairwise-sym-dummies}
		\input{tikz/pairwise-sym-dummies.tikz}
		\vspace{-\baselineskip}
		\subcaption{}
		\label{fig:pairwise-sym-dummies}
	\end{subfigure}
	\hspace{1cm}
	\begin{subfigure}{0.4\textwidth}
		\tikzsetnextfilename{random-pairwise-sym-dummies}
		\input{tikz/random-pairwise-sym-dummies.tikz}
		\vspace{-\baselineskip}
		\subcaption{}
		\label{fig:random-pairwise-sym-dummies}
	\end{subfigure}
	\caption{Pairwise symmetry with contracted dummies, both frustrated and randomized.}
	\label{fig:dummies-pairwise-sym-and-random}
	\tikzsetnextfilename{legend}
	\input{tikz/legend.tikz}
\end{figure}

In the final test, we try a situation which is \emph{not} addressed by our improvements, and in which we expect factorial growth in all cases.  This test uses tensors $T$ and $U$ which have total symmetry among \emph{pairs} of indices,
\begin{equation}
T_{abcdef \dotsm} = T_{cdabef \dotsm} = T_{abefcd \dotsm} = \text{etc.},
\end{equation}
and similarly for $U$.  The results of this test are shown in Figure~\ref{fig:dummies-pairwise-sym-and-random}, again separated into the ``frustrated'' case in Figure~\ref{fig:pairwise-sym-dummies} where each dummy pair is split between $T$ and $U$, and the randomized case in Figure~\ref{fig:random-pairwise-sym-dummies} where the dummy pairs are randomly distributed across both tensors.

As expected, Figure~\ref{fig:pairwise-sym-dummies} shows clear factorial growth, and is also the only case in these tests where Algorithm~\ref{alg:improved} performs worse than \software{Mathematica}, clearly growing at a faster rate.  We expect that this is due to the extra processing done with \variable{Propagated-Symmetries} which is wasted effort in this particular test case.  By contrast, the performance for all three algorithms in Figure~\ref{fig:random-pairwise-sym-dummies} is more equal (because when the labels are randomly distributed, not every pair of slots is necessarily equivalent to every other pair); nevertheless, for high numbers of indices we begin to see the patterns of Figure~\ref{fig:pairwise-sym-dummies} repeated.

We suspect that some adjustments in implementation could bring the performance of Algorithm~\ref{alg:improved} more in line with \software{Mathematica} in the pairwise-symmetric case (and similar) without sacrificing the performance benefits of the prior tests.  However, we point out that real-world occurrences of the pairwise symmetry tested in Figure~\ref{fig:dummies-pairwise-sym-and-random} are somewhat rare.  If one did expect to see them frequently in some given application, it would be worth developing some more sophisticated version of the symmetry-propagation mechanism to bring them into the realm of polynomial time.

%%%%%%%%%%%%%%
\section{Discussion}
\label{sec:discussion}
%%%%%%%%%%%%%%

It is important in many applications of symbolic computation to have algorithms which can efficiently deal with expressions with tensor indices.  The Butler-Portugal algorithm \cite{2002IJMPC..13..859M} has been the state-of-the-art for some time, finding implementations in several popular software packages \cite{MartinGarcia2008597,citeulike:13127953,10.7717/peerj-cs.103,Peeters:2006kp,GRtensorII,2004CoPhC.157..173M}.  We have presented two improvements on the Butler-Portugal algorithm:  First, a faster way to deal with the label-exchange symmetries \emph{implicitly} rather than explicitly; and second, a way to canonicalize contractions over subsets of symmetric or antisymmetric slots in polynomial time (where previously, they took factorial time).

The improvements are evident in the data of Section~\ref{sec:testing} for all of the most common situations actually found in tensor calculations.  These improvements are achieved using a number of different data structures: First we introduce the \variable{Values} and \variable{Label-Groups} arrays which not only allow an efficient \emph{specification} of the problem (i.e., the partitioning of labels into free indices, component indices, and various separate pools of dummies of different types), but also allow one to reduce most label-group computations to simple array lookups.  We also introduce the \variable{Symmetric-Subsets} and \variable{Propagated-Symmetries} arrays which allow efficient tracking of the subsets of slots which are totally (anti)symmetric, and how those symmetries interact with other slots in the expression via propagation over contracted dummies.

In particular, the contractions-over-symmetric-subsets problem is dramatically reduced from factorial to polynomial complexity in both time and memory.  This is achieved via the concept of \emph{symmetry propagation}, wherein the slot symmetries at one end of a contraction can ``travel'' to the other end, where they become label symmetries.  In this work, we have focused on propagating simple (anti)symmetries (i.e., those which are generated by the simple exchange of two slots), because these symmetries are the simplest to describe, and they address the worst-case situations likely to be encountered in real computations.

However, we point out that the symmetry-propagation concept applies to \emph{generic} symmetries as well.  We strongly suggest this approach as an avenue to future improvement.  The main challenge is designing an efficient means to describe such generic symmetries.  One needs a data structure which is both simple to construct and simple to work with, and thus there are trade-offs to consider.  We expect, however, that such an approach will be useful for cases like the pairwise symmetry of Section~\ref{sec:pairwise-symmetry}, provided that the pairs can be provided ahead of time (which is true, if these symmetries derive from the tensor product of many identical factors).  We also expect that one may find efficient algorithms to enumerate Young-tableau-like symmetries, where one has totally (anti)symmetric exchange of totally (anti)symmetric length-$k$ subsets.  Despite these potential future improvements, it is still true that the generic worst-case is $O(n!)$.  However, if it is known in advance how such behavior could arise in a given real-world calculation, we are confident that the ideas presented here will be useful in developing further modifications that can reduce such behavior to polynomial.

One should also consider the broader context in which tensor index canonicalization occurs:  namely, to simplify algebraic expressions with tensor contractions.  As we have pointed out, there are other approaches to this problem, based on \emph{matching} terms rather then canonicalizing them \cite{Bolotin:2013qgr}, which may be faster in certain cases.  Also, there is the largely-unexplored arena of \emph{multi-term} symmetries, for which a graph-theory approach seems to be useful \cite{2017arXiv170108487L}.  A complete tensor computer algebra system should probably combine aspects of all of these approaches.

We look forward to addressing many of these considerations in future work.

%%%%%%%%%%%%%%
\section*{Acknowledgements}
%%%%%%%%%%%%%%

I would like to thank J. M. Mart\'in Garc\'ia for useful discussions and extensive comments on the initial draft.  This work is supported in part by ERC grant ERC-2013-CoG 616732-HoloQosmos.

%%%%%%%%%%%%%
\section*{Appendices}
%%%%%%%%%%%%%

\appendix

%\makeatletter
%\@addtoreset{algorithm}{section}% algorithm counter resets every section
%\makeatother
%\renewcommand{\thealgorithm}{\thesection.\arabic{algorithm}}

%%%%%%%%%%%%%%
\section{Efficient computation with permutation groups}
\label{app:groups}
%%%%%%%%%%%%%%

A permutation group on a sequence of $n$ symbols can in principle be quite large, and one needs an efficient method to manage computations.  In what has now become the standard method \cite{DBLP:Butler91,holt2005handbook,seress2003permutation}, one uses the notion of a \emph{strong generating set} and its corresponding \emph{base}.  This is essentially a way to give a hierarchical structure to a group, which paves the way for doing many (though not all) problems of finite group theory in polynomial time.

To define a base-and-strong-generating-set (or BSGS), let us first define a few group-theory notions.  Suppose $G \subseteq S_n$ is a permutation group acting on a set $\Omega$ of $n$ symbols.  Let $\alpha = \seq{\alpha_1,\alpha_2,\ldots,\alpha_n}$ be a fixed ordering of the points of $\Omega$.  Then for example, a permutation $g \in G$ acts by rearranging the ordering to $\alpha^g = \seq{\alpha_1^g,\alpha_2^g,\ldots,\alpha_n^g}$.  If any of the $\alpha_i$ are not moved by $g$, then we say $g$ \emph{stabilizes} these $\alpha_i$.  Given a subset of points $\set{\alpha_i} \subseteq \Omega$, the set of all $g \in G$ which stabilize these points give a subgroup of $G$ called the \emph{pointwise stabilizer} of $\set{\alpha_i}$.  The standard way to notate a pointwise stabilizer is with a subscript,
\begin{equation}
G_{\alpha_1 \alpha_2 \ldots \alpha_k} \subseteq G,
\end{equation}
which means that $G_{\alpha_1 \alpha_2 \ldots \alpha_k}$ is the pointwise stabilizer of all the $\alpha_i$ listed.

Given the ordering $\alpha$, we can then consider the pointwise stabilizers of the subsequences of $\alpha$:  $G_{\alpha_1}$, $G_{\alpha_1 \alpha_2}$, $G_{\alpha_1 \alpha_2 \alpha_3}$, etc.  Introducing a special notation for the stabilizer of the first $i-1$ points of $\alpha$,
\begin{equation}
G^{(i)} \equiv G_{\alpha_1 \alpha_2 \ldots \alpha_{i-1}},
\end{equation}
we can then observe that each $G^{(i)}$ contains $G^{(i+1)}$ as a subgroup, and thus the group $G$ can be decomposed into the \emph{point stabilizer chain}
\begin{equation}
G = G^{(1)} \supseteq G^{(2)} \supseteq \cdots \supseteq G^{(n-1)} \supseteq G^{(n)} = \set{\mathrm{id}}.
\end{equation}
The stabilizer chain gives a hierarchical way to organize a group which permits many efficient methods of computation.

Given an ordering $\alpha$ and its corresponding stabilizer chain, the \emph{base} of the group $G$ with respect to the ordering $\alpha$ is defined as the shortest subsequence of $\alpha$ whose pointwise stabilizer is trivial.  That is, if $k \leq n$ is the lowest index for which $G^{(k)} = \set{\mathrm{id}}$, then the base with respect to this ordering is $B = \seq{\alpha_1, \alpha_2, \ldots, \alpha_{k-1}}$.  Since the pointwise stabilizer of $B$ is the identity, each member $g$ of the group $G$ is \emph{uniquely} identified by the image $B^g = \seq{\alpha_1^g, \alpha_2^g, \ldots, \alpha_{k-1}^g}$.  Thus a base serves is a sort of index or set of coordinates.

A \emph{strong generating set} of $G$ with respect to a base $B$ is a set of permutations $S = \set{g_1, g_2, \ldots, g_N}$ which generate the group $G$, with an added condition that certain \emph{subsets} of $S$ also generate each of the subsequence stabilizers $G^{(i)}$.  Specifically, $S$ satisfies the property\footnote{Here the notation $\langle \cdot \rangle$ means the group generated by the enclosed set.  We trust this does not cause confusion with the usual meaning of $\seq{\ldots}$ in this article.}
\begin{equation}
\label{sgs prop}
\langle S \cap G^{(i)} \rangle = G^{(i)}, \quad \text{for each} \quad G^{(i)}.
\end{equation}
It is important to note that, as a generating set for $G$, $S$ may have redundant generators which are needed in order to satisfy the property \eqref{sgs prop}.  For example, the symmetric group $S_5$ may be generated by the set (in cyclic notation)
\begin{equation}
\set{ (1,2), \; (1,2,3,4,5) },
\end{equation}
but this is not a strong generating set.  To obtain the property \eqref{sgs prop} for the base $B = [1,2,3,4,5]$, we can use instead the set
\begin{equation}
\set{ (1,2), \; (2,3), \; (3,4), \; (4,5) }.
\end{equation}
When constructing a strong generating set, it is still desirable to eliminate redundant generators, and there are standard algorithms for doing this.

Given a base and a strong generating set, there are many computations that can be done efficiently.  For example, one can test for group membership in $G$, find the order of $G$, or generate a random element of $G$ from a uniform distribution.  To construct a BSGS in the first place requires the \emph{Schreier-Sims} algorithm, whose main variant takes $O(n^5)$ time.  For detailed descriptions of these and other group theory algorithms, we refer the reader to \cite{DBLP:Butler91,holt2005handbook,seress2003permutation}.

%%%%%%%%%%%%%%
\section{Subprocedures of the improved algorithm}
\label{app:subprocedures}
%%%%%%%%%%%%%%

Here we collect the various subprocedures of Algorithm~\ref{alg:improved} whose explicit code does not yield any great insights worthy of discussion in the main text, yet nevertheless whose function is not entirely obvious and ought to be written out.

The first of these is Algorithm~\ref{alg:get-instances-of-least-value} for the subprocedure \textproc{Get-Least-Value-Instances}, which traverses the current level of the search tree and generates the next level, by populating the \variable{Config-Slots-of-Least-Value} array with all of the configurations $(g,s)$ and slot pairs $(p,q)$ at which one can find the least value label.  In particular, in lines \ref{q start} through \ref{q end}, we see how the slot number $q$ is obtained, by starting from $p$ and traversing any \emph{even-numbered} symmetric sets available in \variable{Propagated-Symmetries}.  We only use the even-numbered sets because those ones are the result of symmetry propagation; the odd-numbered sets are already available in the slot symmetry group $S$, and thus should have been reached as one of the slots $p \in \Delta^S_i$.

Next, in Algorithm~\ref{alg:update-propagated-symmetries}, we give the subprocedure \textproc{Update-Propagated-Symmetries}, which updates the \variable{Propagated-Symmetries} array to reflect any new symmetric subsets found thus far in the search tree.  Some effort must be taken to maintain the conditions of Definition~\ref{def:prop-sym} which ensure the correct behavior of Algorithm~\ref{alg:append-non-redundant-instances}.  The function \textproc{Get-Next-Odd-Number} in line \ref{get-next-odd-number} must maintain an internal state, so that it returns a new odd number every time it is called, in sequence, starting from 1.  The function \textproc{Remove-Singletons} in line \ref{singletons} removes any nonzero entry from \variable{Next-Propagated-Symmetries} that occurs only once, thus preserving Rule~\ref{no-singletons} of Definition~\ref{def:prop-sym}.  Alternatively, one could avoid entering singletons in the first place, although this requires a bit more logic.

In Algorithm~\ref{alg:zero-due-to-propagated-symmetries} we give the subprocedure \textproc{Zero-Due-to-Propagated-Symmetries} which looks for conflicts between \variable{Symmetric-Subsets} and \variable{Propagated-Symmetries} which may cause the expression to be zero.  In line \ref{slot symmetry check} is a check whether a propagated slot symmetry conflicts with an original slot symmetry.  In line \ref{label symmetry check} is a check for dummy contractions where both ends lie in the \emph{same} symmetric subset of slots, to see whether the metric (anti)symmetry (if there is a metric) agrees with the slot (anti)symmetry.  We note that the functionality of \textproc{Zero-Due-to-Propagated-Symmetries} could also have been integrated into \textproc{Update-Propagated-Symmetries}, so that zero checks are made as soon as new symmetries are entered; this is the method used in the actual implementation tested in Section~\ref{sec:testing}, although the logic becomes more complex.

Finally, in Algorithm~\ref{alg:update-values-and-label-groups} we give \textproc{Update-Values-and-Label-Groups}, whose purpose is to maintain the conditions of Definition~\ref{def:value} and Table~\ref{tab:group codes} within the \variable{Values} and \variable{Label-Groups} arrays.  Whenever the \variable{least-value} label is consumed, we mark this by increasing the value of every label after it, within the set of labels whose value was \variable{least-value}.  This creates a \emph{new} subset of labels with value $(\variable{least-value}+1)$.  The details of this behavior must be modified depending on whether the label is a component index, or a dummy index with or without metric.  For dummy indices, we must offset the \emph{two} entries at \variable{least-value} and $(\variable{least-value}+1)$, increasing those remaining to $(\variable{least-value}+2)$.  For example, take the \variable{Values} array of \eqref{values 12 slot}:
\begin{equation}
\begin{array}{r|cccccccccccc}
\text{Label}	& a		& b		& g		& h		& c_1	& c_2	& d_1	& d_2	& e_1	& e_2	& f_1	& f_2 \\
\hline
\text{Value}	& 1		& 2		& 3		& 4		& 5		& 5		& 5		& 5		& 5		& 5		& 5		& 5
\end{array}\, .
\end{equation}
If we place the label $c_1$ into a slot, then we must update the \variable{Values} array to
\begin{equation}
\begin{array}{r|cccccccccccc}
\text{Label}	& a		& b		& g		& h		& c_1	& c_2	& d_1	& d_2	& e_1	& e_2	& f_1	& f_2 \\
\hline
\text{Value}	& 1		& 2		& 3		& 4		& 5		& 6		& 7		& 7		& 7		& 7		& 7		& 7
\end{array}\, .
\end{equation}
This effectively marks the \emph{partner} of a previously-used dummy for preferential treatment (as it must automatically precede the rest of the dummies in lexicographical order).  In this case $c_2$ has a lower value than the remaining dummies $\set{d_1, d_2, e_1, e_2, f_1, f_2}$, because $c_2$ is no longer interchangeable with them, and it should be preferred as it comes earlier in order.

%%%%%%%%%%%%%%%
% Get-Least-Value-Instances
%%%%%%%%%%%%%%%
\begin{algorithm}
	\caption{A subprocedure to populate \variable{Config-Slots-of-Least-Value}}
	\label{alg:get-instances-of-least-value}
	\vspace{0.5ex}
	\begin{algorithmic}
		\Input Slot $i$;~ Orbit $\Delta^S_i$;~ \variable{Configs}, \variable{Values}, and \variable{Propagated-Symmetries} arrays
		\Output The \variable{least-value} found in orbit, and the \variable{Config-Slots-of-Least-Value} array
	\end{algorithmic}
	\vspace{0.5ex} \hrule \vspace{0.5ex}
	\begin{algorithmic}[1]\raggedright
		\Procedure{Get-Least-Value-Instances}{$i, \Delta^S_i, \variable{Configs}, \variable{Values}, \variable{Propagated-Symmetries}$}
			\State $\variable{least-value} \gets n$
			\State $\variable{Config-Slots-of-Least-Value} \gets \set{}$
			\ForEach{$(g,s) \in \variable{Configs}$}
				\ForEach{$p \in \Delta^S_i$}
					\State $\variable{value} \gets \variable{Values}\access{g\access{p}}$
					\State $\variable{symmetry} \gets \variable{Propagated-Symmetries}\access{s\access{p}}$
					\State $q \gets p$
					\label{q start}
					\If{\big($\variable{symmetry} \neq 0$ and \variable{symmetry} is even\big)}
						\For{$k \gets i$ to $n$}
							\If{\big($\variable{Propagated-Symmetries}\access{s\access{k}} = \variable{symmetry}$ \linebreak \hspace*{5\continueindent} and $\variable{Values}\access{g\access{k}} < \variable{value}$\big)}
								\State $\variable{value} \gets \variable{Values}\access{g\access{k}}$
								\State $q \gets k$
							\EndIf
						\EndFor
					\EndIf
					\label{q end}
					\If{\big($\variable{value} < \variable{least-value}$\big)}
						\State $\variable{least-value} \gets \variable{value}$
						\State $\variable{Config-Slots-of-Least-Value} \gets \set{}$
					\EndIf
					\If{\big($\variable{value} = \variable{least-value}$\big)}
						\State $\variable{Config-Slots-of-Least-Value} \gets \function{Append}{\variable{Config-Slots-of-Least-Value}, \seq{(g,s), (p,q)}}$
					\EndIf
				\EndFor
			\EndFor
			\State \Return \set{\variable{least-value}, \variable{Config-Slots-of-Least-Value}}
		\EndProcedure
	\end{algorithmic}
\end{algorithm}

%%%%%%%%%%%%%%%
% Update propagated symmetries
%%%%%%%%%%%%%%%
\begin{algorithm}
	\caption{A subprocedure to add new entries to \variable{Propagated-Symmetries}}
	\label{alg:update-propagated-symmetries}
	\vspace{0.5ex}
	\begin{algorithmic}
		\Input The current \variable{Least-Value-Set};~ its configuration $(g,s)$;~ least value for this slot \variable{least-value};~ \linebreak \variable{Label-Groups}, \variable{Symmetric-Subsets}, and \variable{Propagated-Symmetries}
		\Output \variable{Propagated-Symmetries} updated with the provided information
	\end{algorithmic}
	\vspace{0.5ex} \hrule \vspace{0.5ex}
	\begin{algorithmic}[1]\raggedright
		\Procedure{Update-Propagated-Symmetries}{$\variable{Least-Value-Set}, (g,s), \variable{least-value}$, $\hspace*\continueindent \variable{Label-Groups}, \variable{Symmetric-Subsets}, \variable{Propagated-Symmetries}$}
			%\State $\variable{zero-flag} \gets \text{false}$
			\State $\variable{Next-Propagated-Symmetries} \gets \variable{Propagated-Symmetries}$
			\ForEach{$(p,q) \in \variable{Least-Value-Set}$}
				\Comment Note that we use only $q$
				\State $\variable{label} \gets g\access{q}$
				\If{\big($\variable{Label-Groups}\access{\variable{label}} \neq \none$ and $\variable{Symmetric-Subsets}\access{q} \neq 0$ and \linebreak \hspace*\continueindent $\variable{Propagated-Symmetries}\access{s\access{q}} = 0$\big)}
					\State $\variable{current-entry} \gets \variable{Next-Propagated-Symmetries}\access{s\access{q}}$
					\If{\big($\variable{current-entry} \neq 0$ and $\variable{current-entry}$ is even\big)}
						\State $\variable{new-entry} \gets \function{Sign}{\variable{current-entry}} \cdot \big(\abs{\variable{current-entry}} - 1\big)$
					\Else
						\State $\variable{new-entry} \gets \function{Sign}{\variable{Symmetric-Subsets}\access{q}} \cdot \function{Get-Next-Odd-Number}{}$
						\label{get-next-odd-number}
					\EndIf
					\State $\variable{Next-Propagated-Symmerties}\access{s\access{q}} \gets \variable{new-entry}$
					\If{\big($\variable{Label-Groups}\access{\variable{label}}$ is one of $\set{\sdummy, \adummy, \ldummy, \udummy}$\big)}
						%\State $\variable{partner-label} \gets \function{Get-Partner-Label}{\variable{label}, \variable{least-value}, \variable{Label-Groups}\access{\variable{label}}}$
						\If{\big($(\variable{label} - \variable{least-value})$ is odd or $\variable{Label-Groups}\access{\variable{label}} = \udummy$\big)}
							\State $\variable{partner-label} \gets \variable{label} - 1$
						\Else
							\State $\variable{partner-label} \gets \variable{label} + 1$
						\EndIf
						\State $\variable{partner-slot} \gets g^{-1}\access{\variable{partner-label}}$
						\State $\variable{partner-entry} \gets \function{Sign}{\variable{new-entry}} \cdot \big(\abs{\variable{new-entry}} + 1\big)$
						\If{\big($\variable{Next-Propagated-Symmetries}\access{s\access{\variable{partner-slot}}} = 0$ or \linebreak \hspace*\continueindent $\abs{\variable{partner-entry}} < \abs{\variable{Next-Propagated-Symmetries}\access{s\access{\variable{partner-slot}}}}$\big)}
							\State $\variable{Next-Propagated-Symmetries}\access{s\access{\variable{partner-slot}}} \gets \variable{partner-entry}$
						\EndIf
					\EndIf
				\EndIf
			\EndFor
			\State $\variable{Propagated-Symmetries} \gets \function{Remove-Singletons}{\variable{Next-Propagated-Symmetries}}$
			\label{singletons}
			\State \Return \variable{Propagated-Symmetries}
		\EndProcedure
	\end{algorithmic}
\end{algorithm}

%%%%%%%%%%%%%%%
% Zero-Due-to-Propagated-Symmetries
%%%%%%%%%%%%%%%
\begin{algorithm}
	\caption{A subprocedure to do early zero checking}
	\label{alg:zero-due-to-propagated-symmetries}
	\vspace{0.5ex}
	\begin{algorithmic}\raggedright
		\Input Configuration $(g,s)$;~ least value for this slot \variable{least-value};~ \variable{Label-Groups}, \linebreak \variable{Symmetric-Subsets}, and \variable{Propagated-Symmetries}
		\Output true or false
	\end{algorithmic}
	\vspace{0.5ex} \hrule \vspace{0.5ex}
	\begin{algorithmic}[1]\raggedright
		\Procedure{Zero-Due-to-Propagated-Symmetries}{$(g,s), \variable{least-value}$, $\hspace*\continueindent \variable{Label-Groups}, \variable{Symmetric-Subsets}, \variable{Propagated-Symmetries}$}
			\State $\variable{Symmetries-of-Visited-Labels} \gets \seq{0,0,\dotsc,0}_n$
			\For{$p \gets 1$ to $n$}
				\State $\variable{label} \gets g\access{p}$
				\State $\variable{symmetry} \gets \variable{Propagated-Symmetries}\access{s\access{p}}$
				\If{\big($\variable{symmetry} \neq 0$\big)}
					\State $\variable{Symmetries-of-Visited-Labels}\access{\variable{label}} \gets \variable{symmetry}$
					\If{\big($\variable{Label-groups}\access{\variable{label}} = \component$ and $\variable{symmetry} < 0$\big)}
						\State \Return true
					\EndIf
					\If{\big($\variable{Label-groups}\access{\variable{label}}$ is one of $\set{\sdummy, \adummy, \ldummy, \udummy}$\big)}
						\If{\big(\variable{symmetry} is even and \linebreak \hspace*\continueindent $\function{Sign}{\variable{symmetry}} \neq \function{Sign}{\variable{Symmetric-Subsets}\access{p}}$\big)}
						\label{slot symmetry check}
							\State \Return true
						\EndIf
						\If{\big($(\variable{label} - \variable{least-value})$ is odd or $\variable{Label-Groups}\access{\variable{label}} = \udummy$\big)}
							\State $\variable{partner-label} \gets \variable{label} - 1$
						\Else
							\State $\variable{partner-label} \gets \variable{label} + 1$
						\EndIf
						\State $\variable{partner-symmetry} \gets \variable{Symmetries-of-Visited-Labels}\access{\variable{partner-label}}$
						\If{\big($\variable{partner-symmetry} = \variable{symmetry}$\big) and \linebreak \hspace*\continueindent
							\big[\big($\variable{Label-groups}\access{\variable{label}} = \sdummy$ and $\variable{symmetry} < 0$\big) or \linebreak \hspace*{2\continueindent} \big($\variable{Label-groups}\access{\variable{label}} = \adummy$ and $\variable{symmetry} > 0$\big)\big]}
						\label{label symmetry check}
							\State \Return true
						\EndIf
					\EndIf
				\EndIf
			\EndFor
			\State \Return false
		\EndProcedure
	\end{algorithmic}
\end{algorithm}

%%%%%%%%%%%%%%%
% Update-Values-and-Label-Groups
%%%%%%%%%%%%%%%
\begin{algorithm}
	\caption{A subprocedure to maintain the defining conditions of \variable{Values} and \variable{Label-Groups}}
	\label{alg:update-values-and-label-groups}
	\vspace{0.5ex}
	\begin{algorithmic}
		\Input \variable{Values}, \variable{Label-Groups}, and the \variable{least-value} that was just used
		\Output The new \variable{Values} and \variable{Label-Groups} after ``freezing'' the label \variable{least-value} and its dummy partner, if one exists
	\end{algorithmic}
	\vspace{0.5ex} \hrule \vspace{0.5ex}
	\begin{algorithmic}[1]\raggedright
		\Procedure{Update-Values-and-Label-Groups}{$\variable{Values}, \variable{Label-Groups}, \variable{least-value}$}
		\If{$\variable{Label-Groups}\access{\variable{least-value}} = \none$}
			\State \Return \set{\variable{Values}, \variable{Label-Groups}}
		\ElsIf{$\variable{Label-Groups}\access{\variable{least-value}} = \component$}
			\State $\variable{Label-Groups}\access{\variable{least-value}} \gets \none$
			\State $\variable{loop-start} \gets \variable{least-value} + 1$
			\State $\variable{value-threshold} \gets \variable{least-value}$; \qquad $\variable{value-increment} \gets 1$
		\ElsIf{$\variable{Label-Groups}\access{\variable{least-value}}$ is one of $\set{\sdummy, \adummy}$}
			\State $\variable{Values}\access{\variable{least-value} + 1} \gets \variable{Values}\access{\variable{least-value} + 1} + 1$
			\State $\variable{Label-Groups}\access{\variable{least-value}} \gets \none$; \qquad $\variable{Label-Groups}\access{\variable{least-value} + 1} \gets \none$
			\State $\variable{loop-start} \gets \variable{least-value} + 2$
			\State $\variable{value-threshold} \gets \variable{least-value}$; \qquad $\variable{value-increment} \gets 2$
		\ElsIf{$\variable{Label-Groups}\access{\variable{least-value}} = \ldummy$}
			\State $\variable{Label-Groups}\access{\variable{least-value}} \gets \none$; \qquad $\variable{Label-Groups}\access{\variable{least-value} + 1} \gets \none$
			\State $\variable{loop-start} \gets \variable{least-value} + 2$
			\State $\variable{value-threshold} \gets \variable{least-value} + 1$; \qquad $\variable{value-increment} \gets 2$
		\ElsIf{$\variable{Label-Groups}\access{\variable{least-value}} = \udummy$}
			\State $\variable{Label-Groups}\access{\variable{least-value} - 1} \gets \none$; \qquad $\variable{Label-Groups}\access{\variable{least-value}} \gets \none$
			\State $\variable{loop-start} \gets \variable{least-value} + 1$
			\State $\variable{value-threshold} \gets \variable{least-value}$; \qquad $\variable{value-increment} \gets 2$
		\EndIf
		\State $j \gets \variable{loop-start}$
		\While{$j \leq n$ and $\variable{Values}\access{j} \leq \variable{value-threshold}$}
			\State $\variable{Values}\access{j} \gets \variable{Values}\access{j} + \variable{value-increment}$
			\State $j \gets j + 1$
		\EndWhile
		\State \Return \set{\variable{Values}, \variable{Label-Groups}}
		\EndProcedure
	\end{algorithmic}
\end{algorithm}

\clearpage

%%%%%%%%%%
%\end{appendices}
%%%%%%%%%%

%%%%%%%%%%%%%%%%%%%%%%%%%%%%%%%%%

\bibliographystyle{utphys}
\bibliography{compsci}

%\vspace{0.2in}
%\bibliographystyle{JHEP}
%\bibliography{bibliography}

\end{document}

%% file: tikz/theta-abc.tikz
\begin{tikzpicture}[
	baseline=(Thlabel.base),
	scale=0.25,
	text height=1.0ex,
	text depth=0.0ex,
	every node/.style = {font = \footnotesize}]
%\draw [help lines] (-10,-6) grid (10,6)
%	(6,6) -- (-6,-6) (6,-6) -- (-6,6)
%	(0,0) circle [radius=2]
%	(0,0) circle [radius=6]
%	(-8,4) circle [radius=2]
%	(-8,-4) circle [radius=2]
%	(8,4) circle [radius=2]
%	(8,-4) circle [radius=2]
%	(-8,0) circle [radius=2]
%	(8,0) circle [radius=2]
%	(-4,0) circle [radius=2]
%	(4,0) circle [radius=2]
%	(0,4) circle [radius=2]
%	(0,-4) circle [radius=2];

% Tensor \theta
\coordinate (Th) at (0,0);
\path (Th) ++(0,-0.2) node[style = {font = \normalsize}] (Thlabel) {$\theta$};
\draw [thick] (Th) circle [radius={sqrt(2)}];
% Attachment points for indices
\path (Th) +(-1,1) coordinate (t1)
+(1,1) coordinate (t2)
+(0,{-sqrt(2)}) coordinate (t3);

% \theta legs
\coordinate (freeThup) at (0,5);
\path (freeThup) +(-2,0) coordinate (f1)
+(2,0) coordinate (f2);
\coordinate (freeThdown) at (0,-5);
\path (freeThdown) +(0,0) coordinate (f3);

% draw legs
\draw (t1)
	.. controls +(-0.5,0.5) and +(0,-2) ..
	(f1) node [above] {$a$};
\draw (t2)
	.. controls +(0.5,0.5) and +(0,-2) ..
	(f2) node [above] {$b$};
\draw (t3) -- (f3) node [below] {$c$};

\end{tikzpicture}

%% file: tikz/chi-abcd.tikz
\begin{tikzpicture}[
	baseline=(Chlabel.base),
	scale=0.25,
	text height=1.0ex,
	text depth=0.0ex,
	every node/.style = {font = \footnotesize}]
%\draw [help lines] (-10,-6) grid (10,6)
%	(6,6) -- (-6,-6) (6,-6) -- (-6,6)
%	(0,0) circle [radius=2]
%	(0,0) circle [radius=6]
%	(-8,4) circle [radius=2]
%	(-8,-4) circle [radius=2]
%	(8,4) circle [radius=2]
%	(8,-4) circle [radius=2]
%	(-8,0) circle [radius=2]
%	(8,0) circle [radius=2]
%	(-4,0) circle [radius=2]
%	(4,0) circle [radius=2]
%	(0,4) circle [radius=2]
%	(0,-4) circle [radius=2];

% Tensor \chi
\coordinate (Ch) at (0,0);
\path (Ch) +(0,0.1) node[style = {font = \normalsize}] (Chlabel) {$\chi$};
\draw [thick] (Ch) +(0,2) -- +(-2,-1) -- +(2,-1) -- cycle;
% Attachment points for indices
\path (Ch) +(0,2) coordinate (c1)
+(-1.2,-1) coordinate (c2)
+(0,-1) coordinate (c3)
+(1.2,-1) coordinate (c4);

% \theta legs
\coordinate (freeChup) at (0,5);
\path (freeChup) +(0,0) coordinate (f1);
\coordinate (freeChdown) at (0,-5);
\path (freeChdown) +(-1.5,0) coordinate (f2)
+(0,0) coordinate (f3)
+(1.5,0) coordinate (f4);

% draw legs
\draw (c1) -- (f1) node [above] {$a$};
\draw (c2)
	.. controls +(0,-2) and +(0,2) ..
	(f2) node [below] {$b$};
\draw (c3) -- (f3) node [below] {$c$};
\draw (c4)
	.. controls +(0,-2) and +(0,2) ..
	(f4) node [below] {$d$};

\end{tikzpicture}

%% file: tikz/chi-theta-contraction.tikz
\begin{tikzpicture}[
	baseline,
	scale=0.25,
	text height=1.0ex,
	text depth=0.0ex,
	every node/.style = {font = \footnotesize}]
%\draw [help lines] (-10,-10) grid (10,10)
%	(10,10) -- (-10,-10) (10,-10) -- (-10,10)
%	(0,0) circle [radius=2]
%	(0,0) circle [radius=6]
%	(0,0) circle [radius=10]
%	(6,-6) circle [radius=2]
%	(6,-6) circle [radius=4]
%	(6,6) circle [radius=2]
%	(6,6) circle [radius=4]
%	(-6,-6) circle [radius=2]
%	(-6,-6) circle [radius=4]
%	(-6,6) circle [radius=2]
%	(-6,6) circle [radius=4]
%	(0,8) circle [radius=2]
%	(0,-8) circle [radius=2]
%	(-8,0) circle [radius=2]
%	(8,0) circle [radius=2]
%	(-4,0) circle [radius=2]
%	(4,0) circle [radius=2]
%	(0,4) circle [radius=2]
%	(0,-4) circle [radius=2];

% Tensor \chi
\coordinate (Ch) at (1.5,2.5);
\path (Ch) +(0,0.1) node[style = {font = \normalsize}] {$\chi$};
\draw [thick] (Ch) +(0,2) -- +(-2,-1) -- +(2,-1) -- cycle;
% Attachment points for indices
\path (Ch) +(0,2) coordinate (c1)
+(-1.2,-1) coordinate (c2)
+(0,-1) coordinate (c3)
+(1.2,-1) coordinate (c4);

% Tensor \theta
\coordinate (Th) at (-1.5,-2.5);
\path (Th) ++(0,-0.2) node[style = {font = \normalsize}] (Thlabel) {$\theta$};
\draw [thick] (Th) circle [radius={sqrt(2)}];
% Attachment points for indices
\path (Th) +(-1,1) coordinate (t1)
+(1,1) coordinate (t2)
+(0,{-sqrt(2)}) coordinate (t3);

% Draw connecting leg
\draw (t2)
	.. controls +(0.75,0.75) and +(0,-0.5) ..
	(c2) node[midway, left] {$f$};

% Upper frees
\draw (t1)
	.. controls +(-1.5,1.5) and +(0,-2) ..
	(-4,7) node[above] {$a$};
\draw (c1) -- (1.5,7) node[above] {$b$};

% Lower frees
\draw (t3) -- (-1.5,-6) node[below] {$c$};
\draw (c3)
	.. controls +(0,-2) and +(0,2) ..
	(1.5,-6) node[below] {$d$};
\draw (c4)
	.. controls +(0,-2) and +(0,2) ..
	(3.5,-6) node[below] {$e$};

\end{tikzpicture}

%% file: tikz/metric.tikz
\begin{tikzpicture}[
	baseline,
	scale=0.25,
	text height=1.0ex,
	text depth=0.0ex,
	every node/.style = {font = \footnotesize}]
%\draw [help lines] (-10,-10) grid (10,10)
%	(10,10) -- (-10,-10) (10,-10) -- (-10,10)
%	(0,0) circle [radius=2]
%	(0,0) circle [radius=6]
%	(0,0) circle [radius=10]
%	(6,-6) circle [radius=2]
%	(6,-6) circle [radius=4]
%	(6,6) circle [radius=2]
%	(6,6) circle [radius=4]
%	(-6,-6) circle [radius=2]
%	(-6,-6) circle [radius=4]
%	(-6,6) circle [radius=2]
%	(-6,6) circle [radius=4]
%	(0,8) circle [radius=2]
%	(0,-8) circle [radius=2]
%	(-8,0) circle [radius=2]
%	(8,0) circle [radius=2]
%	(-4,0) circle [radius=2]
%	(4,0) circle [radius=2]
%	(0,4) circle [radius=2]
%	(0,-4) circle [radius=2];

% metric
\coordinate (g) (0,0);
\draw (g) ++(-2,-2) node[below] {$a$}
	-- ++(0,2)
	arc [start angle = 180, end angle = 0, radius = 2]
	-- ++(0,-2) node[below] {$b$};

\end{tikzpicture}

%% file: tikz/inverse-metric.tikz
\begin{tikzpicture}[
	baseline,
	scale=0.25,
	text height=1.0ex,
	text depth=0.0ex,
	every node/.style = {font = \footnotesize}]
%\draw [help lines] (-10,-10) grid (10,10)
%	(10,10) -- (-10,-10) (10,-10) -- (-10,10)
%	(0,0) circle [radius=2]
%	(0,0) circle [radius=6]
%	(0,0) circle [radius=10]
%	(6,-6) circle [radius=2]
%	(6,-6) circle [radius=4]
%	(6,6) circle [radius=2]
%	(6,6) circle [radius=4]
%	(-6,-6) circle [radius=2]
%	(-6,-6) circle [radius=4]
%	(-6,6) circle [radius=2]
%	(-6,6) circle [radius=4]
%	(0,8) circle [radius=2]
%	(0,-8) circle [radius=2]
%	(-8,0) circle [radius=2]
%	(8,0) circle [radius=2]
%	(-4,0) circle [radius=2]
%	(4,0) circle [radius=2]
%	(0,4) circle [radius=2]
%	(0,-4) circle [radius=2];

% inverse metric
\coordinate (invg) (0,0);
\draw (invg) ++(-2,2) node[above] {$a$}
	-- ++(0,-2)
	arc [start angle = 180, end angle = 360, radius = 2]
	-- ++(0,2) node[above] {$b$};

\end{tikzpicture}

%% file: tikz/kronecker-delta.tikz
\begin{tikzpicture}[
	baseline,
	scale=0.25,
	text height=1.0ex,
	text depth=0.0ex,
	every node/.style = {font = \footnotesize}]
%\draw [help lines] (-10,-10) grid (10,10)
%	(10,10) -- (-10,-10) (10,-10) -- (-10,10)
%	(0,0) circle [radius=2]
%	(0,0) circle [radius=6]
%	(0,0) circle [radius=10]
%	(6,-6) circle [radius=2]
%	(6,-6) circle [radius=4]
%	(6,6) circle [radius=2]
%	(6,6) circle [radius=4]
%	(-6,-6) circle [radius=2]
%	(-6,-6) circle [radius=4]
%	(-6,6) circle [radius=2]
%	(-6,6) circle [radius=4]
%	(0,8) circle [radius=2]
%	(0,-8) circle [radius=2]
%	(-8,0) circle [radius=2]
%	(8,0) circle [radius=2]
%	(-4,0) circle [radius=2]
%	(4,0) circle [radius=2]
%	(0,4) circle [radius=2]
%	(0,-4) circle [radius=2];

% kronecker delta
\coordinate (delta) (0,0);
\draw (delta) +(-0.5,-2) node[below] {$a$}
	.. controls +(0,2) and +(0,-2) ..
	+(0.5,2) node[above] {$b$};

\end{tikzpicture}

%% file: tikz/metric-inverse-metric.tikz
\begin{tikzpicture}[
	baseline,
	scale=0.25,
	text height=1.0ex,
	text depth=0.0ex,
	every node/.style = {font = \footnotesize}]
%\draw [help lines] (-10,-10) grid (10,10)
%	(10,10) -- (-10,-10) (10,-10) -- (-10,10)
%	(0,0) circle [radius=2]
%	(0,0) circle [radius=6]
%	(0,0) circle [radius=10]
%	(6,-6) circle [radius=2]
%	(6,-6) circle [radius=4]
%	(6,6) circle [radius=2]
%	(6,6) circle [radius=4]
%	(-6,-6) circle [radius=2]
%	(-6,-6) circle [radius=4]
%	(-6,6) circle [radius=2]
%	(-6,6) circle [radius=4]
%	(0,8) circle [radius=2]
%	(0,-8) circle [radius=2]
%	(-8,0) circle [radius=2]
%	(8,0) circle [radius=2]
%	(-4,0) circle [radius=2]
%	(4,0) circle [radius=2]
%	(0,4) circle [radius=2]
%	(0,-4) circle [radius=2];

% metric
\draw (-4,-2) node[below] {$a$}
	-- ++(0,3)
	arc [start angle = 180, end angle = 0, radius = 2]
	-- ++(0,-2)
	node[midway,left] {$c$}
	% inverse metric
	arc [start angle = 180, end angle = 360, radius = 2]
	-- ++(0,3) node[above] {$b$};

\end{tikzpicture}

%% file: tikz/TU-contraction.tikz
\begin{tikzpicture}[
	scale=0.55,
	text height=1.0ex,
	text depth=0.0ex,
	every node/.style = {font = \footnotesize}]

%\draw [help lines] (-10,-6) grid (10,6)
%	(6,6) -- (-6,-6) (6,-6) -- (-6,6)
%	(0,0) circle [radius=2]
%	(0,0) circle [radius=6]
%	(-8,4) circle [radius=2]
%	(-8,-4) circle [radius=2]
%	(8,4) circle [radius=2]
%	(8,-4) circle [radius=2]
%	(-8,0) circle [radius=2]
%	(8,0) circle [radius=2]
%	(-4,0) circle [radius=2]
%	(4,0) circle [radius=2]
%	(0,4) circle [radius=2]
%	(0,-4) circle [radius=2];

% Tensor T
\coordinate (T) at (-1.45,1.75);
\path (T) ++(0.1,0.2) node[style = {font = \normalsize}] {$T$};
\draw [thick] (T) ++(-3.25,-0.75)
	arc [start angle=180, end angle=90, radius=1.5]
	-- ++(2,0)
	arc [start angle=135, end angle=45, radius={sqrt(2)}]
	-- ++(1,0)
	-- ++(0,-1.5)
	-- cycle;
% Attachment points for indices
\path (T) ++(-2.5,-0.75) coordinate (t1)
	++(1,0) coordinate (t2)
	++(1,0) coordinate (t3)
	++(1,0) coordinate (t4)
	++(1,0) coordinate (t5)
	++(1,0) coordinate (t6);
% Write slot numbers
\draw (t1) node[above] {1}
	(t2) node[above] {2}
	(t3) node[above] {3}
	(t4) node[above] {4}
	(t5) node[above] {5}
	(t6) node[above] {6};

% Tensor U
\coordinate (U) at (0.85,-1.75);
\path (U) ++(0.3,-0.35) node[style = {font = \normalsize}] (Ulabel) {$U$};
\draw [thick] (U) ++(-3.25,0.75)
	-- ++(1,-1.5)
	-- ++(4,0)
	arc [start angle = -135, end angle = 45, radius = {0.75*sqrt(2)}]
	-- cycle;
% Attachment points for indices
\path (U) ++(-2.5,0.75) coordinate (u1)
	++(1,0) coordinate (u2)
	++(1,0) coordinate (u3)
	++(1,0) coordinate (u4)
	++(1,0) coordinate (u5)
	++(1,0) coordinate (u6);
% Write slot numbers
\draw (u1) node[below] {7}
	(u2) node[below] {8}
	(u3) node[below] {9}
	(u4) node[below] {10}
	(u5) node[below] {11}
	(u6) node[below] {12};

% Indices (connecting lines)
% Free index attachment points
\coordinate (freeT) at (-4.3,-2);
\path (freeT) coordinate (f1) ++(1,0) coordinate (f2);

\coordinate (freeU) at (2.8,2);
\path (freeU) coordinate (f3) ++(1,0) coordinate (f4);

% Free indices from T
\draw (t1)
	.. controls +(0,-1.25) and +(0,1.25) ..
	(f1) node[below] {$a$};
\draw (t2)
	.. controls +(0,-1.25) and +(0,1.25) ..
	(f2) node[below] {$b$};

% Contracted indices
\draw (t3)
	node[below left] {$c$}
	.. controls +(0,-2) and +(0,2.25) ..
	(u4);
\draw (t4)
	node[below left] {$d$}
	.. controls +(0,-1) and +(0,1) ..
	(u2);
\draw (t5)
	node[below left] {$e$}
	.. controls +(0,-1.5) and +(0,0.75) ..
	(u1);
\draw (t6)
	node[below left] {$f$}
	.. controls +(0,-1) and +(0,1) ..
	(u3);

% Free indices from U
\draw (u5)
	.. controls +(0,1.25) and +(0,-1.25) ..
	(f3) node[above] {$g$};
\draw (u6)
	.. controls +(0,1.25) and +(0,-1.25) ..
	(f4) node[above] {$h$};
\end{tikzpicture}

%% file: tikz/TU-contraction-symmetry.tikz
\begin{tikzpicture}[
	baseline,
	scale=0.55,
	text height=1.0ex,
	text depth=0.0ex,
	every node/.style = {font = \footnotesize}]

%\draw [help lines] (-10,-6) grid (10,6)
%	(6,6) -- (-6,-6) (6,-6) -- (-6,6)
%	(0,0) circle [radius=2]
%	(0,0) circle [radius=6]
%	(-8,4) circle [radius=2]
%	(-8,-4) circle [radius=2]
%	(8,4) circle [radius=2]
%	(8,-4) circle [radius=2]
%	(-8,0) circle [radius=2]
%	(8,0) circle [radius=2]
%	(-4,0) circle [radius=2]
%	(4,0) circle [radius=2]
%	(0,4) circle [radius=2]
%	(0,-4) circle [radius=2];

% Tensor T
\coordinate (T) at (-1.45,1.75);
\path (T) ++(0.1,0.25) node[style = {font = \normalsize}] {$T$};
\draw [thick] (T) ++(-3.1,-0.75)
	arc [start angle=180, end angle=90, radius=1.5]
	-- ++(1.85,0)
	arc [start angle=135, end angle=45, radius={sqrt(2)}]
	-- ++(0.85,0)
	-- ++(0,-1.5)
	-- cycle;
% Attachment points for indices
\path (T) ++(-2.5,-0.75) coordinate (t1)
	++(1,0) coordinate (t2)
	++(1,0) coordinate (t3)
	++(1,0) coordinate (t4)
	++(1,0) coordinate (t5)
	++(1,0) coordinate (t6);
% Write slot numbers
\draw (t1) node[above] {1}
	(t2) node[above] {2}
	(t3) node[above] {3}
	(t4) node[above] {4}
	(t5) node[above] {5}
	(t6) node[above] {6};
% Highlight stripe for symmetry
\path (t3) ++(-0.1,0.46) coordinate (h1) ++(3.2,0) coordinate (h2);
\begin{scope}[on background layer]
	\draw[color = blue!12!white, line width = 10pt, line cap = round] (h1) -- (h2);
\end{scope}

% Tensor U
\coordinate (U) at (0.85,-1.75);
\path (U) ++(0.3,-0.35) node[style = {font = \normalsize}] (Ulabel) {$U$};
\draw [thick] (U) ++(-3.1,0.75)
	-- ++(0.75,-1.5)
	-- ++(3.85,0)
	arc [start angle = -135, end angle = 45, radius = {0.75*sqrt(2)}]
	-- cycle;
% Attachment points for indices
\path (U) ++(-2.5,0.75) coordinate (u1)
	++(1,0) coordinate (u2)
	++(1,0) coordinate (u3)
	++(1,0) coordinate (u4)
	++(1,0) coordinate (u5)
	++(1,0) coordinate (u6);
% Write slot numbers
\draw (u1) node[below] {7}
	(u2) node[below] {8}
	(u3) node[below] {9}
	(u4) node[below] {10}
	(u5) node[below] {11}
	(u6) node[below] {12};

% Indices (connecting lines)
% Free index attachment points
\coordinate (freeT) at (-4.25,-2);
\path (freeT) coordinate (f1) ++(1,0) coordinate (f2);

\coordinate (freeU) at (2.65,2);
\path (freeU) coordinate (f3) ++(1,0) coordinate (f4);

% Free indices from T
\draw (t1)
	.. controls +(0,-1.25) and +(0,1.25) ..
	(f1) node[below] {$a$};
\draw (t2)
	.. controls +(0,-1.25) and +(0,1.25) ..
	(f2) node[below] {$b$};

% Contracted indices
\draw (t3)
	node[below left] {$c$}
	.. controls +(0,-2) and +(0,2.25) ..
	(u4);
\draw (t4)
	node[below left] {$d$}
	.. controls +(0,-1) and +(0,1) ..
	(u2);
\draw (t5)
	node[below left] {$e$}
	.. controls +(0,-1.5) and +(0,0.75) ..
	(u1);
\draw (t6)
	node[below left] {$f$}
	.. controls +(0,-1) and +(0,1) ..
	(u3);

% Free indices from U
\draw (u5)
	.. controls +(0,1.25) and +(0,-1.25) ..
	(f3) node[above] {$g$};
\draw (u6)
	.. controls +(0,1.25) and +(0,-1.25) ..
	(f4) node[above] {$h$};
\end{tikzpicture}

%% file: tikz/TU-contraction-unwind.tikz
\begin{tikzpicture}[
	baseline,
	scale=0.55,
	text height=1.0ex,
	text depth=0.0ex,
	every node/.style = {font = \footnotesize}]

%\draw [help lines] (-10,-6) grid (10,6)
%	(6,6) -- (-6,-6) (6,-6) -- (-6,6)
%	(0,0) circle [radius=2]
%	(0,0) circle [radius=6]
%	(-8,4) circle [radius=2]
%	(-8,-4) circle [radius=2]
%	(8,4) circle [radius=2]
%	(8,-4) circle [radius=2]
%	(-8,0) circle [radius=2]
%	(8,0) circle [radius=2]
%	(-4,0) circle [radius=2]
%	(4,0) circle [radius=2]
%	(0,4) circle [radius=2]
%	(0,-4) circle [radius=2];

% Tensor T
\coordinate (T) at (-1.45,1.75);
\path (T) ++(0.1,0.25) node[style = {font = \normalsize}] {$T$};
\draw [thick] (T) ++(-3.1,-0.75)
	arc [start angle=180, end angle=90, radius=1.5]
	-- ++(1.85,0)
	arc [start angle=135, end angle=45, radius={sqrt(2)}]
	-- ++(0.85,0)
	-- ++(0,-1.5)
	-- cycle;
% Attachment points for indices
\path (T) ++(-2.5,-0.75) coordinate (t1)
	++(1,0) coordinate (t2)
	++(1,0) coordinate (t3)
	++(1,0) coordinate (t4)
	++(1,0) coordinate (t5)
	++(1,0) coordinate (t6);
% Write slot numbers
\draw (t1) node[above] {1}
	(t2) node[above] {2}
	(t3) node[above] {3}
	(t4) node[above] {4}
	(t5) node[above] {5}
	(t6) node[above] {6};
% Highlight stripe for symmetry
\path (t3) ++(-0.1,0.46) coordinate (h1) ++(3.2,0) coordinate (h2);
\begin{scope}[on background layer]
	\draw[color = blue!12!white, line width = 10pt, line cap = round] (h1) -- (h2);
\end{scope}

% Tensor U
\coordinate (U) at (0.85,-1.75);
\path (U) ++(0.3,-0.35) node[style = {font = \normalsize}] (Ulabel) {$U$};
\draw [thick] (U) ++(-3.1,0.75)
	-- ++(0.75,-1.5)
	-- ++(3.85,0)
	arc [start angle = -135, end angle = 45, radius = {0.75*sqrt(2)}]
	-- cycle;
% Attachment points for indices
\path (U) ++(-2.5,0.75) coordinate (u1)
	++(1,0) coordinate (u2)
	++(1,0) coordinate (u3)
	++(1,0) coordinate (u4)
	++(1,0) coordinate (u5)
	++(1,0) coordinate (u6);
% Write slot numbers
\draw (u1) node[below] {7}
	(u2) node[below] {8}
	(u3) node[below] {9}
	(u4) node[below] {10}
	(u5) node[below] {11}
	(u6) node[below] {12};

% Indices (connecting lines)
% Free index attachment points
\coordinate (freeT) at (-4.25,-2);
\path (freeT) coordinate (f1) ++(1,0) coordinate (f2);

\coordinate (freeU) at (2.65,2);
\path (freeU) coordinate (f3) ++(1,0) coordinate (f4);

% Free indices from T
\draw (t1)
	.. controls +(0,-1.25) and +(0,1.25) ..
	(f1) node[below] {$a$};
\draw (t2)
	.. controls +(0,-1.25) and +(0,1.25) ..
	(f2) node[below] {$b$};

% Contracted indices
\draw (t3)
	node[below left] {$e$}
	.. controls +(0,-1) and +(0,1) ..
	(u1);
\draw (t4)
	node[below left] {$d$}
	.. controls +(0,-1) and +(0,1) ..
	(u2);
\draw (t5)
	node[below left] {$f$}
	.. controls +(0,-1) and +(0,1) ..
	(u3);
\draw (t6)
	node[below left] {$c$}
	.. controls +(0,-1) and +(0,1) ..
	(u4);

% Free indices from U
\draw (u5)
	.. controls +(0,1.25) and +(0,-1.25) ..
	(f3) node[above] {$g$};
\draw (u6)
	.. controls +(0,1.25) and +(0,-1.25) ..
	(f4) node[above] {$h$};
\end{tikzpicture}

%% file: tikz/TU-contraction-rename.tikz
\begin{tikzpicture}[
	baseline,
	scale=0.55,
	text height=1.0ex,
	text depth=0.0ex,
	every node/.style = {font = \footnotesize}]

%\draw [help lines] (-10,-6) grid (10,6)
%	(6,6) -- (-6,-6) (6,-6) -- (-6,6)
%	(0,0) circle [radius=2]
%	(0,0) circle [radius=6]
%	(-8,4) circle [radius=2]
%	(-8,-4) circle [radius=2]
%	(8,4) circle [radius=2]
%	(8,-4) circle [radius=2]
%	(-8,0) circle [radius=2]
%	(8,0) circle [radius=2]
%	(-4,0) circle [radius=2]
%	(4,0) circle [radius=2]
%	(0,4) circle [radius=2]
%	(0,-4) circle [radius=2];

% Tensor T
\coordinate (T) at (-1.45,1.75);
\path (T) ++(0.1,0.25) node[style = {font = \normalsize}] {$T$};
\draw [thick] (T) ++(-3.1,-0.75)
	arc [start angle=180, end angle=90, radius=1.5]
	-- ++(1.85,0)
	arc [start angle=135, end angle=45, radius={sqrt(2)}]
	-- ++(0.85,0)
	-- ++(0,-1.5)
	-- cycle;
% Attachment points for indices
\path (T) ++(-2.5,-0.75) coordinate (t1)
	++(1,0) coordinate (t2)
	++(1,0) coordinate (t3)
	++(1,0) coordinate (t4)
	++(1,0) coordinate (t5)
	++(1,0) coordinate (t6);
% Write slot numbers
\draw (t1) node[above] {1}
	(t2) node[above] {2}
	(t3) node[above] {3}
	(t4) node[above] {4}
	(t5) node[above] {5}
	(t6) node[above] {6};
% Highlight stripe for symmetry
\path (t3) ++(-0.1,0.46) coordinate (h1) ++(3.2,0) coordinate (h2);
\begin{scope}[on background layer]
	\draw[color = blue!12!white, line width = 10pt, line cap = round] (h1) -- (h2);
\end{scope}

% Tensor U
\coordinate (U) at (0.85,-1.75);
\path (U) ++(0.3,-0.35) node[style = {font = \normalsize}] (Ulabel) {$U$};
\draw [thick] (U) ++(-3.1,0.75)
	-- ++(0.75,-1.5)
	-- ++(3.85,0)
	arc [start angle = -135, end angle = 45, radius = {0.75*sqrt(2)}]
	-- cycle;
% Attachment points for indices
\path (U) ++(-2.5,0.75) coordinate (u1)
	++(1,0) coordinate (u2)
	++(1,0) coordinate (u3)
	++(1,0) coordinate (u4)
	++(1,0) coordinate (u5)
	++(1,0) coordinate (u6);
% Write slot numbers
\draw (u1) node[below] {7}
	(u2) node[below] {8}
	(u3) node[below] {9}
	(u4) node[below] {10}
	(u5) node[below] {11}
	(u6) node[below] {12};

% Indices (connecting lines)
% Free index attachment points
\coordinate (freeT) at (-4.25,-2);
\path (freeT) coordinate (f1) ++(1,0) coordinate (f2);

\coordinate (freeU) at (2.65,2);
\path (freeU) coordinate (f3) ++(1,0) coordinate (f4);

% Free indices from T
\draw (t1)
	.. controls +(0,-1.25) and +(0,1.25) ..
	(f1) node[below] {$a$};
\draw (t2)
	.. controls +(0,-1.25) and +(0,1.25) ..
	(f2) node[below] {$b$};

% Contracted indices
\draw (t3)
	node[below left] {$c$}
	.. controls +(0,-1) and +(0,1) ..
	(u1);
\draw (t4)
	node[below left] {$d$}
	.. controls +(0,-1) and +(0,1) ..
	(u2);
\draw (t5)
	node[below left] {$e$}
	.. controls +(0,-1) and +(0,1) ..
	(u3);
\draw (t6)
	node[below left] {$f$}
	.. controls +(0,-1) and +(0,1) ..
	(u4);

% Free indices from U
\draw (u5)
	.. controls +(0,1.25) and +(0,-1.25) ..
	(f3) node[above] {$g$};
\draw (u6)
	.. controls +(0,1.25) and +(0,-1.25) ..
	(f4) node[above] {$h$};
\end{tikzpicture}

%% file: tikz/TU-contraction-propagate.tikz
\begin{tikzpicture}[
	baseline,
	scale=0.55,
	text height=1.0ex,
	text depth=0.0ex,
	every node/.style = {font = \footnotesize}]

%\draw [help lines] (-10,-6) grid (10,6)
%	(6,6) -- (-6,-6) (6,-6) -- (-6,6)
%	(0,0) circle [radius=2]
%	(0,0) circle [radius=6]
%	(-8,4) circle [radius=2]
%	(-8,-4) circle [radius=2]
%	(8,4) circle [radius=2]
%	(8,-4) circle [radius=2]
%	(-8,0) circle [radius=2]
%	(8,0) circle [radius=2]
%	(-4,0) circle [radius=2]
%	(4,0) circle [radius=2]
%	(0,4) circle [radius=2]
%	(0,-4) circle [radius=2];

% Tensor T
\coordinate (T) at (-1.45,1.75);
\path (T) ++(0.1,0.2) node[style = {font = \normalsize}] {$T$};
\draw [thick] (T) ++(-3.1,-0.75)
	arc [start angle=180, end angle=90, radius=1.5]
	-- ++(1.85,0)
	arc [start angle=135, end angle=45, radius={sqrt(2)}]
	-- ++(0.85,0)
	-- ++(0,-1.5)
	-- cycle;
% Attachment points for indices
\path (T) ++(-2.5,-0.75) coordinate (t1)
	++(1,0) coordinate (t2)
	++(1,0) coordinate (t3)
	++(1,0) coordinate (t4)
	++(1,0) coordinate (t5)
	++(1,0) coordinate (t6);
% Write slot numbers
\draw (t1) node[above] {1}
	(t2) node[above] {2}
	(t3) node[above] {3}
	(t4) node[above] {4}
	(t5) node[above] {5}
	(t6) node[above] {6};

% Tensor U
\coordinate (U) at (0.85,-1.75);
\path (U) ++(0.3,-0.35) node[style = {font = \normalsize}] (Ulabel) {$U$};
\draw [thick] (U) ++(-3.1,0.75)
	-- ++(0.75,-1.5)
	-- ++(3.85,0)
	arc [start angle = -135, end angle = 45, radius = {0.75*sqrt(2)}]
	-- cycle;
% Attachment points for indices
\path (U) ++(-2.5,0.75) coordinate (u1)
	++(1,0) coordinate (u2)
	++(1,0) coordinate (u3)
	++(1,0) coordinate (u4)
	++(1,0) coordinate (u5)
	++(1,0) coordinate (u6);
% Write slot numbers
\draw (u1) node[below] {7}
	(u2) node[below] {8}
	(u3) node[below] {9}
	(u4) node[below] {10}
	(u5) node[below] {11}
	(u6) node[below] {12};
% Highlight stripe for symmetry
\path (u1) ++(-0.5,0.43) coordinate (h1) ++(3.2,0) coordinate (h2);
\begin{scope}[on background layer]
	\draw[color = red!15!white, line width = 10pt, line cap = round] (h1) -- (h2);
\end{scope}

% Indices (connecting lines)
% Free index attachment points
\coordinate (freeT) at (-4.25,-2);
\path (freeT) coordinate (f1) ++(1,0) coordinate (f2);

\coordinate (freeU) at (2.65,2);
\path (freeU) coordinate (f3) ++(1,0) coordinate (f4);

% Free indices from T
\draw (t1)
	.. controls +(0,-1.25) and +(0,1.25) ..
	(f1) node[below] {$a$};
\draw (t2)
	.. controls +(0,-1.25) and +(0,1.25) ..
	(f2) node[below] {$b$};

% Contracted indices
\draw (t3)
	.. controls +(0,-1) and +(0,2.25) ..
	(u4) node[above left] {$c$};
\draw (t4)
	.. controls +(0,-1) and +(0,1) ..
	(u2) node[above left] {$d$};
\draw (t5)
	.. controls +(0,-1) and +(0,1.5) ..
	(u1) node[above left] {$e$};
\draw (t6)
	.. controls +(0,-1) and +(0,1) ..
	(u3) node[above left] {$f$};

% Free indices from U
\draw (u5)
	.. controls +(0,1.25) and +(0,-1.25) ..
	(f3) node[above] {$g$};
\draw (u6)
	.. controls +(0,1.25) and +(0,-1.25) ..
	(f4) node[above] {$h$};
\end{tikzpicture}

%% file: tikz/TU-contraction-unwind2.tikz
\begin{tikzpicture}[
	baseline,
	scale=0.55,
	text height=1.0ex,
	text depth=0.0ex,
	every node/.style = {font = \footnotesize}]

%\draw [help lines] (-10,-6) grid (10,6)
%	(6,6) -- (-6,-6) (6,-6) -- (-6,6)
%	(0,0) circle [radius=2]
%	(0,0) circle [radius=6]
%	(-8,4) circle [radius=2]
%	(-8,-4) circle [radius=2]
%	(8,4) circle [radius=2]
%	(8,-4) circle [radius=2]
%	(-8,0) circle [radius=2]
%	(8,0) circle [radius=2]
%	(-4,0) circle [radius=2]
%	(4,0) circle [radius=2]
%	(0,4) circle [radius=2]
%	(0,-4) circle [radius=2];

% Tensor T
\coordinate (T) at (-1.45,1.75);
\path (T) ++(0.1,0.2) node[style = {font = \normalsize}] {$T$};
\draw [thick] (T) ++(-3.1,-0.75)
	arc [start angle=180, end angle=90, radius=1.5]
	-- ++(1.85,0)
	arc [start angle=135, end angle=45, radius={sqrt(2)}]
	-- ++(0.85,0)
	-- ++(0,-1.5)
	-- cycle;
% Attachment points for indices
\path (T) ++(-2.5,-0.75) coordinate (t1)
	++(1,0) coordinate (t2)
	++(1,0) coordinate (t3)
	++(1,0) coordinate (t4)
	++(1,0) coordinate (t5)
	++(1,0) coordinate (t6);
% Write slot numbers
\draw (t1) node[above] {1}
	(t2) node[above] {2}
	(t3) node[above] {3}
	(t4) node[above] {4}
	(t5) node[above] {5}
	(t6) node[above] {6};

% Tensor U
\coordinate (U) at (0.85,-1.75);
\path (U) ++(0.3,-0.35) node[style = {font = \normalsize}] (Ulabel) {$U$};
\draw [thick] (U) ++(-3.1,0.75)
	-- ++(0.75,-1.5)
	-- ++(3.85,0)
	arc [start angle = -135, end angle = 45, radius = {0.75*sqrt(2)}]
	-- cycle;
% Attachment points for indices
\path (U) ++(-2.5,0.75) coordinate (u1)
	++(1,0) coordinate (u2)
	++(1,0) coordinate (u3)
	++(1,0) coordinate (u4)
	++(1,0) coordinate (u5)
	++(1,0) coordinate (u6);
% Write slot numbers
\draw (u1) node[below] {7}
	(u2) node[below] {8}
	(u3) node[below] {9}
	(u4) node[below] {10}
	(u5) node[below] {11}
	(u6) node[below] {12};
% Highlight stripe for symmetry
\path (u1) ++(-0.5,0.43) coordinate (h1) ++(3.2,0) coordinate (h2);
\begin{scope}[on background layer]
	\draw[color = red!15!white, line width = 10pt, line cap = round] (h1) -- (h2);
\end{scope}

% Indices (connecting lines)
% Free index attachment points
\coordinate (freeT) at (-4.25,-2);
\path (freeT) coordinate (f1) ++(1,0) coordinate (f2);

\coordinate (freeU) at (2.65,2);
\path (freeU) coordinate (f3) ++(1,0) coordinate (f4);

% Free indices from T
\draw (t1)
	.. controls +(0,-1.25) and +(0,1.25) ..
	(f1) node[below] {$a$};
\draw (t2)
	.. controls +(0,-1.25) and +(0,1.25) ..
	(f2) node[below] {$b$};

% Contracted indices
\draw (t3)
	.. controls +(0,-1) and +(0,1) ..
	(u1) node[above left] {$c$};
\draw (t4)
	.. controls +(0,-1) and +(0,1) ..
	(u2) node[above left] {$d$};
\draw (t5)
	.. controls +(0,-1) and +(0,1) ..
	(u3) node[above left] {$e$};
\draw (t6)
	.. controls +(0,-1) and +(0,1) ..
	(u4) node[above left] {$f$};

% Free indices from U
\draw (u5)
	.. controls +(0,1.25) and +(0,-1.25) ..
	(f3) node[above] {$g$};
\draw (u6)
	.. controls +(0,1.25) and +(0,-1.25) ..
	(f4) node[above] {$h$};
\end{tikzpicture}

%% file: tikz/TU-contraction-propagate-shift.tikz
\begin{tikzpicture}[
	scale=0.55,
	text height=1.0ex,
	text depth=0.0ex,
	every node/.style = {font = \footnotesize}]

%\draw [help lines] (-10,-6) grid (10,6)
%	(6,6) -- (-6,-6) (6,-6) -- (-6,6)
%	(0,0) circle [radius=2]
%	(0,0) circle [radius=6]
%	(-8,4) circle [radius=2]
%	(-8,-4) circle [radius=2]
%	(8,4) circle [radius=2]
%	(8,-4) circle [radius=2]
%	(-8,0) circle [radius=2]
%	(8,0) circle [radius=2]
%	(-4,0) circle [radius=2]
%	(4,0) circle [radius=2]
%	(0,4) circle [radius=2]
%	(0,-4) circle [radius=2];

% Tensor T
\coordinate (T) at (-1.45,1.75);
\path (T) ++(0.1,0.2) node[style = {font = \normalsize}] {$T$};
\draw [thick] (T) ++(-3.25,-0.75)
	arc [start angle=180, end angle=90, radius=1.5]
	-- ++(2,0)
	arc [start angle=135, end angle=45, radius={sqrt(2)}]
	-- ++(1,0)
	-- ++(0,-1.5)
	-- cycle;
% Attachment points for indices
\path (T) ++(-2.5,-0.75) coordinate (t1)
	++(1,0) coordinate (t2)
	++(1,0) coordinate (t3)
	++(1,0) coordinate (t4)
	++(1,0) coordinate (t5)
	++(1,0) coordinate (t6);
% Write slot numbers
\draw (t1) node[above] {1}
	(t2) node[above] {2}
	(t3) node[above] {3}
	(t4) node[above] {4}
	(t5) node[above] {5}
	(t6) node[above] {6};

% Tensor U
\coordinate (U) at (0.85,-1.75);
\path (U) ++(0.3,-0.35) node[style = {font = \normalsize}] (Ulabel) {$U$};
\draw [thick] (U) ++(-3.25,0.75)
	-- ++(1,-1.5)
	-- ++(4,0)
	arc [start angle = -135, end angle = 45, radius = {0.75*sqrt(2)}]
	-- cycle;
% Attachment points for indices
\path (U) ++(-2.5,0.75) coordinate (u1)
	++(1,0) coordinate (u2)
	++(1,0) coordinate (u3)
	++(1,0) coordinate (u4)
	++(1,0) coordinate (u5)
	++(1,0) coordinate (u6);
% Write slot numbers
\draw (u1) node[below] {7}
	(u2) node[below] {8}
	(u3) node[below] {9}
	(u4) node[below] {10}
	(u5) node[below] {11}
	(u6) node[below] {12};

% Highlight stripe for symmetry
\path (u1)
	++(-0.5,0.43) coordinate (h1)
	++(1.2,0) coordinate (h2)
	++(2.75,0) coordinate (h3)
	++(1.2,0) coordinate (h4);
\begin{scope}[on background layer]
	\draw[color = red!15!white, line width = 10pt, line cap = round]
		(h1) -- (h2) (h3) -- (h4);
\end{scope}

% Indices (connecting lines)
% Free index attachment points
\coordinate (freeT) at (-4.3,-2);
\path (freeT) coordinate (f1) ++(1,0) coordinate (f2);

\coordinate (freeU) at (2.8,2);
\path (freeU) coordinate (f3) ++(1,0) coordinate (f4);

% Free indices from T
\draw (t1)
	.. controls +(0,-1.25) and +(0,1.25) ..
	(f1) node[below] {$a$};
\draw (t2)
	.. controls +(0,-1.25) and +(0,1.25) ..
	(f2) node[below] {$b$};

% Contracted indices
\draw (t3)
	.. controls +(0,-1) and +(0,2.6) ..
	(u6) node[above left] {$c$};
\draw (t4)
	.. controls +(0,-1) and +(0,1) ..
	(u2) node[above left] {$d$};
\draw (t5)
	.. controls +(0,-1) and +(0,1.5) ..
	(u1) node[above left] {$e$};
\draw (t6)
	.. controls +(0,-1.25) and +(0,1.5) ..
	(u5) node[above left] {$f$};

% Free indices from U
\draw (u3)
	.. controls +(0,2) and +(0,-2.25) ..
	(f3) node[above] {$g$};
\draw (u4)
	.. controls +(0,2.12) and +(0,-2) ..
	(f4) node[above] {$h$};
\end{tikzpicture}

%% file: tikz/sym-frees.tikz
%\tikzpicturedependsonfile{data/sym-frees-builtin.table}
%\tikzpicturedependsonfile{data/sym-frees-xperm.table}
%\tikzpicturedependsonfile{data/sym-frees-improved.table}
\begin{tikzpicture}
\pgfplotsset{
	compat = {1.3},
	ylabel shift = -10,
	log base 10 number format code/.code = {
		\pgfmathsetmacro{\x}{#1}
		\pgfmathparse{int(and(-3 < \x, \x < 3))}
		\ifnum\pgfmathresult=1
			\pgfmathparse{10^\x}
			$\pgfmathprintnumber[assume math mode=true]{\pgfmathresult} \, \mathrm{s}$
		\else
			$10^{\pgfmathprintnumber[assume math mode=true]{\x}} \, \mathrm{s}$
		\fi
	}
}

\begin{semilogyaxis}[
	title = Free indices with total symmetry,
	xlabel = Number of indices,
	ylabel = Running time,
	legend entries = {\software{Mathematica}, \software{xPerm}, Algorithm~\theimprovedalgorithm},
	legend style = {
		legend pos = outer north east,
		cells = {anchor = west},
		draw = none
	},
	xtick = {1,50,100,...,300},
	each nth point = 3,
	enlarge x limits = 0.05,
	enlarge y limits = 0.05,
	width = 7.5cm
]

\addplot[
	only marks,
	mark = o,
	color = darkgray,
	error bars/y dir = both,
	error bars/y explicit
]
	table [
		x = size,
		y = time,
		y error plus = error+,
		y error minus = error-
	]
	{data/sym-frees-builtin.table};

\addplot[
	only marks,
	mark = triangle*,
	color = red,
	error bars/y dir = both,
	error bars/y explicit
]
	table [
		x = size,
		y = time,
		y error plus = error+,
		y error minus = error-
	]
	{data/sym-frees-xperm.table};

\addplot[
	only marks,
	mark = *,
	mark size = 1.5pt,
	color = blue,
	error bars/y dir = both,
	error bars/y explicit
]
	table [
		x = size,
		y = time,
		y error plus = error+,
		y error minus = error-
	]
	{data/sym-frees-improved.table};

\end{semilogyaxis}
\end{tikzpicture}

%% file: tikz/no-sym-dummies.tikz
%\tikzpicturedependsonfile{data/no-sym-dummies-builtin.table}
%\tikzpicturedependsonfile{data/no-sym-dummies-xperm.table}
%\tikzpicturedependsonfile{data/no-sym-dummies-improved.table}
\begin{tikzpicture}
\pgfplotsset{
	compat = {1.3},
	ylabel shift = -10,
	log base 10 number format code/.code = {
		\pgfmathsetmacro{\x}{#1}
		\pgfmathparse{int(and(-3 < \x, \x < 3))}
		\ifnum\pgfmathresult=1
			\pgfmathparse{10^\x}
			$\pgfmathprintnumber[assume math mode=true]{\pgfmathresult} \, \mathrm{s}$
		\else
			$10^{\pgfmathprintnumber[assume math mode=true]{\x}} \, \mathrm{s}$
		\fi
	}
}

\begin{semilogyaxis}[
	title = No symmetry,
	xlabel = Number of dummy pairs,
	ylabel = Running time,
	xtick = {1,50,100,...,200},
	each nth point = 2,
	enlarge x limits = 0.08,
	enlarge y limits = 0.08,
	width = 7cm
]

\addplot[
	only marks,
	mark = o,
	color = darkgray,
	error bars/y dir = both,
	error bars/y explicit
]
	table [
		x = size,
		y = time,
		y error plus = error+,
		y error minus = error-
	]
	{data/no-sym-dummies-builtin.table};

\addplot[
	only marks,
	mark = triangle*,
	color = red,
	error bars/y dir = both,
	error bars/y explicit
]
	table [
		x = size,
		y = time,
		y error plus = error+,
		y error minus = error-
	]
	{data/no-sym-dummies-xperm.table};

\addplot[
	only marks,
	mark = *,
	mark size = 1.5pt,
	color = blue,
	error bars/y dir = both,
	error bars/y explicit
]
	table [
		x = size,
		y = time,
		y error plus = error+,
		y error minus = error-
	]
	{data/no-sym-dummies-improved.table};

\end{semilogyaxis}
\end{tikzpicture}

%% file: tikz/cyclic-dummies.tikz
%\tikzpicturedependsonfile{data/cyclic-dummies-builtin.table}
%\tikzpicturedependsonfile{data/cyclic-dummies-xperm.table}
%\tikzpicturedependsonfile{data/cyclic-dummies-improved.table}
\begin{tikzpicture}
\pgfplotsset{
	compat = {1.3},
	ylabel shift = -10,
	log base 10 number format code/.code = {
		\pgfmathsetmacro{\x}{#1}
		\pgfmathparse{int(and(-3 < \x, \x < 3))}
		\ifnum\pgfmathresult=1
			\pgfmathparse{10^\x}
			$\pgfmathprintnumber[assume math mode=true]{\pgfmathresult} \, \mathrm{s}$
		\else
			$10^{\pgfmathprintnumber[assume math mode=true]{\x}} \, \mathrm{s}$
		\fi
	}
}

\begin{semilogyaxis}[
	title = Cyclic symmetry,
	xlabel = Number of dummy pairs,
	%ylabel = Running time,
	xtick = {1,50,100,...,200},
	ytickten = {-5,...,1},
	ymax = 1.5,
	each nth point = 2,
	enlarge x limits = 0.08,
	enlarge y limits = 0.08,
	width = 7cm
]

\addplot[
	only marks,
	mark = o,
	color = darkgray,
	error bars/y dir = both,
	error bars/y explicit
]
	table [
		x = size,
		y = time,
		y error plus = error+,
		y error minus = error-
	]
	{data/cyclic-dummies-builtin.table};

\addplot[
	only marks,
	mark = triangle*,
	color = red,
	error bars/y dir = both,
	error bars/y explicit
]
	table [
		x = size,
		y = time,
		y error plus = error+,
		y error minus = error-
	]
	{data/cyclic-dummies-xperm.table};

\addplot[
	only marks,
	mark = *,
	mark size = 1.5pt,
	color = blue,
	error bars/y dir = both,
	error bars/y explicit
]
	table [
		x = size,
		y = time,
		y error plus = error+,
		y error minus = error-
	]
	{data/cyclic-dummies-improved.table};

\end{semilogyaxis}
\end{tikzpicture}

%% file: tikz/legend.tikz
\begin{tikzpicture}

\node at (0,0) [darkgray, label={[black, label distance = -3pt]right:\software{Mathematica}}] {\pgfuseplotmark{o}};
\node at (3.6,0) [red, label={[black, label distance = -3pt]right:\software{xPerm}}] {\pgfuseplotmark{triangle*}};
\node at (6,0) [blue, mark size = 1.5pt, label={[black, label distance = -3pt]right:Algorithm~\theimprovedalgorithm}] {\pgfuseplotmark{*}};

\end{tikzpicture}

%% file: tikz/nonzero-riemanns.tikz
\begin{tikzpicture}
%\tikzpicturedependsonfile{data/nonzero-riemanns-builtin.table}
%\tikzpicturedependsonfile{data/nonzero-riemanns-xperm.table}
%\tikzpicturedependsonfile{data/nonzero-riemanns-improved.table}

\pgfplotsset{
	compat = {1.3},
	ylabel shift = -10,
	log base 10 number format code/.code = {
		\pgfmathsetmacro{\x}{#1}
		\pgfmathparse{int(and(-3 < \x, \x < 3))}
		\ifnum\pgfmathresult=1
			\pgfmathparse{10^\x}
			$\pgfmathprintnumber[assume math mode=true]{\pgfmathresult} \, \mathrm{s}$
		\else
			$10^{\pgfmathprintnumber[assume math mode=true]{\x}} \, \mathrm{s}$
		\fi
	}
}

\begin{semilogyaxis}[
	title = Nonzero results,
	xlabel = Number of Riemann tensors,
	ylabel = Running time,
	xtick = {1,10,20,...,50},
	ytickten = {-5,...,2},
	ymax = 12,
	each nth point = 2,
	enlarge x limits = 0.08,
	enlarge y limits = 0.08,
	width = 7cm
]

\addplot[
	only marks,
	mark = o,
	color = darkgray,
	error bars/y dir = both,
	error bars/y explicit
]
	table [
		x = size,
		y = time,
		y error plus = error+,
		y error minus = error-
	]
	{data/nonzero-riemanns-builtin.table};

\addplot[
	only marks,
	mark = triangle*,
	color = red,
	error bars/y dir = both,
	error bars/y explicit
]
	table [
		x = size,
		y = time,
		y error plus = error+,
		y error minus = error-
	]
	{data/nonzero-riemanns-xperm.table};

\addplot[
	only marks,
	mark = *,
	mark size = 1.5pt,
	color = blue,
	error bars/y dir = both,
	error bars/y explicit
]
	table [
		x = size,
		y = time,
		y error plus = error+,
		y error minus = error-
	]
	{data/nonzero-riemanns-improved.table};

\end{semilogyaxis}
\end{tikzpicture}

%% file: tikz/zero-riemanns.tikz
\begin{tikzpicture}
%\tikzpicturedependsonfile{data/zero-riemanns-builtin.table}
%\tikzpicturedependsonfile{data/zero-riemanns-xperm.table}
%\tikzpicturedependsonfile{data/zero-riemanns-improved.table}

\pgfplotsset{
	compat = {1.3},
	ylabel shift = -10,
	log base 10 number format code/.code = {
		\pgfmathsetmacro{\x}{#1}
		\pgfmathparse{int(and(-3 < \x, \x < 3))}
		\ifnum\pgfmathresult=1
			\pgfmathparse{10^\x}
			$\pgfmathprintnumber[assume math mode=true]{\pgfmathresult} \, \mathrm{s}$
		\else
			$10^{\pgfmathprintnumber[assume math mode=true]{\x}} \, \mathrm{s}$
		\fi
	}
}

\begin{semilogyaxis}[
	title = Zero results,
	xlabel = Number of Riemann tensors,
	%ylabel = Running time,
	xtick = {1,10,20,...,50},
	each nth point = 2,
	enlarge x limits = 0.08,
	enlarge y limits = 0.08,
	width = 7cm
]

\addplot[
	only marks,
	mark = o,
	color = darkgray,
	error bars/y dir = both,
	error bars/y explicit
]
	table [
		x = size,
		y = time,
		y error plus = error+,
		y error minus = error-
	]
	{data/zero-riemanns-builtin.table};

\addplot[
	only marks,
	mark = triangle*,
	color = red,
	error bars/y dir = both,
	error bars/y explicit
]
	table [
		x = size,
		y = time,
		y error plus = error+,
		y error minus = error-
	]
	{data/zero-riemanns-xperm.table};

\addplot[
	only marks,
	mark = *,
	mark size = 1.5pt,
	color = blue,
	error bars/y dir = both,
	error bars/y explicit
]
	table [
		x = size,
		y = time,
		y error plus = error+,
		y error minus = error-
	]
	{data/zero-riemanns-improved.table};

\end{semilogyaxis}
\end{tikzpicture}

%% file: tikz/total-sym-dummies.tikz
%\tikzpicturedependsonfile{data/total-sym-dummies-builtin.table}
%\tikzpicturedependsonfile{data/total-sym-dummies-xperm.table}
%\tikzpicturedependsonfile{data/total-sym-dummies-improved.table}
\begin{tikzpicture}
\pgfplotsset{
	compat = {1.3},
	ylabel shift = -10,
	log base 10 number format code/.code = {
		\pgfmathsetmacro{\x}{#1}
		\pgfmathparse{int(and(-3 < \x, \x < 3))}
		\ifnum\pgfmathresult=1
			\pgfmathparse{10^\x}
			$\pgfmathprintnumber[assume math mode=true]{\pgfmathresult} \, \mathrm{s}$
		\else
			$10^{\pgfmathprintnumber[assume math mode=true]{\x}} \, \mathrm{s}$
		\fi
	}
}

\begin{semilogyaxis}[
	title = Frustrated,
	xlabel = Number of dummy pairs,
	ylabel = Running time,
	xtick = {1,5,10,...,30},
	each nth point = 1,
	enlarge x limits = 0.05,
	enlarge y limits = 0.05,
	width = 7cm
]

\addplot[
	only marks,
	mark = o,
	color = darkgray,
	error bars/y dir = both,
	error bars/y explicit
]
	table [
		x = size,
		y = time,
		y error plus = error+,
		y error minus = error-
	]
	{data/total-sym-dummies-builtin.table};

\addplot[
	only marks,
	mark = triangle*,
	color = red,
	error bars/y dir = both,
	error bars/y explicit
]
	table [
		x = size,
		y = time,
		y error plus = error+,
		y error minus = error-
	]
	{data/total-sym-dummies-xperm.table};

\addplot[
	only marks,
	mark = *,
	mark size = 1.5pt,
	color = blue,
	error bars/y dir = both,
	error bars/y explicit
]
	table [
		x = size,
		y = time,
		y error plus = error+,
		y error minus = error-
	]
	{data/total-sym-dummies-improved.table};

\end{semilogyaxis}
\end{tikzpicture}

%% file: tikz/random-total-sym-dummies.tikz
%\tikzpicturedependsonfile{data/random-total-sym-dummies-builtin.table}
%\tikzpicturedependsonfile{data/random-total-sym-dummies-xperm.table}
%\tikzpicturedependsonfile{data/random-total-sym-dummies-improved.table}
\begin{tikzpicture}
\pgfplotsset{
	compat = {1.3},
	ylabel shift = -10,
	log base 10 number format code/.code = {
		\pgfmathsetmacro{\x}{#1}
		\pgfmathparse{int(and(-3 < \x, \x < 3))}
		\ifnum\pgfmathresult=1
			\pgfmathparse{10^\x}
			$\pgfmathprintnumber[assume math mode=true]{\pgfmathresult} \, \mathrm{s}$
		\else
			$10^{\pgfmathprintnumber[assume math mode=true]{\x}} \, \mathrm{s}$
		\fi
	}
}

\begin{semilogyaxis}[
	title = Randomized,
	xlabel = Number of dummy pairs,
	%ylabel = Running time,
	xtick = {1,5,10,...,30},
	each nth point = 1,
	enlarge x limits = 0.05,
	enlarge y limits = 0.05,
	width = 7cm
]

\addplot[
	only marks,
	mark = o,
	color = darkgray,
	error bars/y dir = both,
	error bars/y explicit
]
	table [
		x = size,
		y = time,
		y error plus = error+,
		y error minus = error-
	]
	{data/random-total-sym-dummies-builtin.table};

\addplot[
	only marks,
	mark = triangle*,
	color = red,
	error bars/y dir = both,
	error bars/y explicit
]
	table [
		x = size,
		y = time,
		y error plus = error+,
		y error minus = error-
	]
	{data/random-total-sym-dummies-xperm.table};

\addplot[
	only marks,
	mark = *,
	mark size = 1.5pt,
	color = blue,
	error bars/y dir = both,
	error bars/y explicit
]
	table [
		x = size,
		y = time,
		y error plus = error+,
		y error minus = error-
	]
	{data/random-total-sym-dummies-improved.table};

\end{semilogyaxis}
\end{tikzpicture}

%% file: tikz/pairwise-sym-dummies.tikz
%\tikzpicturedependsonfile{data/pairwise-sym-dummies-builtin.table}
%\tikzpicturedependsonfile{data/pairwise-sym-dummies-xperm.table}
%\tikzpicturedependsonfile{data/pairwise-sym-dummies-improved.table}
\begin{tikzpicture}
\pgfplotsset{
	compat = {1.3},
	ylabel shift = -10,
	log base 10 number format code/.code = {
		\pgfmathsetmacro{\x}{#1}
		\pgfmathparse{int(and(-3 < \x, \x < 3))}
		\ifnum\pgfmathresult=1
			\pgfmathparse{10^\x}
			$\pgfmathprintnumber[assume math mode=true]{\pgfmathresult} \, \mathrm{s}$
		\else
			$10^{\pgfmathprintnumber[assume math mode=true]{\x}} \, \mathrm{s}$
		\fi
	}
}

\begin{semilogyaxis}[
	title = Frustrated,
	xlabel = Total number of indices,
	ylabel = Running time,
	xtick = {4,12,...,36},
	each nth point = 1,
	enlarge x limits = 0.08,
	enlarge y limits = 0.08,
	width = 7cm
]

\addplot[
	only marks,
	mark = o,
	color = darkgray,
	error bars/y dir = both,
	error bars/y explicit
]
	table [
		x = size,
		y = time,
		y error plus = error+,
		y error minus = error-
	]
	{data/pairwise-sym-dummies-builtin.table};

\addplot[
	only marks,
	mark = triangle*,
	color = red,
	error bars/y dir = both,
	error bars/y explicit
]
	table [
		x = size,
		y = time,
		y error plus = error+,
		y error minus = error-
	]
	{data/pairwise-sym-dummies-xperm.table};

\addplot[
	only marks,
	mark = *,
	mark size = 1.5pt,
	color = blue,
	error bars/y dir = both,
	error bars/y explicit
]
	table [
		x = size,
		y = time,
		y error plus = error+,
		y error minus = error-
	]
	{data/pairwise-sym-dummies-improved.table};

\end{semilogyaxis}
\end{tikzpicture}

%% file: tikz/random-pairwise-sym-dummies.tikz
%\tikzpicturedependsonfile{data/random-pairwise-sym-dummiess-builtin.table}
%\tikzpicturedependsonfile{data/random-pairwise-sym-dummies-xperm.table}
%\tikzpicturedependsonfile{data/random-pairwise-sym-dummies-improved.table}
\begin{tikzpicture}
\pgfplotsset{
	compat = {1.3},
	ylabel shift = -10,
	log base 10 number format code/.code = {
		\pgfmathsetmacro{\x}{#1}
		\pgfmathparse{int(and(-3 < \x, \x < 3))}
		\ifnum\pgfmathresult=1
			\pgfmathparse{10^\x}
			$\pgfmathprintnumber[assume math mode=true]{\pgfmathresult} \, \mathrm{s}$
		\else
			$10^{\pgfmathprintnumber[assume math mode=true]{\x}} \, \mathrm{s}$
		\fi
	}
}

\begin{semilogyaxis}[
	title = Randomized,
	xlabel = Total number of indices,
	%ylabel = Running time,
	xtick = {4,20,40,...,60},
	each nth point = 1,
	enlarge x limits = 0.08,
	enlarge y limits = 0.08,
	width = 7cm
]

\addplot[
	only marks,
	mark = o,
	color = darkgray,
	error bars/y dir = both,
	error bars/y explicit
]
	table [
		x = size,
		y = time,
		y error plus = error+,
		y error minus = error-
	]
	{data/random-pairwise-sym-dummies-builtin.table};

\addplot[
	only marks,
	mark = triangle*,
	color = red,
	error bars/y dir = both,
	error bars/y explicit
]
	table [
		x = size,
		y = time,
		y error plus = error+,
		y error minus = error-
	]
	{data/random-pairwise-sym-dummies-xperm.table};

\addplot[
	only marks,
	mark = *,
	mark size = 1.5pt,
	color = blue,
	error bars/y dir = both,
	error bars/y explicit
]
	table [
		x = size,
		y = time,
		y error plus = error+,
		y error minus = error-
	]
	{data/random-pairwise-sym-dummies-improved.table};

\end{semilogyaxis}
\end{tikzpicture}